%% file: crrr-arxiv.tex
\documentclass[12pt,reqno]{amsart}

\usepackage{tgpagella}
\usepackage{amsmath,amssymb,amsthm}
\usepackage{graphicx,epsfig,epstopdf}
\usepackage{url,mathtools}
\usepackage{colortbl,framed}
\usepackage{enumitem}
\usepackage[authoryear]{natbib}
\usepackage{colortbl}

\usepackage{subcaption}
\usepackage{caption}
\usepackage{booktabs}
\usepackage{threeparttable}
\usepackage{centernot}

\usepackage{accents}

\renewcommand{\qed}{\hfill{\tiny \ensuremath{\blacksquare} }}%

\newcommand{\R}{\mathbb{R}}

\newcommand{\RR}{\mathcal{R}}

\newcommand{\mW}{\mathcal{W}}

\newcommand{\mX}{\mathcal{X}}

\newcommand{\dd}{\mathrm{d}}

\newcommand{\indep}{\mathop{\perp\!\!\!\!\perp}}
\newcommand{\notindep}{\not\perp\!\!\!\!\perp}

\newcommand{\ci}{\perp\!\!\!\perp}
\renewcommand{\qed}{\hfill {\tiny {\ensuremath{\blacksquare}}}}

\newtheorem{theorem}{Theorem}[section]

\newtheorem{lemma}{Lemma}[section]

\newtheorem{assumption}{Assumption}[section]

\theoremstyle{definition}

\newtheorem{algorithm}{Algorithm}

\newtheorem{remark}{Comment}[section]
\numberwithin{remark}{section}

\numberwithin{equation}{section}
\numberwithin{theorem}{section}

\newcommand{\eps}{\epsilon}

\usepackage{url}
\newcommand{\citen}{\cite}

\usepackage{pagecolor,lipsum}% http://ctan.org/pkg/{pagecolor,lipsum}

\newcommand{\Ep}{{\mathrm{E}}}

\renewcommand{\Pr}{{\mathrm{P}}}

\renewcommand{\hat}{\widehat}

\renewcommand{\Pr}{{\mathrm{P}}}

\renewcommand{\hat}{\widehat}
\renewcommand{\leq}{\leqslant}
\renewcommand{\geq}{\geqslant}

\DeclareMathOperator{\argmax}{arg max}

\DeclareMathOperator{\Corr}{Cor}
\DeclareMathOperator{\Cov}{Cov}
\DeclareMathOperator{\Var}{Var}



\usepackage{wallpaper}
\usepackage{graphicx}

\usepackage{xr}
\begin{document}
%\pagecolor{blue!1!white}

\title[CRRR]{Conditional Rank-Rank Regression$^*$}\thanks{$^*$ We thank Andrew Chesher, Toru Kitagawa, Patrick Kline, Michal Kolesar, Essie Maasoumi, Konrad Menzel, Ulrich Muller, Daniele Paserman and seminar participants at Alicante, Brandeis, Brown,  BU, Cambridge, Cemfi, CUHK-SZ, Emory, Georgetown, Groningen, IAAE 2025,  International Panel Data Conference 2025, IESR, Kansas Econometrics Workshop, NYU, Princeton, UCL, Warwick and Oxford Workshop on Recent Advances in Panel and Network Data  for helpful comments, and Matt Hong, Stella Hong, ChatGPT, Claude and Refine for able research assistance.}
\author[Chernozhukov, Fern\'andez-Val, Meier, van Vuuren \and Vella]{Victor Chernozhukov \and Iv\'an Fern\'andez-Val \and Jonas Meier \and Aico van Vuuren \and Francis Vella$^\dag$}\thanks{$^\dag$ Chernozhukov: Department of Economics and Center for Statistics and Data Science, MIT; Fern\'andez-Val: Department of Economics, Boston University; Meier: Swiss National Bank; van Vuuren: Faculty of Economics and Business, University of Groningen; Vella: Department of Economics, Georgetown University} 
%\pubMonth{Month}
\date{\today}
%\pubYear{Year}
%\pubVolume{Vol}
%\pubIssue{Issue}

\begin{abstract}

Rank-rank regression is commonly employed in economic research as a way of capturing the relationship between two economic variables. The slope of this regression is the Spearman rank correlation, a classical measure of association. However, in many applications it is common practice to include covariates to account for differences in association levels between groups as defined by the values of these covariates. This is either done by including the 
covariates or by modeling the residuals 
obtained after partialing out the impact of the covariates.
In each of these instances the resulting rank-rank regression coefficients can be difficult to interpret. We propose the conditional rank-rank regression, which uses conditional ranks instead of unconditional ranks, to measure average within-group  persistence. The coefficient of this new regression corresponds to the average Spearman rank correlation conditional on the covariates, a natural summary measure of within-group association.
We develop a flexible estimation approach using distribution regression and establish a theoretical framework for large sample inference. An empirical study on intergenerational income mobility in Switzerland demonstrates the advantages of this approach. The study reveals stronger intergenerational persistence between fathers and sons compared to fathers and daughters, with the within-group persistence being 62\% of the overall income persistence for sons and 52\% for daughters. Smaller families and those with  highly educated fathers exhibit greater persistence in economic status.

\end{abstract}
\maketitle

% \begin{footnotesize}
% \textbf{Abstract.}  
% Empirical investigations of intergenerational mobility typically regress an individual's
% location in their generation's income distribution on their parent's rank in their generation's income distribution to estimate the intergenerational association in ranks captured as the slope coefficient. While this estimate is straightforward to interpret in the bivariate setting it is frequently necessary to account for additional influences via the inclusion of additional covariates. We show that including covariates in these rank-rank regressions can lead to incorrect conclusions regarding the level of intergenerational mobility. We show that the bias in rank rank regressions with covariates can be overcome via the use of conditional ranks. We introduce the conditional rank-rank regression and proposes an estimator based on distribution regression. We provide the associated  theory and simulation evidence which illustrates the utility of our approach. We conlcude with an empirical example based on Swiss data.  
% \end{footnotesize}

\bigskip

\newpage

\section{Introduction}

The linear regression of the rank of a variable $Y$ on the rank of another variable $W$ is commonly referred to as a rank-rank regression (RRR). 
It 
%This type of regression 
has become increasingly popular in empirical investigations in economics to analyze policy relevant issues such as, for example,  mobility and sorting behavior \citep[e.g., ][]{beller2006intergenerational,dahl2008association,chetty2014land,adermon2018intergenerational,murphy2020top}.\footnote{\cite{chetverikov2025inferencerankrankregressions} have recently documented that 40 articles published between January 2013 and February 2024 in \textit{American Economic Review},
\textit{Journal of Political Economy}, \textit{Quarterly Journal of Economics}, and \textit{Review of Economic Studies} employ RRRs in their empirical analysis.} A desirable feature of RRR is that its slope coefficient corresponds to the Spearman rank correlation coefficient between $Y$ and $W$, a classical measure of association with desirable properties such as invariance to monotone transformations of the variables and robustness to outliers and heavy tails \citep{spearman1904proof,kendall1948rank}.\footnote{\cite{maasoumi2022generalized} questioned the economic interpretation of the RRR as a measure of mobility due to the use of linear regression and proposed alternative measures based on nonparametric regression.}

We propose a general method to control for covariates $X$ in RRRs. Covariates are commonly accounted for in RRRs either via their inclusion  as additional regressors (RRRX), or by conducting RRR on residuals after partialing out their influence (RRR.res). We show that these two approaches can have undesirable properties. In particular, their resulting coefficients can be difficult to interpret and produce counterintuitive results. For example, \cite{chetverikov2025inferencerankrankregressions} noted that the slope of RRRX is no longer the Spearman correlation and might lie outside the interval $[-1,1]$. Moreover, the ranks of the residuals in RRR.res can only be related  to meaningful ranks of the original variables under restrictive assumptions on the distribution of the variables of interest conditional on the covariates.\footnote{For example, the ranks of the residuals of the linear regression of $Y$ on $X$  correspond to the ranks of $Y$ conditional on $X$ under the location-shift model $Y = X'\beta + \varepsilon$ where $\varepsilon$ is independent of $X$. However this relationship generally does not hold otherwise.} 
We call our proposal conditional rank-rank regression (CRRR) because it employs conditional ranks. Intuitively, we replace the linear partialing out of the covariates implicit or explicit in the existing approaches by a nonlinear partialing out. We show that our approach always delivers coefficients that are easy to interpret and which do not have the undesirable properties of the existing approaches noted above.

%CRRR 
%is flexible in that it 
%can be adapted to different empirical settings. 
In its canonical form, CRRR regresses the rank of $Y$ conditional on $X$ on the rank of $W$ conditional on $X$. It provides an alternative to RRRX and RRR.res and produces estimates that are easy to interpret.  Indeed, we show that the CRRR slope is equal to the Spearman correlation of $Y$ and $W$ conditional on $X$, averaged over the distribution of $X$, which is a natural summary measure of within-group association with similar properties to the Spearman rank correlation. In Statistics, this type of  measure of association is so-called \textit{covariate-adjusted} and it has been frequently used in biostatistic applications \citep[e.g.,][]{gijbels2011conditional,liu2018covariate,eden2022nonparametric,wei2023partial}. 
%For example, Liu et al. (2018) investigate the correlation of biomarkers that indicate risks for cardiovascular diseases for a group of HIV-positive patients. These biomarkers may be only correlated to each other because of some underlying patient characteristics such as age, sex, race or body mass index, making it necessary to correct for these characteristics. Another example that they investigate is the correlation between subjective answers to livelihood such as the quality of life and nutrition from a set of female respondents in Mozambique. Again, it is possible that such indicators of livelihood are only correlated because of underlying demographic characteristics such as age. 
Another attractive feature of CRRR is that it is suitable for subgroup analysis,  i.e. to run separate CRRR by groups defined by values of a categorical variable. If this variable is included in $X$, the CRRR slope for each group has the interpretation of the average conditional Spearman rank correlation for that group.
%and thefore lies in the interval $[-1,1]$. 

CRRR is also attractive for economic applications. For example, \cite{hagedorn2017identifying} used ranks of residuals of wages on worker and establishment characteristics to analyze labor market sorting. However, ranks based on residuals are statistical constructions that, in general, do not have a natural interpretation in terms of the original wages. Our proposal is to employ conditional wage ranks, which reflect wage ranks within groups of workers with the same characteristics, and are accordingly closely related to the observed wages. CRRR can also be applied to the analysis of sorting in other matching settings such as marriage and worker-firm markets \citep[e.g., ][]{haltiwanger2015cyclical,fagereng2022assortative,haner2024marry}.

%Another example is the Adermon et al. (2018) investigation of the role of grandparents' wealth in the rank-rank correlation of wealth between father and son. That paper examines whether the coefficient of father's wealth in the RRR of son's wealth on father's wealth is driven by some unobservable passed across generations.  To address this, they include the rank of the grandfather's wealth via RRRX. Including this variable in RRRX is associated with the challenges regarding how it should be incorporated and the interpretation of the resulting coefficients. CRRR provides the Spearman rank accounting for the impact of grandfather's wealth and the coefficients from CRRR and RRR are comparable.[IFV: I DO NOT FOLLOW THE POINT OF THIS EXAMPLE. I DOES NOT LOOK CLEAN TO ME]

The interpretation of the CRRR slope is different from the RRR slope. Assume, for example, in an investigation of sorting behavior of married couples that $Y$ is husband's wage, $W$ is wife's wage and $X$ is number of children. The RRR slope without covariates is the correlation between husband's and wife's wage ranks where the ranks are relative to the entire wage distribution.  The CRRR slope is the rank correlation where the ranks are relative to the wage distribution of those households with the same number of children. The slope of RRRX is the regression slope of the husband's wage rank on wife's wage rank, where the ranks are relative to the entire wage distribution and the wife's rank is centered to have the same mean within households with the same number of children.
%\textcolor{red}{For example, in Adermon setting et al. (2018), $Y$ is son's wealth, $W$ is father's wealth and $X$ is grandfather's wealth. The RRR slope without covariates is the correlation between son's and father's wealth ranks where the ranks are relative to the entire wealth distribution.  The CRRR slope is the rank correlation where the ranks are relative to the wealth distribution of those son/father pairs with the same level of grandfather wealth.The slope of RRRX is the regression slope of the son's wage wealth rank on father's wealth rank, where the ranks are relative to the entire wealth distribution and the father's rank is centered to have the same mean within pairs with the same level of grandfather wealth.}
%The slope of RRRX is the regression slope of the husband's wage rank on wife's wage rank, where the ranks are relative to the entire wage distribution and the wife's rank is centered to have the same mean within households with the same number of children. 
The slope of this regression might be difficult to interpret as the centered rank is no longer a rank.   We believe that in many settings the CRRR slope better reflects the relationship researchers are intending to capture when they control for covariates as it is closer to the ceteris paribus effect that economic models typically predict. Mathematically, the difference between CRRR and RRRX is the order in the application of the rank and covariate partialing out operators. RRRX first obtains ranks and then partials out the covariates, whereas CRRR reverses the order. The final outcome differs across procedures as the two operators do not commute due to the nonlinearity of the rank operator.

CRRR is different from RRR because they have different dependent variables, conditional versus unconditional (marginal) ranks. In some instances the explicit objective is to predict the marginal rank of $Y$ using $W$ and $X$. For example, in intergenerational mobility applications one might wish to predict  the child's income rank using father's income and family household characteristics such as father's immigration status \citep[e.g., ][]{abramitzky2021intergenerational}. Here, the marginal rank of $Y$ (child's rank in the entire distribution) might be more interesting than the rank of $Y$ conditional on $X$ (child's rank among those who have the same father's immigration status indicator). The commonly employed  approach in these instances is to include $X$ as a covariate in the RRR or to perform subgroup analysis for different values of $X$. We refer to both approaches as RRRX as the subgroup analysis can also be implemented using RRR by including suitable interactions. As noted above,  \cite{chetverikov2025inferencerankrankregressions} observed that the coefficient of the rank of $W$ in RRRX might be difficult to interpret. In Section \ref{sec:example} we show through an example  that both forms of RRRX can deliver counterintuitive economic results in terms of mobility and predict ranks of $Y$ outside $[0,1]$.\footnote{Another way to avoid this problem is to use a fractional response regression to model the ranks of $Y$ conditional on the ranks of $W$ and $X$ %, instead of the linear regression model
\citep[e.g., ][]{papke1996econometric}. However, this approach requires parametric assumptions and the model coefficients lack a natural interpretation beyond their signs.} The underlying cause of these problems is that the rank of $W$ on the right hand side of the RRRX does not satisfy the properties of a rank after partialing out the effect of the covariates.

A variant of CRRR (CRRR') can be used in these situations. CRRR' regresses the marginal rank of $Y$ on the rank of $W$ conditional on $X$ and possibly $X$. Similar to CRRR, CRRR' bypasses the conceptual problems associated with RRRX. We show that the CRRR' slope corresponds to a correlation and therefore always lies between $-1$ and $1$. The square of this slope indeed corresponds to the fraction of the variance of the marginal rank of $Y$ explained by the conditional rank of $W$, net of the impact of the covariates $X$. This measure is useful for performing variance decompositions. CRRR' is also suitable for subgroup analysis. The CRRR' slope indeed can be decomposed as the average of the CRRR' slopes in each group, weighted by the size of the group. Moreover,   CRRR'  always predicts ranks in $[0,1]$ when $X$ is not included in the regression. RRRX does not satisfy any of the previous properties. This is illustrated in the example presented in Section \ref{sec:example}.
%,  and is suitable for between-within group decompositions of the variance of the marginal ranks, unlike RRRX.  
There are also more nuanced situations where we have two groups of covariates, $X=(X_1,X_2)$, and we are interested in the ranks of $Y$ conditional on $X_1$, but want to use $X_2$ to improve prediction or for some other purposes. For example, in an intergenerational mobility application, we might be interested in predicting regional child's income ranks using father's income and household characteristics such as father's immigration status. Our approach can also be applied to these settings. Here we would regress the ranks of $Y$ conditional on $X_1$ (region) on the ranks of $W$ conditional on $X$  (region and father's immigration status) and possibly $X$.

    We provide an estimator of the CRRR coefficients based on distribution regression (DR). Here we do not distinguish between the different variations of CRRR because they all can be treated jointly by allowing different conditioning covariates or equivalently restricting some coefficients of the DR to be zero. Like the estimator of the RRR coefficients, our estimator consists of two steps. In the RRR, the first step estimates the marginal ranks of $Y$ and $W$ using the empirical distribution, and the second step runs the linear regression of the estimated ranks of $Y$ on the estimated ranks of $W$ or computes the sample correlation between these ranks. Both versions of the second step produce numerically identical results if there are no ties in the observed values of $Y$ and $W$. In CRRR, the first step estimates the conditional ranks by running logit or probit DRs of $Y$ on $X$ and $W$ on $X$ at multiple values of $Y$ and $W$ to trace the entire conditional distributions. The second steps are identical to RRR, but the linear regression and  correlation versions are no longer numerically identical even if there are no ties. 
    They are both, however, consistent estimators of the CRRR slope. 
    The  CRRR estimator is computationally tractable, albeit somewhat more demanding than RRR.
    %The  CRRR estimator is computationally more demanding than the RRR estimator, but it is still highly tractable. The second step is trivial in both cases. The first step of the CRRR estimator is more intensive than the RRR estimator as it involves numerical optimization, but all the DR programs are convex and the estimation algorithm can be run in parallel at the different values of $Y$ and $W$.
%To illustrate the utility of CRRR, and to provide a means to illustrate empirical and simulation evidence, we employ intergenerational mobility as our working example. We do so as the Spearman correlation is often an object of interest in this literature (for a recent example see Kenedi and Sirugue 2023). %although we stress many empirical questions are appropriately addressed by RRRX estimates. For example,
%\footnote{Deutscher and Mazumder (2023) note that Spearman rank correlation is frequently employed to describe "global" intergenerational mobility, where "global" reflects the relationship between $Y$ and $W$ over the entire distribution of $W$.}
%the investigations of Chetty et al. (2014) are appropriately addressed by RRRX

    We derive the asymptotic distribution of the CRRR estimator and provide feasible inference theory. \cite{chetverikov2025inferencerankrankregressions} noted that standard inference methods for linear regression do not apply to the RRR estimator because both the independent and dependent variables, namely the estimated ranks, are generated. The same problem applies to CRRR.  The theory for the RRR estimator was derived using U-statistic theory \citep{hoeffding1948class} or the delta method \citep{ren1995hadamard} when $Y$ and $W$ are continuous. We employ the functional delta method approach to derive the theory because the CRRR estimator does not have a U-statistic structure. The application of the functional delta method to the CRRR estimator presents several differences with respect to RRR. The first ingredient of both approaches consists of writing the parameter of interest as a functional of inputs: the joint distribution of $Y$ and $W$ for RRR, or the conditional distributions of $Y$ and $W$ conditional on $X$ and the joint distribution of $Y$, $W$ and $X$, for CRRR. The CRRR functional is more complicated than the RRR functional and the inputs for CRRR live in more complex spaces than those for RRR. As a result, the Hadamard differentiability of the RRR functional established by \cite{ren1995hadamard} does not cover the CRRR functional. We establish the Hadamard differentiability of the CRRR functional in the relevant spaces.

    The second ingredient required for the application of the functional delta method is to establish functional central limit theorems for the estimators of the inputs. For RRR, this follows from the now classical large sample theory of the empirical distribution function. A challenge for CRRR is that existing theory for DR estimators of conditional distributions excludes the tails. In particular, the available functional central limit theorems only hold for compact strict subsets of the support. We deal with this problem by imposing assumptions on the tail behavior of the DR model that allows us to estimate the conditional distribution in the tails. We then obtain functional central limit theorems for DR estimators of conditional distributions over the entire support. 
    
    Combining these two ingredients we establish that the CRRR estimator follows a normal distribution around the CRRR slope in large samples via the delta method. The asymptotic variance has a complicated expression that might be difficult to estimate analytically. We develop the use of the exchangeable bootstrap to obtain standard errors and construct confidence intervals. Exchangeable bootstrap includes the most common forms of bootstrap such as empirical, weighted and subsampling bootstrap as special cases. We establish its validity in large samples from the functional delta method for the bootstrap.  Like \cite{hoeffding1948class}  and \cite{ren1995hadamard}, our theory covers the case where $Y$ and $W$ are continuous. The theory can be extended to noncontinuous variables following the analysis of \cite{chetverikov2025inferencerankrankregressions} for the RRR estimator. We leave this extension to future research.

    We apply the CRRR estimator to analyze intergenerational income mobility in Switzerland using the Economic Well-Being of the Working and Retirement Age Population Data (WiSiER) from 1982 to 2016 for 11 cantons. This dataset contains rich information on socioeconomic, demographic and other variables merged from tax records, social insurance, unemployment records and surveys, 
    %Fortunately for our purposes, 
    and can be linked for fathers and children. We uncover a gender gap in intergenerational income mobility as the persistence between fathers and sons is stronger than between fathers and daughters, both with and without controlling for covariates. We also find that the within-group income persistence is about $62\%$ and $52\%$ of the overall (unconditional) income persistence  for sons and daughters, respectively; where groups are defined by child's and father's marital status, Swiss citizenship, high school graduation, experience, number of children and canton and year fixed effects. We also provide evidence supporting greater persistence for fathers with higher education and fathers with less than two children. 
    These results uncover the substantial role of both within-group and between-group persistence in explaining the intergenerational transmission of income.

Our methodology complements related methodological developments in the econometric and statistical literature. While \cite{chetverikov2025inferencerankrankregressions} provide the inferential theory for RRR and RRRX with marginal ranks, their approach does not apply to conditional ranks employed in CRRR, as we explained earlier. \cite{kitagawa2018measurement} studied measurement error in RRRs in the context of intergenerational income mobility. Another new development is \cite{lei2024causal} which examines causal underpinnings of the marginal rank regressions. More closely related to our work, \cite{liu2018covariate} introduced a covariate-adjusted Spearman coefficient based on the probability scale residuals of \cite{li2012new} and \cite{shepherd2016probability}, which we further discuss in Section \ref{sec:crrr}. They proposed a modelling strategy and an estimator based on a monotone transformation of a location-shift model, which is a special case of the distribution regression model \citep{chernozhukov+13inference}.\footnote{\cite{liu2018covariate} did not develop inference theory for their estimator in the case where $Y$ and $W$ are continuous, although they conjectured that it is possible; see Section 5 ibid.} The DR approach is more flexible and comprehensive, in that it can approximate the true conditional distribution function arbitrarily well by considering rich sets of basis functions with respect to the covariates. This is not generally possible with transformations of location models.\footnote{A transformation of a location-shift model takes the form $H(Y) = b(X)'\beta + \epsilon$, where $\epsilon$ is independent of $X$ and has a known distribution, $b(X)$ is a basis of functions of $X$, and $H$ is an unknown monotone transformation function. The model is more flexible than just a location model, but it does not allow the covariates to affect the distribution of $H(Y)$ other than through its location. No matter how rich the basis functions $b(X)$ are, the model is not guaranteed to cover the true conditional distribution, even in the limit where the dimension of $b(X)$ grows large.} 
\cite{gijbels2011conditional} and \cite{veraverbeke2011estimation} developed estimators of conditional measures of association using copulas, relying on the representation of these measures in terms of the conditional copula for continuous outcomes. They derived distribution theory for estimators of conditional versions of Kendall's tau and Blomqvist's beta \citep{blomqvist1950measure} with scalar covariates. In contrast, our modelling and estimators are based on conditional distributions and apply to multivariate covariates.
%Thus, our methodological contributions complement the existing literature. 

\medskip

\paragraph{\textbf{Outline}} The rest of the paper is organized as follows. Section \ref{sec:example} contrasts CRRR with  RRR via a simple conceptual example. Section \ref{sec:crrr} introduces formally CRRR and derives its properties.  Section \ref{sec:estim} describes the estimation procedure based on DR and a bootstrap algorithm to make inference. Section \ref{sec:empirics} discusses an application examining the relationship between fathers' and their children's labor income using Swiss data, while Section \ref{sec:theory} provides  asymptotic theory. 
%Section \ref{sec:simul} reports some simulation evidence. 
% Additional theoretical and numerical simulation results, and proofs are reported in the Appendix.
Section \ref{sec:conclusion} concludes with some remarks and  Appendix \ref{app:reg-estimators} reports results for regression-based estimators not included in the main text. 
%Section \ref{sec:simul} reports some simulation evidence. 
Additional theoretical, empirical and numerical simulation results, and proofs are reported in the Online Supplementary Material \citep[][OSM]{supp}.

\section{CRRR vs RRR: A Conceptual Example}\label{sec:example} To contrast the relative performances of CRRR and different forms of RRR in capturing the relationship between $Y$ and $W$ in the presence of covariates $X$,
we examine a simple conceptual example in which $X$ is binary. This example is convenient because with a binary covariate there is no concern that the differences between the methods are driven by particular modeling strategies to specify various regression functions.

%To make the example more concrete, 
Let $Y$ be daughter's height (in cm), $W$ be father's height (in cm) and $X$ be a country indicator, say $X=0$ for the Netherlands  and $X=1$ for Ireland. Conditional on $X$, $Y$ and $W$ follow a bivariate normal distribution with mean parameters that may depend on the value of $X$ and constant covariance matrix. More specifically,
\begin{equation}\label{eq:design}
\left(\begin{array}{c}
     Y  \\
     W 
\end{array}\right) \mid X = x \sim N_2 \left( \left(\begin{array}{c}
     165  \\
     180-\delta x 
\end{array}\right), 4^2 \left(\begin{array}{cc}
     1 & .6  \\
     .6 & 1 
\end{array}\right)\right),
\end{equation}
where $\Pr(X=0) = \Pr(X=1) = 1/2$. For example, $\delta$ can be a negative country shock such as the Irish Famine that affects the father's height in Ireland, but not in the Netherlands. We consider two cases depending on the extent of the effect of the shock as measured by $\delta$:
\begin{itemize}
    \item No shock: $\delta = 0$.
    \item Negative shock: $\delta = 12$.
\end{itemize}

Table \ref{table:comparison} compares measures of overall and within-country intergenerational height persistence  based on rank correlation with the estimands of RRR, CRRR and two versions of RRRX.  Thus, $\rho_{Y,W}$ and $\bar \rho_{Y,W \mid X}$ are the Spearman rank correlation coefficient between $Y$ and $W$ and the expected Spearman rank correlation between $Y$ and $W$ conditional on $X$;  RRR is the RRR slope; CRRR is the CRRR slope; RRRX-A is the slope of the RRRX, that is, $\beta_1$ in
$$
\tilde U = \beta_0 + \beta_1 \tilde V + \beta_2 X + \epsilon, \quad \Ep[\epsilon] = \Ep[\tilde V \epsilon] = \Ep[X \epsilon] = 0,
$$
where $\tilde U$ and $\tilde V$ denote the marginal ranks of $Y$ and $W$, respectively; and RRRX-I is the average slope of the RRRs run separately by the values of $X$, that is, $\beta_1$ in
$$
\tilde U = \beta_0 + \beta_1 \tilde V + \beta_2 [X-.5] + \beta_3 [X-.5] \tilde V + \epsilon, \quad \Ep[\epsilon] = \Ep[\tilde V \epsilon] = \Ep[X \epsilon] = \Ep[X \tilde V \epsilon] = 0
$$

%The results of the following comparison are obtained by 2 million simulations:
% latex table generated in R 4.2.1 by xtable 1.8-4 package
% Fri Jan 12 17:03:16 2024
\begin{table}[ht]\caption{Mobility Measures and  Estimands}\label{table:comparison}
\centering
\begin{threeparttable}[b]
\begin{tabular*}{12cm}{@{\extracolsep{\fill}} lcccccc}
\toprule\toprule
  & \multicolumn{2}{c}{Overall} & \multicolumn{4}{c}{Within-Country} \\
  \cmidrule{2-3} \cmidrule{4-7}
%  & True & Estimand & True  &\multicolumn{3}{c}{Estimand} \\ 
%  \cmidrule{5-7}
  & $\rho_{Y,W}$ & RRR & $\bar \rho_{Y,W \mid X}$  & RRRX-A & RRRX-I & \textbf{CRRR} \\ 
  \hline
$\delta=0$ & 0.58 & 0.58 & 0.58 & 0.58 & 0.58 & \textbf{0.58} \\  
$\delta=12$ & 0.32 & 0.32 & 0.58  & 1.07 & 1.07 & \textbf{0.58} \\
\bottomrule\bottomrule
\end{tabular*}
\begin{tablenotes}[flushleft]
\footnotesize
\item \textit{Notes: based on 2,000,000 simulations.} 
\end{tablenotes}

\end{threeparttable}

\end{table}

%\textsc{[VC: Introduce exta row in table, after Uncondition and Conitional line; and Put "True" above $\rho_{Y,W}$ and above $\bar \rho_{Y,W|X}$. And Put "Estimators" above estimators.]}

We find that all the methods give the same answer when $\delta=0$, that is when the joint distribution of daughter's and father's heights is the same in both countries.\footnote{Up to numerical error, all the slopes are equal to the rank correlation of the bivariate normal with correlation $c=.6$, $\rho_S(Y,W)=6\arcsin{(c/2)}/\pi = .58$ \citep{cramer1999mathematical}.} When $\delta=12$, RRR gives the overall mobility $\rho_{Y,W}$, whereas CRRR gives the within-country mobility $\bar \rho_{Y,W \mid X}$. Both forms of RRRX produce measures greater than one, which do not correspond to any rank correlation and might be difficult to interpret. Whether RRR or CRRR is the right measure depends on the application. In this case, CRRR measures average intergenerational mobility within each country whereas RRR measures intergenerational mobility pooling the two countries. They would lead to different conclusions about the effect of the Irish Famine. According to RRR, the famine reduces overall height persistence, whereas it does not have any effect within each country according to CRRR. Both versions of RRRX lead to the opposite conclusion that the famine increases height persistence. This conclusion does not correspond to a change in either overall or within-country mobility as measured by Spearman correlations.

\begin{figure}[h]
\caption{Predicted Daughter's Height Rank when $\delta=0$} \label{fig:aum0}
\vspace{.2cm}
%\begin{scriptsize}
\begin{center}
\includegraphics[width=\textwidth]{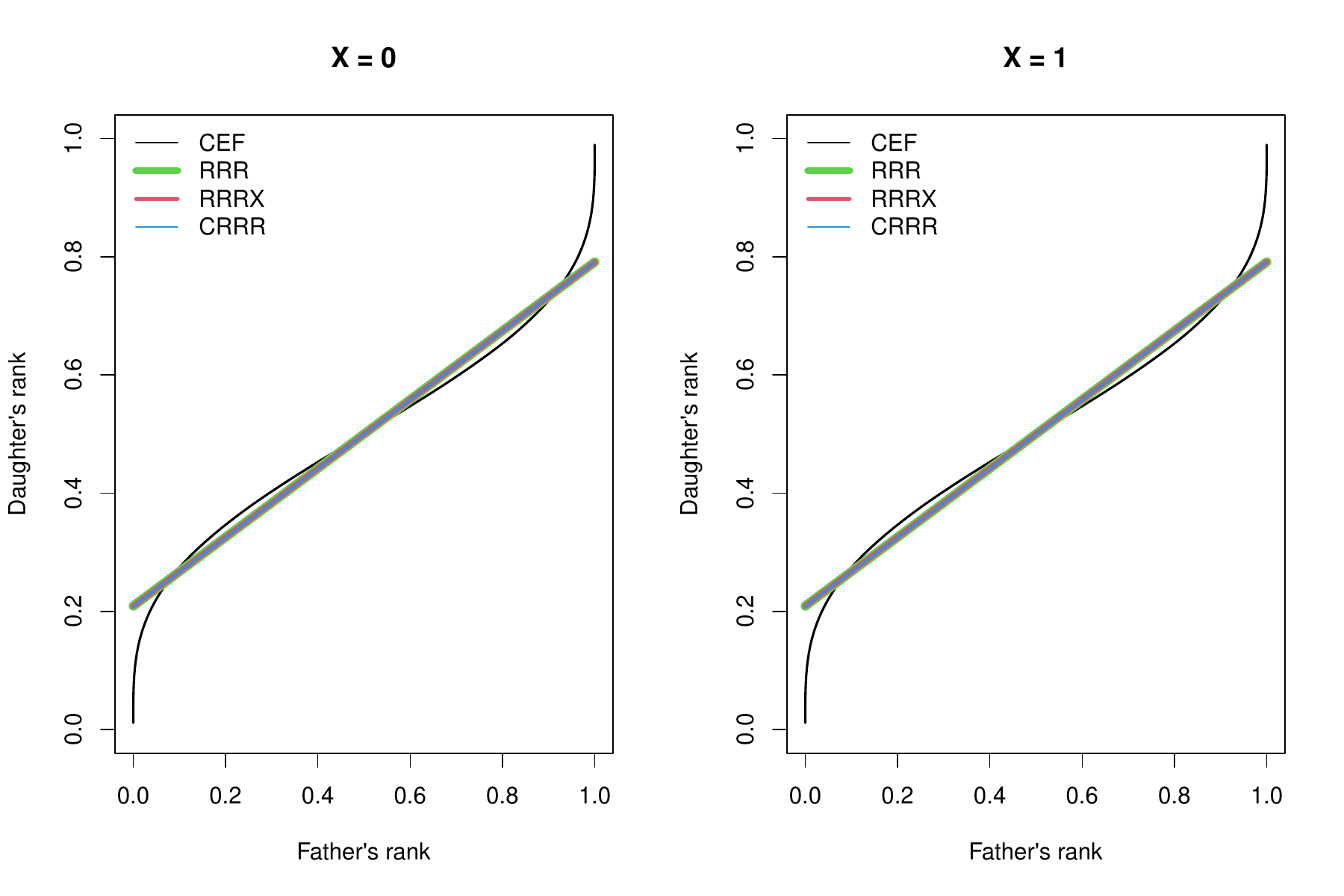}
\end{center}
%\end{scriptsize}
\vspace{-.75cm}
\caption*{\footnotesize \textit{Notes:  based on 2,000,000 simulations.} }
\end{figure}

\begin{figure}[h]
\caption{Predicted Daughter's Height Rank when  $\delta=12$} \label{fig:aum12}
\vspace{.2cm}
%\begin{scriptsize}
\begin{center}
\includegraphics[width=\textwidth]{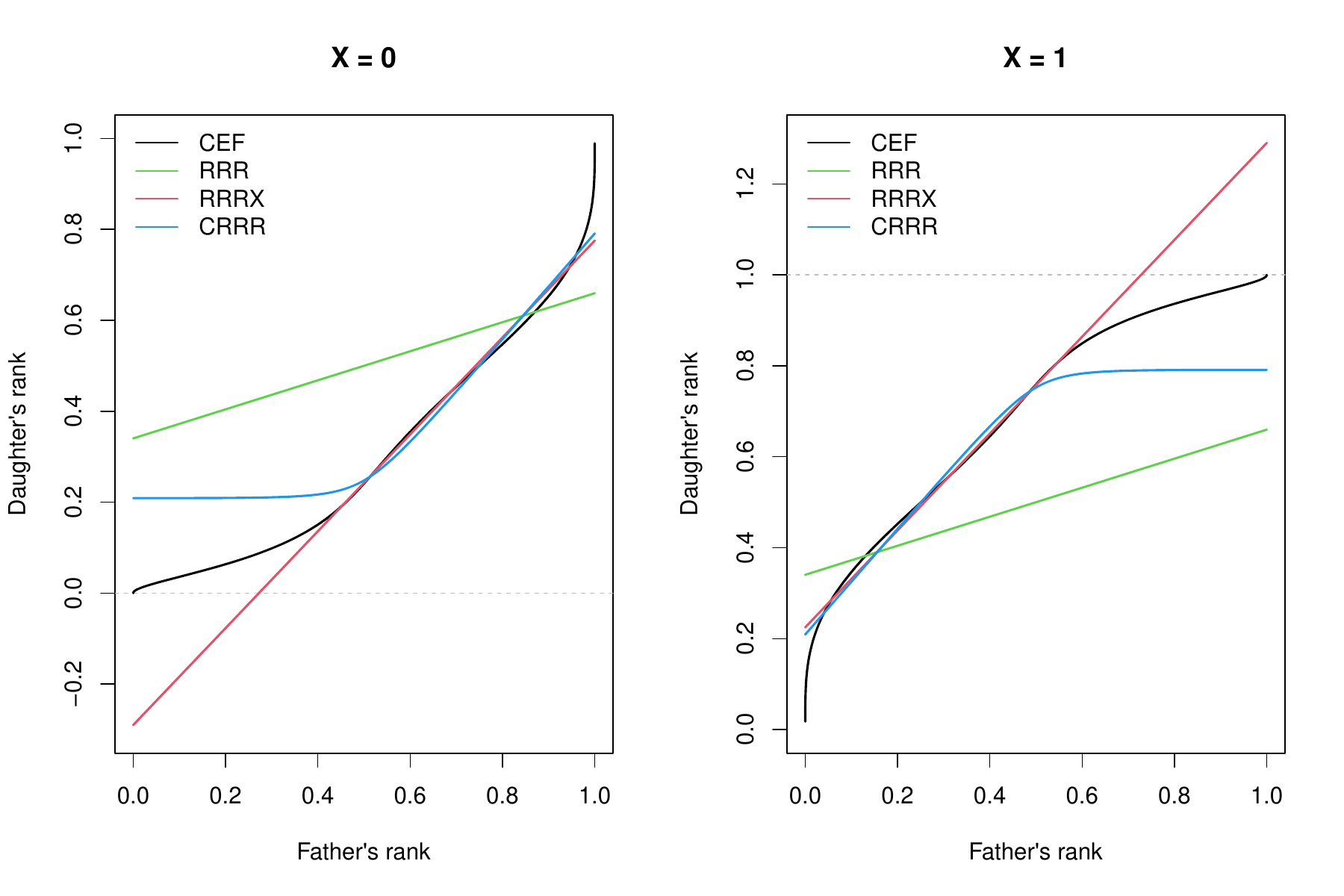}
\end{center}
%\end{scriptsize}
\vspace{-.75cm}
\caption*{\footnotesize \textit{Notes:  based on 2,000,000 simulations.} }
\end{figure}

While RRR and RRRX are frequently employed in intergenerational mobility analyses, the rank correlation is not always the object of interest. Many empirical studies in this literature focus on the so-called level of absolute upward mobility. This is defined as the expected marginal rank of a child with a father at a specified percentile, typically the 25th percentile \citep{chetty2014land}. More generally, it is the conditional expectation function (CEF) of the child's rank given values for the father's rank and the covariates. Figures \ref{fig:aum0} and \ref{fig:aum12}  report  predicted daughter's height (marginal) ranks obtained from CRRR', the variant of CRRR that regresses marginal ranks of $Y$ on conditional ranks of $W$, and different versions of RRR  for the scenarios $\delta=0$ and $\delta=12$, respectively. In this case CRRR' is the same as CRRR as the marginal and conditional ranks of $Y$ coincide because $Y$ is independent of $X$.\footnote{We also examined a variant of the data generating process of the model in which the conditional mean of $Y$ was affected by a different binary variable such that CRRR' was different from CRRR. It did not produce any qualitative differences from the conclusions that follow.} Accordingly, we shall refer to CRRR' as CRRR. RRR only uses the father's height $W$, whereas RRRX and CRRR also employ the value of the covariate $X$.\footnote{We can also include $X$ as an additional regressor in CRRR, but in this case it does not change the results because $Y$ is independent of $X$ conditional on the conditional rank of $W$.} We do not distinguish between RRRX-A and RRRX-I because they produce very similar predictions. To make the methods comparable, CRRR evaluates the prediction at the father's conditional rank corresponding to the marginal quantile of the father's (marginal) rank.\footnote{More specifically, if the father's marginal rank is $v$ and $X=x$, then the corresponding father's conditional rank is $V_{w,x} = F_{W \mid X}(F_W^{-1}(v) \mid x)$, where $F_W$ and $F_{W \mid X}$ are the marginal and conditional distributions of $W$.} As a benchmark of comparison, we report the conditional expectation function (CEF) of the marginal rank of $Y$ given $X$ and $W$ which, using the properties of the multivariate normal, is:
\begin{equation}\label{eq:cef}
    \Ep(\tilde U \mid W=w,X=x) = \Phi\left(\frac{.6*\Phi^{-1}(V_{w,x})}{\sqrt{2-.6^2}} \right),\quad V_{w,x}=\Phi\left(\frac{w-180+\delta x}{4}\right)
\end{equation}
where $\Phi$ is the distribution of the standard normal and $V_{w,x}$ is the conditional rank of $W$ at $W=w$ and $X=x$.\footnote{Note that it is equivalent conditioning on $W$ to conditioning on the marginal rank of $W$ because there is a one-to-one relationship between them. Indeed,
$$
\Ep(\tilde U \mid \tilde V = \tilde v, X=x) = \Phi \left( \frac{.6 *\left( F^{-1}_W(\tilde v) - 180 + \delta x\right) }{4 \sqrt{2 - 0.6^2}} \right),
$$
where $F_W$ is the distribution of $W$.
} The true CEF can be computed in this example
because we know the joint distribution of $(Y,W)$ conditional on $X$. 

When $\delta =0$, $X$ does not help %with 
predict because $(Y,W)$ and $X$ are independent. All the methods yield the same prediction function in both countries, which is a linear approximation to the CEF. When $\delta = 12$, $X$ helps predict because the CEF is different in both countries. In this case, RRRX and CRRR give better approximations to the CEF than RRR, because they make use of the information in $X$. RRRX provides a linear approximation to the CEF in each country, but delivers predicted daughter's ranks lower than $0$ for father's ranks  below the 28th percentile when $X=0$ (Netherlands), and  greater than $1$ for father's ranks above the 72th percentile when $X=1$ (Ireland). RRRX therefore produces unreasonable predictions  for many relevant values of the father's rank in half of the population. CRRR does not suffer from this problem. The implicit nonlinearity introduced by the use of the conditional father's rank, instead of the marginal rank, ensures the predicted daughter's ranks are between $0$ and $1$. In other words, while CRRR  provides a linear approximation to the CEF with respect to the conditional father's rank, it provides a nonlinear approximation in terms of marginal father's rank. This approximation is bounded in $[0,1]$ because the absolute value of the slope is less than $1$. There is still approximation error, however, because the CEF is also a nonlinear function of the conditional father's rank in this case; see \eqref{eq:cef}. 
% Indeed, using the properties of the multivariate normal, it can be shown that 
% $$
% \Ep(\tilde U \mid W=w,X=x) = \Phi\left(\frac{.6*\Phi^{-1}(V_{w,x})}{\sqrt{2-.6^2}} \right),\quad V_{w,x}=\Phi\left(\frac{w-180+\delta x}{4}\right)
% $$
% where $\tilde U$ is the marginal rank of $Y$, $\Phi$ is the distribution of the standard normal and $V_{w,x}$ is the conditional rank of $W$ at $W=w$ and $X=x$.

When $\delta=12$, the CEF is convex when $X=0$ (Netherlands) and concave when $X=1$ (Ireland). Intuitively, a low marginal father's rank corresponds to a very low conditional father's rank in the Netherlands, which is associated with a very low conditional and marginal daughter's rank due to the high positive within-country correlation. Conversely, a high marginal father's rank corresponds to a very high conditional father's rank in Ireland, which is associated with a very high conditional and marginal daughter's rank. The CRRR predictions agree with these shape restrictions, whereas RRRX imposes linearity by construction. 
 In both scenarios for $\delta$, the CRRR slope of $0.58$ indicates that about $34\% (=0.58^2)$ of the variability of the daughter's rank is explained by the father's rank net of covariates (country indicator), which is also the R-squared of CRRR. In this case, this explained fraction is the same in both countries. The RRRX slope of $1.07$ when $\delta=12$ does not have an interpretation as either a partial or overall  R-squared.

% latex table generated in R 4.2.1 by xtable 1.8-4 package
% Tue Apr  2 11:30:01 2024
\begin{table}[ht]\caption{Subgroup Analysis by Country using RRRX and CRRR}\label{table:comparison2}
\centering
\begin{threeparttable}[b]
\begin{tabular*}{12cm}{@{\extracolsep{\fill}}lcccccc}
  \toprule\toprule
 %    & \multicolumn{2}{c}{True} & \multicolumn{4}{c}{Estimand}\\
 %    \cmidrule{4-7}
   & \multicolumn{2}{c}{$\rho_{Y,W \mid X}$} & \multicolumn{2}{c}{RRRX} & \multicolumn{2}{c}{\textbf{CRRR}}\\
     \cmidrule{2-3}   \cmidrule{4-5} \cmidrule{6-7}
   & $X=0$ & $X=1$  & $X=0$ & $X=1$  & $X=0$  & $X=1$  \\ 
  \hline
$\delta=0$ & 0.58 & 0.58 & 0.58 & 0.58 & \textbf{0.58} & \textbf{0.58} \\ 
$\delta=12$ & 0.58 & 0.58 & 1.06 & 1.07 & \textbf{0.58} & \textbf{0.58}  \\ 
   \bottomrule\bottomrule
\end{tabular*}
\begin{tablenotes}[flushleft]
\footnotesize
\item \textit{Notes: based on 2,000,000 simulations.} 
\end{tablenotes}

\end{threeparttable}
\end{table}

%\textsc{[VC:  I would give CRRR results after RRR results, to make it parallel to Table 1. I ould also put "True : $rho_{Y,X \mid X}$". This would make it easy for the reader to navigate.]}

Finally, we conduct a subgroup analysis using CRRR and RRR.  Table \ref{table:comparison2} compares the conditional Spearman rank correlation, CRRR slope and RRR slope  for each value of  $X$.
%; whereas the entries for RRR report the RRR slopes for each value of $X=x$. 
CRRR produces measures that are invariant to both $X$ and $\delta$, which correspond to the rank correlations between $Y$ and $W$ conditional on $X$. If $Y$ and $X$ were not independent, then CRRR and CRRR' would give different results, but still both would produce slopes for each country that would average to the overall CRRR or CRRR' slope.\footnote{In the case of CRRR, the slopes for each country would also correspond to conditional Spearman rank correlations.} The RRR slopes are the same as the CRRR slopes when $X$ is irrelevant. RRR, however, delivers different slopes for the different values of $X$ when $\delta=12$, and also across the different values of $\delta$. The RRR slopes are greater than one when $\delta=12$, confirming that they do not correspond to correlations and making them hard to interpret. Moreover, the within-group RRR slopes do not average to the overall RRR slope in general \citep{hertz2008group}.

While the above example is simple, it illustrates a number of important features which apply, or are likely to apply, in more complicated settings. First, in the presence of covariates, the RRRX slope does not correspond to any measure of rank correlation and can be difficult to interpret. In contrast, the CRRR slope corresponds to a specific form of rank correlation. Moreover, while neither CRRR nor RRRX exactly predict the CEF for all values of the father's rank, the results here suggest that CRRR provides a more sensible and accurate approximation. Moreover, it directly produces a measure of the proportion of variability in the ranks of $Y$ explained by the ranks of $W$ net of the covariate $X$ and is suitable for subgroup analysis.

\section{Conditional Rank-Rank Regression}\label{sec:crrr}

Let $(Y,W)$ be a bivariate random variable with joint distribution $F_{Y,W}$ and marginal distributions $F_Y$ and $F_W$ for $Y$ and $W$, respectively. For example, $Y$ is child's income and $W$ is father's income.  We assume that $Y$ and $W$ are continuous.

\subsection{Canonical RRR}   We start by reviewing the canonical rank-rank regression (RRR). Let $\tilde U := F_Y(Y)$ and $\tilde V:=F_W(W)$ denote the (marginal) ranks of $Y$ and $W$. By continuity of $Y$ and $W$, ranks are uniformly distributed, $\tilde U \sim U(0,1)$ and $\tilde V \sim U(0,1)$. The RRR of $Y$ on $W$ is defined as the correlation between $\tilde U$ and $\tilde V$ or the slope of the linear regression of $\tilde U$ on $\tilde V$ (or vice versa):
$$
\rho := \mathrm{Cor}(\tilde U, \tilde V)= \frac{\Cov(\tilde U,\tilde V)}{\Var(\tilde U)} = \frac{\Cov(\tilde U,\tilde V)}{\Var(\tilde V)} =
12 \ \Ep[(\tilde U-.5)(\tilde V - .5)],
%\mathrm{Cov}(\tilde U, \tilde V).
$$
where all the equalities follow from the uniform distribution of $\tilde U$ and $\tilde V$.
In statistics this correlation measure is the celebrated Spearman rank correlation between $Y$ and $W$, and is widely used to measure dependence between variables. It is invariant to rescaling and all increasing monotone transformations of the variables, and has gained prominence for that reason. The rank correlation has become popular in economics in studies of income and wealth mobility due to its  interpretability as a measure of persistence and scale-free nature.
\medskip

\subsection{Conditional RRR} We introduce now the conditional rank-rank regression (CRRR). Let $X$ denote a vector of covariates related to $Y$ and $W$ including, for example, child's and father's education, age, marital status and nationality. Let $F_{Y\mid X}$ and $F_{W \mid X}$ denote the distributions of $Y$ and $W$ conditional on $X$. Then, $U := F_{Y \mid X}(Y \mid X)$ and $V:=F_{W \mid X}(W \mid X)$ are the conditional ranks of $Y$ and $W$, where conditioning is on $X$. For example, $U$ and $V$ would be child's and father's income ranks among families with the same composition in terms of covariates. By continuity of $Y$ and $W$ conditional on $X$ almost surely, the conditional ranks follow the uniform distribution, conditional on $X$: 
$$U \mid X \sim U(0,1) \text{ and } V \mid X \sim U(0,1),$$
and also unconditionally. This implies the constant variance property, 
$$\Var(V) = \Var(U) =\Var(V \mid X) = \Var(U \mid X) = 1/12$$ and the constant mean property,
$$\Ep V = \Ep U = \Ep(V \mid X) = \Ep (U \mid X) =.5.$$
Note that both $U$ and $V$ are marginally independent of $X$ , but not necessarily jointly independent so the correlation between $U$ and $V$ can depend on $X$.\footnote{That is, $U \indep X$ and $V \indep X$, but generally $(U,V) \notindep X$, where $\indep$ denotes stochastic independence.}

The CRRR of $Y$ on $W$ given $X$ is defined as either the correlation between $U$ and $V$ or the slope of the linear regression of $U$ on $V$ (or vice versa):
\begin{equation}\label{eq:crrr-def}
\rho_{C} = \mathrm{Cor}(U,V) = \frac{\Cov(U,V)}{\Var(V)} = \frac{\Cov(U,V)}{\Var(U)}.
\end{equation}
CRRR is the average conditional correlation between conditional ranks:
\begin{equation}\label{eq: link}
\rho_{C}= \Ep[\rho_{Y,W\mid X}], \quad \rho_{Y,W\mid X} := \mathrm{Cor}(U,V\mid X),
\end{equation}
where $\rho_{Y,W\mid X}$ denotes the conditional Spearman rank correlation between $Y$ and $W$ conditional on $X$, which is equal to $\mathrm{Cor}(U,V\mid X)$ by definition. Equation \eqref{eq: link} follows from $\Cov(U,V) = \Ep [\Cov(U,V\mid X)]$ by the law of total covariance since $\Cov[\Ep(U\mid X), \Ep(V\mid X)] = 0$; moreover, the conditional variance of $U$ and $V$ is equal to the unconditional variance.  In summary, CRRR is the average Spearman rank correlation between $Y$ and $W$ conditional on $X$, averaged over the distribution of $X$, which is a summary measure of within-group persistence.

By the properties of $U$ and $V$, the CRRR can also be represented as the rescaled covariance of conditional ranks:
\begin{equation}\label{eq:fully-restricted}
    \rho_C= 12 \ \Ep[(U-.5)(V-.5)],
\end{equation}
a formula convenient for estimation. Moreover, in the regression version of the CRRR, the intercept, $\alpha_C$, is mechanically related to the slope, $\rho_C$,  through: 
\begin{equation}\label{eq:abs_mob}
\alpha_C = \Ep(U) - \rho_C \Ep(V) = \frac{1 - \rho_C}{2}.    
\end{equation}
This relationship raises concerns about the interpretation of intercepts and slopes in the regression versions of the RRR used in intergenerational mobility studies as measures of absolute and relative mobility, respectively.\footnote{\citen{chetty2014land}  noted a relationship analogous to \eqref{eq:abs_mob} for the regression version of the canonical RRR.}

\subsection{Rank-rank regression with covariates (RRRX)} CRRR is different from RRR with covariates $X$ (RRRX) where $X$ is included additively (or non-additively) in the regression of marginal ranks, $\tilde U$ on $\tilde V$. We believe that our proposal is a more natural and adequate way to incorporate covariates. In fact, RRRX with additive covariates is no longer related to a rank correlation nor has to lie in the interval $[-1,1]$. RRRX is also more difficult to interpret as it does not correspond to a meaningful  measure  of within-group persistence  and can predict ranks outside $[0,1]$. Making RRRX more flexible by including interactions between $X$ and $\tilde V$ does not mitigate any of these problems.  In fact, making RRRX fully nonparametric also does not alleviate the problem, as we showed in Section \ref{sec:example}.

\subsection{Subgroup Analysis} When $X$ is discrete, it is common to run RRRs separately for each value of $X$ instead of including $X$ as an additive control. For example, \cite{abramitzky2021intergenerational} run separate RRR of child's income on father's income  by  father's immigration status. The slopes of these regressions cannot be interpreted in terms of rank correlations or even as conditional correlations between the marginal ranks. To see this, note that the slope of the regression of $\tilde U$ on $\tilde V$ conditional on $X=x$, is not equal to the conditional correlation of  $\tilde U$ and $\tilde V$:
$$
\frac{\Cov(\tilde U,\tilde V \mid X=x)}{\Var(\tilde V \mid X=x)} \neq \frac{\Cov(\tilde U,\tilde V \mid X=x)}{\sqrt{\Var(\tilde V \mid X=x) \Var(\tilde U \mid X=x)}},
$$
because marginal ranks have different conditional distributions, i.e. $\tilde U \overset{d}{\not\sim} \tilde V \mid X=x$, in general. The slope therefore does not generally correspond to the conditional correlation of the marginal ranks   nor the conditional rank correlation between $Y$ and $W$. Section \ref{sec:example} provided an example where this slope is greater than one. 

Consider now the CRRR. Assume we are interested in conducting a subgroup analysis of intergenerational mobility with respect to  father's  high school diploma or immigration status.  Let $X_1 \subseteq X$ be a set of variables that define the subpopulation of interest such as an indicator for high school diploma and/or Swiss nationality. Then, the CRRR slope conditional on $X_1=x_1$ is:\footnote{Indeed, by the law of total covariance with respect to $X$ and uniformity of $U$ and $V$ conditional on $X$,
$$
\Cov(U,V \mid X_1) = \Ep[\Cov(U,V \mid X) \mid X_1] + \Cov[\Ep(U\mid X),\Ep(V\mid X) \mid X_1] = \Ep[\Cov(U,V \mid X) \mid X_1],
$$
and
$$
\Var(V \mid X_1) = \Var(V \mid X) = \Var(U \mid X),
$$
almost surely.}
\begin{multline}\label{eq:subgroup}
\rho_{C}(x_1)  = \frac{\Cov(U,V \mid X_1 = x_1)}{\Var(V \mid X_1=x_1)} = \Ep\left[\frac{\Cov(U,V \mid X)}{\sqrt{\Var(V \mid X)\Var(U \mid X)}} \mid X_1=x_1\right]  \\ = \Ep [ \rho_{Y,W \mid X} \mid X_1 = x_1].  
\end{multline}
%where $\rho_C(x) :=  \Corr(U,V \mid X=x)$. 
Hence, the CRRR slope for the subgroup defined by $X_1=x_1$ corresponds to the average conditional rank correlation between $Y$ and $W$, where the average is taken with respect to the distribution of $X$ conditional on $X_1=x_1$. This allows us, for example, to measure intergenerational mobility separately for families with fathers with and without a high school diploma.\footnote{If $X_1 \not\subseteq X$,  the slope no longer has an interpretation as average conditional rank correlation because $V \not\sim U \mid X_1$ in general.}
%$\Var(V \mid X_1=x_1) \neq \Var(V)$, $\Cov(U,V \mid X_1=x_1) = \Ep[\Cov(U,V \mid X,X_1) \mid X_1=x_1] + \Cov[\Ep(U\mid X,X_1), \Ep(V\mid X,X_1) \mid X_1=x_1]$, and $\Cov[\Ep(U\mid X,X_1), \Ep(V\mid X,X_1) \mid X_1=x_1] \neq 0$ in general. }

\subsection{Variations of CRRR (CRRR')}
There are applications where the researcher might want to use different sets of covariates to obtain the conditional ranks $U$ and $V$. In the intergenerational mobility application, for example, we might not want to control for son's education to obtain the father's income rank. In this case the CRRR' slope still corresponds to an average correlation between the ranks. To see this, let $U_1:=F_{Y \mid X_1}(Y \mid X_1)$ and $V_2:=F_{W \mid X_2}(W \mid X_2)$ with $X_1 \neq X_2$ and $X= X_1 \cap X_2$, the set of covariates included in both $X_1$ and $X_2$, then:
$$
\rho_C' = \frac{\Cov(U_1,V_2)}{\Var(V_2)}  = \Ep\left[\frac{\Cov(U_1,V_2 \mid X) }{\sqrt{\Var(V_2 \mid X)\Var(U_1 \mid X)}}\right],
$$
where we use the law of total covariance with respect to $X$,  $U_1 \indep X$, $V_2 \indep X$ and iterated expectations. The CRRR' slope therefore corresponds to the correlation between the ranks $U_1$ and $V_2$ conditional on the common covariates $X$, averaged over the distribution of $X$. Note, however, that $\rho_C'$ now does not correspond to an average conditional rank correlation between $Y$ and $W$. The source of the difference is that $U_1 \neq F_{Y \mid X}(Y \mid X)$ and  $V_2 \neq F_{W \mid X}(W \mid X)$ in general.\footnote{This rank correlation can be obtained by constructing the conditional ranks as $U=F_{Y \mid X}(Y \mid X)$ and $V=F_{W \mid X}(W \mid X)$.} An exception occurs when $Y$ is independent of the components of $X_2$ not included in $X_1$ conditional on $X_1$, and $W$ is independent of the components of $X_1$ not included in $X_2$ conditional on $X_2$. In that case,
$$
\rho_C' = \Ep\left[\frac{\Cov(U_1,V_2 \mid \bar X) }{\sqrt{\Var(V_2 \mid \bar X)\Var(U_1 \mid \bar X)}}\right] = \Ep [ \rho_{Y,W \mid \bar X} ],
$$
where $\bar X = X_1 \cup X_2$. This result follows by the law of total covariance with respect to $\bar X$ and uniformity of $V_2$ and $U_1$ conditional on $\bar X$.\footnote{Note that if $V \mid X_1 \sim U(0,1)$ and $V \indep X_2 \mid X_1$, then $V \mid X \sim U(0,1)$.}

Like CRRR, CRRR' is suitable for subgroup analysis in the following sense. Let $\rho_C'(x_2)$ be the CRRR' slope in the group defined by $X_2=x_2$, that is:
$$
\rho_C'(x_2) = \frac{\Cov(U_1,V_2 \mid X_2 = x_2)}{\Var(V_2 \mid X_2 = x_2)}.
$$
Then, by the law of total covariance and $V_2 \ci X_2$,
$$
\rho_C' = \frac{\Ep\left[\Cov(U_1,V_2 \mid X_2)\right]}{\Var(V_2)} = \Ep[\rho_C'(X_2)],
$$
that is, the CRRR' slope can be decomposed as the average of the CRRR' slopes in each group defined by $X_2$, weighted by the size of the group.

An interesting example occurs when $X_1 = \emptyset$ and $X_2 = X$. In this case, $U_1 = \tilde U$, $V_2 = V$ and $\rho_C'$ is the correlation between the marginal ranks of $Y$ and the conditional ranks of $W$. Unlike RRRX, the inclusion of covariates in this CRRR' does not affect the coefficient of $V$ because $V \indep X$,  and can be used to perform a variance decomposition of $\tilde U$. Let
$$
\tilde{U} = V \rho_C' + X'\beta_C + \varepsilon, \quad \Ep[(V; X) \varepsilon] = 0,
$$
be the extended CRRR with covariates, where the first term of $X$ is a constant. Then,
$$
\text{Var}(\tilde U) =  \text{Var}(V)(\rho_C')^2 + \text{Var}(X'\beta_C) + \text{Var}(\varepsilon),
$$
where the first two terms of the right-hand-side correspond to the contribution of $V$ and $X$ to the variance of $\tilde U$, and the third term to the unobserved component. Indeed, $(\rho_C')^2$ measures the fraction of the variance of $\tilde U$ explained by $V$ since $\text{Var}(\tilde U) = \text{Var}(V)$.

\subsection{Properties of CRRR} We conclude this section by gathering the properties of the CRRR slope in the following lemma.
\begin{lemma}[CRRR Properties]\label{lemma:crrr-prop} Assume that $Y$ and $W$ are continuous random variables conditional on $X$, $X$ is a vector of covariates, $U := F_{Y \mid X}(Y \mid X)$ and $V := F_{W \mid X}(W \mid X)$. Then, (1) The CRRR slope, $\rho_C$, has the representations given in \eqref{eq:crrr-def} and \eqref{eq:fully-restricted}. (2) The slope $\rho_C$ is the expected conditional Spearman rank correlation between $Y$ and $W$:
    $
    \rho_C = \Ep[\rho_{Y,W \mid X}].
    $ (3) Subgroup analysis: if $X_1 \subseteq X$, then \eqref{eq:subgroup} holds.
    % $$
    % \rho_{C}(x_1)  := \Corr(U,V \mid X_1 = x_1)
    % = \Ep [ \Corr(U,V \mid X = x) \mid X_1 = x_1].$$
   Therefore, $ \rho_C(x_1)$  is the average conditional rank correlation between $Y$ and $W$ in the group defined by $X_1= x_1$. (4) Let $U_1:=F_{Y \mid X_1}(Y \mid X_1)$ and $V_2:=F_{W \mid X_2}(W \mid X_2)$ with $X_1 \neq X_2$ and $X= X_1 \cap X_2$, then
    $$
    \rho_C' = \Corr(U_1,V_2) = \Ep[\Corr(U_1,V_2 \mid X)],
    $$
    that is $\rho_C'$ is the average conditional correlation between $U_1$ and $V_2$ given the set of common covariates $X$.
\end{lemma}

\begin{remark} This lemma simply records the observations given above.  It is useful to highlight the relationship between our results with those in
\cite{liu2018covariate} who introduced the covariate-adjusted Spearman correlation coefficient as the correlation between the probability scale residuals of $Y$ and $W$. These residuals are defined as $ \nu(Y,F_{Y\mid X})$ and $\nu(W,F_{W\mid X})$, where $\nu(r,F_{R\mid X}) = F_{R\mid X}(r \mid X) + F_{R\mid X}(r- \mid X) -1$ and $F_{R \mid X}(r- \mid x) = \lim_{u \nearrow r}F_{R \mid X}(u \mid x)$, for $R \in \{Y,W\}$. In the case where $Y$ and $W$ are continuous, the probability scale residuals are affine transformations of the conditional ranks because $F_{R \mid X}(r- \mid x) = F_{R \mid X}( r \mid x)$, e.g., $\nu(Y,F_{Y\mid X}) = 2 U -1$, and the covariate-adjusted Spearman correlation equals to the CRRR slope.  The properties in Lemma \ref{lemma:crrr-prop}(2) and (3) then  follow from  results in  \cite{liu2018covariate} when $Y$ and $W$ are continuous. The conceptual difference is that our definition and derivations are  based on the characterization of the Spearman correlation as the correlation between ranks or grade correlation \citep{kruskal1958ordinal}, while theirs are based on the characterization of the Spearman correlation in terms of concordance-discordance probabilities.\end{remark}

\section{Distribution Regression Estimator of CRRR}\label{sec:estim}

\subsection{DR Model for Conditional Distributions} For estimation purposes, it is convenient to model the conditional distributions $F_{Y\mid X}$ and $F_{W \mid X}$ using the distribution regression (DR) model:
$$
F_{R \mid X}(r \mid x) = \Lambda(x'\beta_R(r)), \quad R \in \{Y,W\}, \quad r \in \RR, 
$$
where $\Lambda$ is the standard normal or logistic distribution, $\RR$ is the support of $R$ and the first component of $x$ is a constant. The specification can be made more flexible by replacing $x$ by a vector of transformations of $x$ with good approximating properties.  

As the data to estimate the conditional distribution function in the tails are sparse, it is necessary to impose some structure. We assume that the conditional distribution far in the tails can be extrapolated from the conditional distribution not too far in the tails.\footnote{This is in line with approaches used in extreme value theory that impose restrictions on the tail behavior allowing similar extrapolations. For example, see \cite{embrechts:1997}  for a broad reference on the theory of extremes and \cite{victor:annals} or \cite{chernozhukov:2011}  for similar approaches in the context of extremal quantile regression.} We formalize this approach by imposing restrictions on the coefficient of the DR model in the tails. 

%For technical reasons that would be explained in Section \ref{sec:theory}, we impose restrictions on the DR model at the tails. 
Let $\bar{\RR}$ be a compact strict subset of $\RR$, for $\RR \in \{\mathcal{Y},\mathcal{W}\}$, where $\mathcal{Y}$ and $\mathcal{W}$ are the supports of $Y$ and $W$, respectively. Then, we assume:
$$
F_{R\mid X}(r \mid x) = \Lambda((r-\bar r)\alpha_R(\bar r)  + x'\beta_R(\bar{r})), \quad  R \in \{Y,W\}, \quad r \in \mathcal{R}\setminus\bar{\mathcal{R}},
$$
where $\bar r := \arg \min_{r' \in \bar{\mathcal{R}}} |r-r'|$ and $\alpha_R(\bar r) > 0$. 
That is, we postulate that the random variable $R$ behaves in the tails like a random variable with distribution $\Lambda$, after subtracting the location shift $x'\beta_R(\bar{r})$ and dividing by the scale 
$\alpha_R(\bar r)$, which are different at the upper and lower tails. Thus, the DR coefficient is restricted  in the tails by:
$$\beta_{R,1}(r) = \beta_{R,1}(\bar r) + (r - \bar r)\alpha_R(\bar r), \quad \beta_{R,-1}(r) = \beta_{R,-1}(\bar r),\quad  R \in \{Y,W\}, \quad r \in \mathcal{R}\setminus\bar{\mathcal{R}},$$
where $\beta_R(r)$ is  partitioned into $(\beta_{R,1}(r),\beta_{R,-1}(r)')'$ where $\beta_{R,1}(r)$ is the intercept and $\beta_{R,-1}(r)$ are the slope components.  That is,  $r \mapsto \beta_{R,1}(r)$ is a linear function and $r \mapsto \beta_{R,-1}(r) $ is constant on $\mathcal{R}\setminus\bar{\mathcal{R}}$. 

Under the DR model, the conditional ranks can be expressed as the following functionals of the parameters:
$$
U = \Lambda(X'\beta_Y(Y)), \quad V = \Lambda(X'\beta_W(W)).
$$

\subsection{Estimation} We provide several estimators of the CRRR slope based on the different representations of $\rho_C$ in \eqref{eq:crrr-def} and \eqref{eq:fully-restricted}. This section presents correlation-based and fully-restricted estimators. Regression-based estimators are given in Appendix \ref{app:reg-estimators}.  
%We do not discuss explicitly the reverse regression-based estimators because they can be constructed using the same algorithms as the regression-based estimators after switching the names of the variables $Y$ and $W$. 
We recommend the use of at least the  correlation-based and fully-restricted estimators. The fully-restricted estimator, based on \eqref{eq:fully-restricted}, uses all the information available and is the simplest to compute, but it might be sensitive to misspecification of the model for the conditional distributions. In particular, it can deliver estimates outside the interval $[-1,1]$ under misspecification. The correlation-based estimator is more robust in the sense that it is the only estimator that guarantees estimates in the interval $[-1,1]$ under misspecification.\footnote{While we impose correct specification of the DR model for the conditional distributions, the derivation of the theoretical results does not rely fundamentally on correct specification. We conjecture that the probability limit of the correlation-based estimator still has an interpretation as correlation of pseudo-ranks under misspecification, but leave the formal analysis to future research.} 
We show in Appendix \ref{app:reg-estimators} that the correlation-based estimator is asymptotically equivalent to the average of the regression-based and reversed regression-based estimators.

%All  these estimators can be combined to improve theoretical efficiency. We discuss this possibility in Section \ref{sec:theory}.

Let $\{Z_i:=(Y_i,W_i,X_i)\}_{i=1}^n$ be a random sample of $Z:=(Y,W,X)$. The following algorithms describe the estimators of $\rho_{C} $. All of them are based on DR. 

\begin{algorithm}[Correlation-based and Fully-Restricted Estimators]\label{alg:crrr-reg} Let  $d_x := \dim X$,  $\RR_n$   denote the set consisting of the observed values of $R$  and $\bar{\RR}_n = \RR_n \cap \bar{\RR}$, for $R \in \{Y,W\}$.
\begin{enumerate}
\item Estimate $\beta_R(r)$ at $r \in \bar{\RR}_n \cup \partial \bar \RR$ by DR, that is, 
$$
\hat \beta_R(r) \in \argmax_{b \in \mathbb{R}^{d_x}}  \sum_{i=1}^n \left[1(R_i \leq r)  \log \Lambda(X_i'b)  + 1(R_i > r)  \log \Lambda(-X_i'b)\right]. 
$$
\item Estimate $\beta_R(r)$ at $r \in \RR_n \setminus \bar{\RR}_n$ by restricted DR, that is, 
$$
\hat \beta_R(r) = (r-\bar r)\hat \alpha_R(\bar r) e_1  + \hat \beta_R(\bar{r}),$$ where $e_1$ is the first unitary $d_x$-vector, $\bar r := \arg \min_{r' \in \bar{\RR}} |r-r'|$, 
\begin{eqnarray*}
\hat \alpha_R(\bar r) \in \argmax_{a \in \mathbb{R}}  \sum_{i=1}^n  \left[  1(R_i \leq r_0)  \log \Lambda((r_0-\bar r)a  + X_i'\hat \beta_R(\bar{r})) \right. \\     \left. + 1(R_i > r_0)  \log \Lambda(-(r_0-\bar r)a  - X_i'\hat \beta_R(\bar{r}))\right],
\end{eqnarray*}
and $r_0$ is a fixed tail anchor, with $r_0 > \bar r$ for the upper tail and $r_0 < \bar r$ for the lower tail.
%, and (ii) $\hat \alpha_R(\bar r) > 0$.}
\item Obtain plug-in estimators of the conditional ranks:
$$
\hat U_i = \Lambda(X_i'\hat \beta_Y(Y_i)), \quad \hat V_i = \Lambda(X_i'\hat \beta_W(W_i)).
$$
\item Estimate $\rho_C$ as either (a) the sample correlation between $\widehat U_i$ and $\widehat V_i$, that is:
$$
\hat \rho_{C}  = \frac{\sum_{i=1}^n (\hat U_i - \overline{\hat U}) (\hat V_i - \overline{\hat V}) }{\sqrt{\sum_{i=1}^n (\hat V_i - \overline{\hat V})^2 \sum_{i=1}^n (\hat U_i - \overline{\hat U})^2}}, \quad \overline{\hat V} = \frac{1}{n} \sum_{i=1}^n \hat V_i, \quad \overline{\hat U} = \frac{1}{n} \sum_{i=1}^n \hat U_i;
$$
%or (b) the restricted sample correlation between $\widehat U_i$ and $\widehat V_i$, that is
%$$
%\tilde \rho_{C}  = %\frac{\sum_{i=1}^n (\hat U_i - .5)(\hat V_i - .5) }{\sqrt{\sum_{i=1}^n (\hat V_i - .5)^2 \sum_{i=1}^n (\hat U_i - .5)^2}};
%$$
or (b) the sample analog of \eqref{eq:fully-restricted}, that is
$
\breve \rho_C = 12 \sum_{i=1}^n (\hat U_i - .5)(\hat V_i - .5)/n.
$

% This step can be replace by estimating $
% \rho_{Y,W \mid X}
% $ with the sample correlation between $\widehat U_i$ and $\widehat V_i$, that is
% $$
% \tilde \rho_{Y,W \mid X}  = \frac{\sum_{i=1}^n \hat U_i (\hat V_i - \overline{\hat V}) }{\sqrt{\sum_{i=1}^n (\hat V_i - \overline{\hat V})^2 \sum_{i=1}^n (\hat U_i - \overline{\hat U})^2}}, \quad \overline{\hat U} = \frac{1}{n} \sum_{i=1}^n \hat U_i.
% $$
\end{enumerate}
\end{algorithm}

\begin{remark}[Computation] If the set $\mathcal{R}_n$ contains many elements, in step (1) we can either replace it by a smaller fine mesh or use a computationally fast method similar to \cite{chernozhukov2022fast} to speed-up computation.\footnote{By stochastic equicontinuity of the conditional distribution processes $(r,x) \mapsto \sqrt{n} \left[\Lambda(x'\hat \beta_R(r)) - \Lambda(x' \beta_R(r))\right]$, $R \in \{Y,W\}$, in Lemma \ref{lemma:crrr}, the meshwidth $\delta$ should be such that $\delta \sqrt{n} \to 0$ as $n \to \infty$.} Note that the optimization program to obtain $\hat \alpha_R(\bar r)$ in step (2) only needs to be solved twice, once for $r_0$ in the upper tail and once for $r_0$ in the lower tail.  Our asymptotic analysis treats the tail anchors $r_0$ and the thresholds $\bar r$ used to estimate the tail slopes as fixed. In applications, it is useful to check that $\hat \alpha_R(\bar r) > 0$ and there are sufficiently many observations both between each anchor and its corresponding threshold and beyond that anchor; we suggest at least $m=30$ in each region as a practical diagnostic. Developing data-driven threshold selection procedures and establishing their inferential validity are left to future work.
% In practice, we recommend choosing $r_0$ such that there are at least $m \geq 30$ observations between $\bar r$ and $r_0$, and greater than $r_0$ if $r_0 > \bar r$ (upper tail) or less than $r_0$ if $r_0 < \bar r$ (lower tail). 
%Also, we require $m \geq 30$,  which is thought to be the minimal sample size required to estimate one parameter.
\end{remark}

% \begin{algorithm}[Regression-based Restricted Estimator]\label{alg:crrr2} Steps (1)--(3) are the same as in Algorithm \ref{alg:crrr}. In step (4) estimate $\rho_C$ as the slope of the restricted linear regression of $\hat U_i$ on $\hat V_i$, that is
% $$
% \tilde \rho_{C}  = \frac{\sum_{i=1}^n (\hat U_i - .5)(\hat V_i - .5) }{\sum_{i=1}^n (\hat V_i - .5)^2}.
% $$
% \end{algorithm}

\subsection{Bootstrap Inference} Section \ref{sec:theory} shows that the estimators described in Algorithm \ref{alg:crrr-reg} follow normal distributions in large samples. The variances of these distributions, however, have complicated forms and are difficult to estimate.  Section \ref{sec:theory} also shows that the asymptotic distributions can be consistently estimated using the exchangeable bootstrap. Exchangeable bootstrap is a general resampling method that includes empirical and weighted bootstrap as special cases; see Comment \ref{remark:bw}. The following algorithm describes how to obtain bootstrap draws of the estimators of $\rho_C$.

\begin{algorithm}[Exchangeable Bootstrap Draws of Estimators] \label{alg:eb} \hfill 
\begin{enumerate}
\item Draw a realization of the weights $(\omega_{n1},\ldots,\omega_{nn})$ from a distribution that satisfies Assumption \ref{ass:eb} in Section \ref{sec:theory}. Normalize the weights to add-up to $n$.
\item Obtain a bootstrap draw of $\hat \beta_R(r)$ at $r \in \bar{\RR}_n \cup \partial \bar \RR$ by weighted DR, that is, 
$$
\hat \beta_R^*(r) \in \argmax_{b \in \mathbb{R}^{d_x}}  \sum_{i=1}^n \omega_{ni} \left[1(R_i \leq r)  \log \Lambda(X_i'b)  + 1(R_i > r)  \log \Lambda(-X_i'b)\right]. 
$$
\item Obtain a bootstrap draw of $\hat \beta_R(r)$ at $r \in \RR_n \setminus \bar{\RR}_n$ by restricted weighted DR, that is,
$$
\hat \beta_R^*(r) = (r-\bar r)\hat \alpha^*_R(\bar r) e_1 + \hat \beta^*_R(\bar{r}),
$$
where $e_1$ is the first unitary $d_x$-vector,  $\bar r := \arg \min_{r' \in \bar{\RR}} |r-r'|$,  
\begin{eqnarray*}
\hat \alpha^*_R(\bar r) \in \argmax_{a \in \mathbb{R}}  \sum_{i=1}^n  \omega_{ni} \left[  1(R_i \leq r_0)  \log \Lambda((r_0-\bar r)a  + X_i'\hat \beta^*_R(\bar{r})) \right. \\     \left. + 1(R_i > r_0)  \log \Lambda(-(r_0-\bar r)a  - X_i'\hat \beta^*_R(\bar{r}))\right],
\end{eqnarray*}
and $r_0 \in \RR_n \setminus \bar{\RR}_n$ is the same as in Algorithm \ref{alg:crrr-reg}. 
\item Obtain bootstrap draws of the estimators of the conditional ranks:
$$
\hat U_i^* = \Lambda(X_i'\hat \beta^*_Y(Y_i)), \quad \hat V^*_i = \Lambda(X_i'\hat \beta^*_W(W_i)).
$$
\item Obtain a bootstrap draw of the estimator of $\rho_C$ as either  (a) the weighted sample correlation, that is:
$$
\hat \rho^*_{C}  = \frac{\sum_{i=1}^n \omega_{ni} (\hat U^*_i - \overline{\hat U}^*)(\hat V^*_i - \overline{\hat V}^*) }{\sqrt{\sum_{i=1}^n \omega_{ni} (\hat V^*_i - \overline{\hat V}^*)^2\sum_{i=1}^n \omega_{ni} (\hat U^*_i - \overline{\hat U}^*)^2}},$$ 
where $\overline{\hat V}^* =   \sum_{i=1}^n \omega_{ni} \hat V^*_i/n$ and $ \overline{\hat U}^* =    \sum_{i=1}^n \omega_{ni} \hat U^*_i/n;
$
%or (b) the restricted sample weighted correlation between $\widehat U_i$ and $\widehat V_i$, that is
%$$
%\tilde \rho^*_{C}  = \frac{\sum_{i=1}^n \omega_{ni} (\hat U^*_i - .5)(\hat V^*_i - .5) }{\sqrt{\sum_{i=1}^n \omega_{ni} (\hat V^*_i - .5)^2\sum_{i=1}^n \omega_{ni} (\hat U^*_i - .5)^2}};
%$$
or (b) the weighted sample analog of \eqref{eq:fully-restricted}, 
$$
\breve \rho_C^* = \frac{12}{n} \sum_{i=1}^n \omega_{ni} (\hat U_i^* - .5)(\hat V_i^* - .5).
$$
% (a) the slope of the weighted linear regression of $\hat U^*_i$ on $\hat V^*_i$, that is
% $$
% \hat \rho^*_{C}  = \frac{\sum_{i=1}^n \omega_{ni} \hat U^*_i (\hat V^*_i - \overline{\hat V}^*) }{\sum_{i=1}^n \omega_{ni} (\hat V^*_i - \overline{\hat V}^*)^2}, \quad \overline{\hat V}^* =  \sum_{i=1}^n \omega_{ni} \hat V^*_i;
% $$
% or (b) the slope of the weighted restricted linear regression of $\hat U^*_i$ on $\hat V^*_i$, that is
% $$
% \tilde \rho^*_{C}  = \frac{\sum_{i=1}^n \omega_{ni} (\hat U^*_i - .5)(\hat V^*_i - .5) }{\sum_{i=1}^n \omega_{ni} (\hat V^*_i - .5)^2}.
% $$
% (c) for $\hat \varrho_C$ of Algorithm \ref{alg:crrr3} as the  weighted sample correlation between $\widehat U^*_i$ and $\widehat V^*_i$, that is
% $$
% \hat \varrho^*_{C}  = \frac{\sum_{i=1}^n \omega_{ni} \hat U^*_i (\hat V^*_i - \overline{\hat V}^*) }{\sqrt{\sum_{i=1}^n \omega_{ni} (\hat V^*_i - \overline{\hat V}^*)^2 \sum_{i=1}^n \omega_{ni} (\hat U^*_i - \overline{\hat U}^*)^2}}, \quad \overline{\hat V}^* =  \sum_{i=1}^n \omega_{ni} \hat V^*_i, \quad \overline{\hat U}^* =  \sum_{i=1}^n \omega_{ni} \hat U^*_i;
% $$
% and (d) for $\tilde \varrho_C$ of Algorithm \ref{alg:crrr4} as the  weighted restrcited sample correlation between $\widehat U^*_i$ and $\widehat V^*_i$, that is
% $$
% \tilde \varrho^*_{C}  = \frac{\sum_{i=1}^n \omega_{ni} (\hat U^*_i - .5) (\hat V^*_i - .5) }{\sqrt{\sum_{i=1}^n \omega_{ni} (\hat V^*_i - .5)^2 \sum_{i=1}^n \omega_{ni} (\hat U^*_i - .5^2})}.
% $$
\end{enumerate}
\end{algorithm}

\begin{remark}[Bootstrap Weights]\label{remark:bw}
    \cite{van1996weak} notes that by appropriately
selecting the distribution of the weights, the exchangeable bootstrap covers the
most common bootstrap schemes as special cases. The empirical bootstrap
corresponds to where $(\omega_{n1},...,\omega_{nn})$ is a multinomial
vector with parameter $n$ and probabilities $(1/n,...,1/n)$. The
weighted bootstrap corresponds to where $\omega_{n1},...,\omega_{nn}$ are
i.i.d. nonnegative random variables with $\Ep(\omega_{n1})=\Var(\omega_{n1})=1$, e.g.
standard exponential. 
%The Bayesian or weighted bootstrap corresponds to the case where $%
%U_{k1}, ..., U_{kn_{k}}$ are i.i.d. nonnegative random variables, e.g. unit
%exponential, with $E| U_{k1}^{2+\varepsilon}] < \infty$ for some $%
%\varepsilon> 0$, and $w_{ki} = U_{ki} / \bar U_{k},$ where $\bar U_{k} = {%
%n_{k}}^{-1} \sum_{i=1}^{n_{k}} U_{ki}$.
%The wild bootstrap corresponds to where $w_{n1}, ...,
%w_{nn}$ are i.i.d. vectors with $\Ep(w_{n1}^{2+\varepsilon}) < \infty$
%for some $\varepsilon > 0$, and $\Var(w_{n1})=1$.
% The $m$ out of $n$ bootstrap corresponds to letting $(w_{n1},...,w_{nn})$
%  be equal to $\sqrt{n/m}$ times multinomial vectors with parameter $%
%  m$ and probabilities $(1/n,...,1/n)$. 
% The subsampling bootstrap
% corresponds to letting $(w_{n1},...,w_{nn})$ be a row in which the
% number $n(n-m)^{-1/2}m^{-1/2}$ appears $m$ times and $0$
% appears $n-m$ times ordered at random, independent of the data. 
\end{remark}

We now show how to use the exchangeable bootstrap to obtain standard errors for the estimators of $\rho_C$ and construct asymptotic confidence intervals for $\rho_C$. Algorithm \ref{alg:infer} describes the procedure for $\hat \rho_C$. A similar algorithm applies to  $\breve \rho_C$.  Let $B$ be a prespecified number of bootstrap repetitions and $\alpha$ be the significance level for the confidence intervals. For example, $B=500$ and $\alpha=0.05$.

\begin{algorithm}[Inference on $\rho_{C}$ based on $\hat \rho_C$]\label{alg:infer} \hfill
\begin{enumerate}
\item  Draw $\{\hat Z_{\rho,b}^*: 1 \leq b \leq B \} $ as i.i.d. realizations of $\hat Z_{\rho}^* = \sqrt{n}\left(\hat \rho^*_{C} - \hat \rho_{C} \right)$ using Algorithms \ref{alg:crrr-reg} and \ref{alg:eb}. 
\item Compute a bootstrap estimate of the asymptotic standard deviation of $\hat \rho_{C}$, $\sigma_{\rho}$, such as the bootstrap interquartile range rescaled by the normal distribution:
$$
\hat \sigma_{\rho} = \frac{q_{.75} - q_{.25}}{z_{.75} - z_{.25}},
$$
where $q_p$ is the $p$-th quantile of $\{\hat Z_{\rho,b}^*: 1 \leq b \leq B \}$ and $z_p$ is the $p$-th quantile of $N(0,1)$.
\item Compute $B$ bootstrap draws of the T-statistic, $\{T_b : 1 \leq b \leq B  \}$, where
$
T_b = |\hat Z_{\rho,b}^*|/\hat \sigma_{\rho}$.
\item Construct an asymptotic $(1-\alpha)$-confidence interval for $\rho_{C}$ as:
$$
\textrm{ACI}_{1-\alpha}(\rho_{C}) = \hat \rho_{C} \pm \hat t_{1-\alpha} \hat \sigma_{\rho}/\sqrt{n},
$$
where $\hat t_{1-\alpha}$ is the $(1-\alpha)$-quantile of $\{T_b : 1 \leq b \leq B  \}$.
\end{enumerate}
\end{algorithm}

\section{Empirical Application}\label{sec:empirics}

We analyze intergenerational income mobility in Switzerland using the Economic Well-Being of the Working and Retirement Age Population Data (WiSiER). 

% Since the seminal work by \cite{beller2006intergenerational} and \cite{dahl2008association}, several studies extended the unconditional rank-rank framework by adding covariates to the regression \citep[e.g.,][]{chetty2014land,abramitzky2021intergenerational,adermon2018intergenerational,chetty2018impacts}. The covariates typically include location and cohort fixed effects as well as father's and child's education level. 

\subsection{Data}\label{s:data} WiSiER data include Swiss individuals from 11 Cantons from 1982 to 2016. The Swiss Federal Statistical Office merged data from tax records, social insurance, unemployment data, and surveys, creating a unique opportunity to analyze mobility. An ID can match parents and children. While many approaches seem feasible, we compare fathers and children at the same age of 35. As a result, the observations stem from different periods, with most of our successful matches coming from 1982-1990 (fathers) and 2000-2016 (children). The primary outcome variable is yearly real insured labor income (AHV) in $1,000$ Swiss francs (CHF) at the age of 35. The following covariates are available for both fathers and children: months of experience, indicators for high-education (12 or more years of schooling), Swiss citizenship, and being single, and number of own children. Further, we include the father's age at child's birth, and year and canton fixed effects for the children. Finally, for the analysis we exclude the following observations: (i) children where there is no parent in the data, (ii) observations with no information on the child's or father's birth year, and (iii) whenever the father was younger than 15 at the birth of the child. We focus on the conditional child's ranks (CRRR) for the main analysis and provide results for marginal child's ranks (CRRR') in Appendix \ref{app:empirics} of the OSM. We conduct separate analyses for the relationships with sons and daughters. Table \ref{table:ds} reports descriptive statistics for the data used in the analysis. It shows that father's characteristics are similar in families with sons and daughters. This alleviates a potential concern about endogenous selection in the comparison between  sons and daughters.

\begin{table}[h!]\caption{Descriptive Statistics}\label{table:ds}
\begin{center}
\begin{threeparttable}[b]
\setlength{\tabcolsep}{0pt}
%\caption{Intergenerational Mobility in Switzerland: Descriptives} 
\begin{tabular*}{12cm}{ @{\extracolsep{\fill}} lrrrrrrrr} %
\toprule\toprule
&  \multicolumn{4}{c}{Father-Son}  & \multicolumn{4}{c}{Father-Daughter}  \\
\cmidrule{2-5}  \cmidrule{6-9} 
&  \multicolumn{2}{c}{Son}  & \multicolumn{2}{c}{Father} & \multicolumn{2}{c}{Daughter} & \multicolumn{2}{c}{Father} \\
 & Mean & SD & Mean & SD & Mean & SD & Mean & SD \\
\midrule
Income (1,000 CHF) &   91 &   43 &   80 &   42 &   52 &   35 &   80 &   42 \\ 
Age at birth &    &     &   26.8 &    3.3 &    &     &   26.9 &    3.3 \\ 
Higher Education &    0.56 &    0.50 &    0.34 &    0.47 &    0.50 &    0.50 &    0.34 &    0.48 \\ 
Months of Experience &  191 &   30 &   50 &   34 &  184 &   31 &   51 &   34 \\ 
Swiss Citizen &    0.96 &    0.20 &    0.88 &    0.33 &    0.96 &    0.21 &    0.87 &    0.34 \\ 
Single &    0.46 &    0.50 &    0.16 &    0.37 &    0.43 &    0.49 &    0.17 &    0.38 \\ 
Number of Children &    1.20 &    1.11 &    2.45 &    0.86 &    1.35 &    1.10 &    2.46 &    0.87 \\ 
%Year & 2012.14 &    3.19 & 2012.14 &    3.19 & 2012.32 &    3.14 & 2012.32 &    3.14 \\ 

\bottomrule\bottomrule

\end{tabular*}
\begin{tablenotes}[flushleft]
\footnotesize 
\item \textit{Notes: sample size is $10,363$ for father-son and $9,581$ for father-daughter.} 
\end{tablenotes}
\end{threeparttable}
\end{center}
\end{table}

%\subsection{Descriptive Statistics} Table \ref{t:desc} presents descriptive statistics for our estimation sample. 

\subsection{Rank-Rank Regressions} 
Table \ref{table:crrr} reports the results of RRR and CRRR. The CRRR results are obtained using Algorithms \ref{alg:crrr-reg} and \ref{alg:infer} for the correlation-based estimator with a logistic link function and a mesh of 200 points located at sample quantiles in a sequence of orders from $0.01$ to $0.99$ with increments of $0.98/199$. We use linear interpolation to obtain estimates of the conditional ranks corresponding to intermediate points in the mesh. The standard errors (SE) and 95\% confidence intervals (95\% CI) are computed by the empirical bootstrap with 500 repetitions. Based on the results of numerical simulations reported in Appendix \ref{sec:simul}, we do not impose  tail restrictions. In results not reported, we find very similar estimates, standard errors and confidence intervals for regression-based and fully restricted estimators.\footnote{These results are available from the authors upon request.} We show the robustness of the results to the choice of  link function in Section \ref{subsec:probit}.

We find significant positive income persistence in both father-son and father-daughter relationships, with and without covariates. However, the persistence is much stronger for sons than for daughters suggesting the presence of a gender gap in the intergenerational transmission of income  even after controlling for the father's and child's characteristics. 
% To compare RRR and CRRR, it is useful to decompose the overall income persistence ($\rho$) as follows:
% $$
% \rho = \rho_C + (\rho - \rho_C),
% $$
% where $\rho_C$ represents the average \textit{within-group} or unexplained income persistence, and the remainder term $(\rho - \rho_C)$ can be interpreted as \textit{between-group} income persistence or income persistence explained by covariates. Thus, the overall income persistence is equal to the sum of the average within-group income persistence and the between-group income persistence. 
Comparing RRR and CRRR,  we find that within-group persistence is about 62\%  of the overall income persistence for sons and  52\% for daughters. These results highlight the substantial role of both within-group and between-group differences in explaining intergenerational mobility.

A subgroup analysis reveals relatively more mobility in families with a larger number of children and with a low educated father. In particular, we find relatively less persistence for sons in large families and more for daughters of high educated fathers. This is consistent with decreasing returns of intergenerational transfers with respect to family size and increasing  with respect to father's education. This heterogeneity, however, is not statistically significant. We do not find differences in intergenerational mobility for families with immigrant fathers in Switzerland, unlike the results of \cite{abramitzky2021intergenerational} for the U.S. This difference might be due to the small fraction of immigrant fathers in the sample, see Table \ref{table:ds}.

% In the following, we compare four models. We (i) run the RRR without covariates and (ii) with the full set of covariates. Next, we run a reduced CRRR where only the children- and father-specific regressors are used in their univarate DRs, respectively. Finally, we allow all regressors to affect both univariate DRs. Table \ref{table:results} presents the results. We observe that within the RRR model, accounting for covariates lowers the correlation between fathers and children. Further, considering only the approaches using covariates, there is no statistically significant difference between the estimates. At the margin, the estimated correlation is lower for the CRRR models as likely, they account more precisely for the effect of the covariates. 

\begin{table}[ht]\caption{Intergenerational Income Mobility in Switzerland}\label{table:crrr}
\begin{center}
\begin{threeparttable}[b]
\setlength{\tabcolsep}{0pt}
%\caption{Intergenerational Mobility in Switzerland} 
\begin{tabular*}{15cm}{ @{\extracolsep{\fill}} lcccccccc} %
\toprule\toprule
&  \multicolumn{4}{c}{Father-Son}  & \multicolumn{4}{c}{Father-Daughter}  \\
\cmidrule{2-5}  \cmidrule{6-9} 
 & Coef. & SE & \multicolumn{2}{c}{95\% CI} & Coef. & SE & \multicolumn{2}{c}{95\% CI}  \\
\midrule
\input{"Tables/crrr_l_cor_2001"}
\multicolumn{9}{l}{CRRR, by Father:}\\ 
\input{"Tables/crrr_l_cor_2002"}
\bottomrule\bottomrule

\end{tabular*}
\begin{tablenotes}[flushleft]
\footnotesize 
\item \textit{Notes: Correlation-based estimator with logistic link function and a mesh of 200 points. SE and 95\% CI obtained by empirical bootstrap with 500 repetitions. Covariates include father's and child's months of experience, higher education, Swiss citizenship, single and number of own children; father's age  at birth; and child's year and canton fixed effects. Sample size is $10,363$ for father-son  and $9,581$ for  father-daughter data.}
\end{tablenotes}
\end{threeparttable}
\end{center}
\end{table}

\subsection{Transition Matrices} 
\begin{figure}[h]\caption{Transition matrices: Father-Son} \label{fig:tm-fs}
\vspace{.2cm}
\begin{scriptsize}
\begin{center}
	\begin{subfigure}{0.5\textwidth}
		\centering
		\includegraphics[scale=0.43,keepaspectratio]{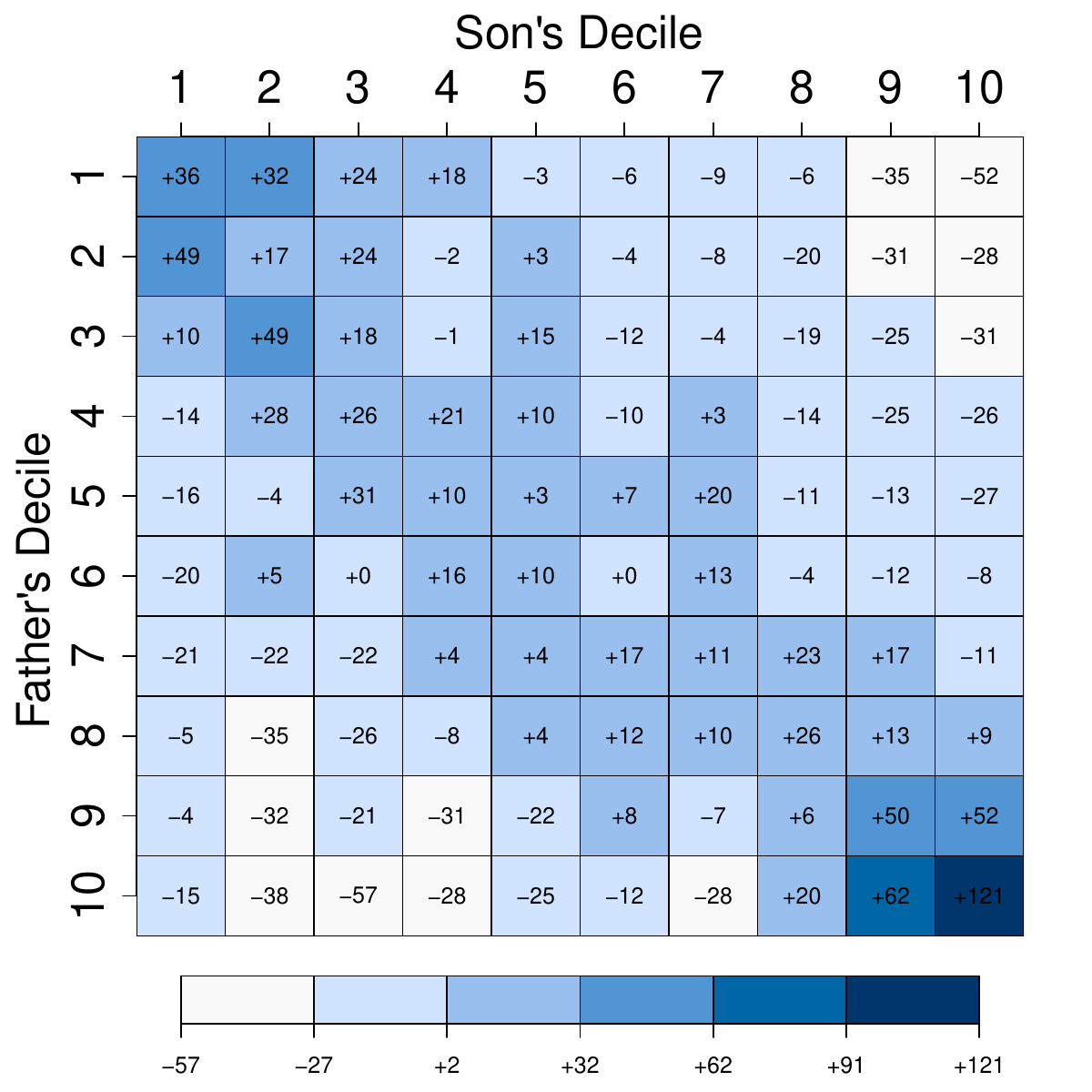}
		\caption{Marginal ranks}
	\end{subfigure}%
	\begin{subfigure}{0.5\textwidth}
		\centering
		\includegraphics[scale=0.43,keepaspectratio]{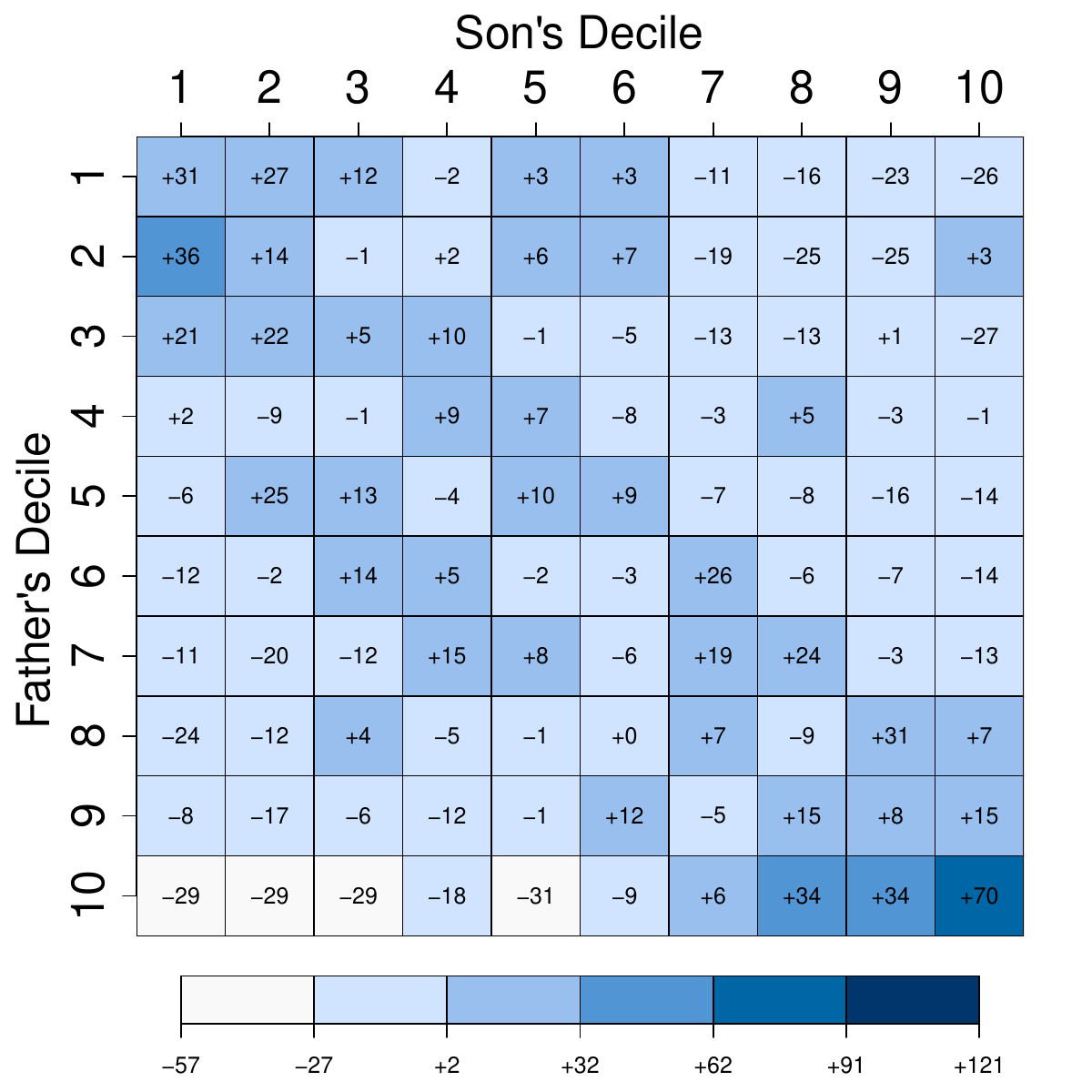}
		\caption{Conditional ranks}
	\end{subfigure} 
\caption*{\footnotesize \textit{Notes: Entries are in percent deviations from $0.01$. Sample size is $10,363$ for father-son and $9,581$ for father-daughter. } }
\end{center}
\end{scriptsize}
\end{figure}

\begin{figure}[h]
\caption{Transition matrices: Father-Daughter} \label{fig:tm-fd}
\vspace{.2cm}
\begin{scriptsize}
\begin{center}
	\begin{subfigure}{0.5\textwidth}
		\centering
		\includegraphics[scale=0.43,keepaspectratio]{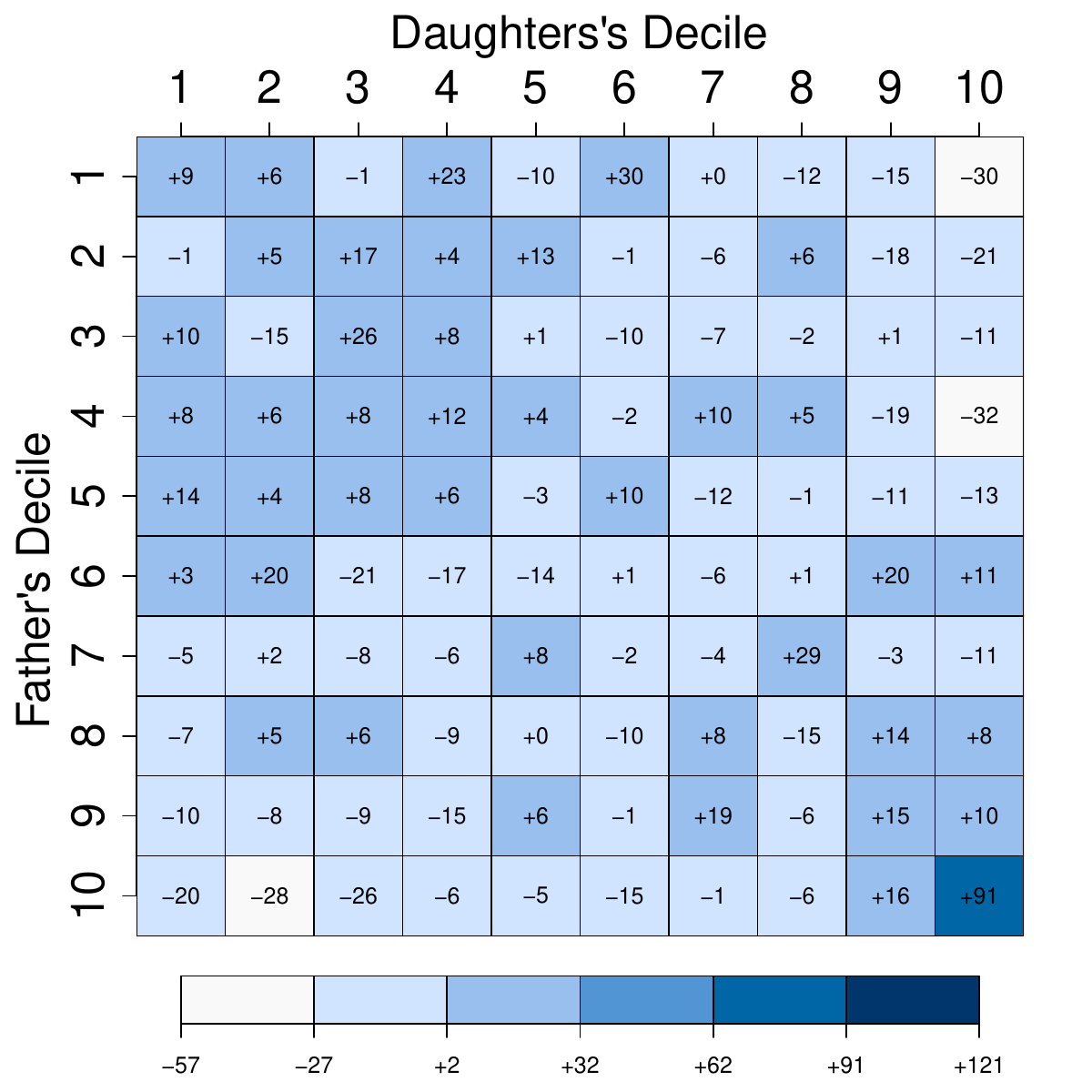}
		\caption{Marginal ranks}
	\end{subfigure}%
	\begin{subfigure}{0.5\textwidth}
		\centering
		\includegraphics[scale=0.43,keepaspectratio]{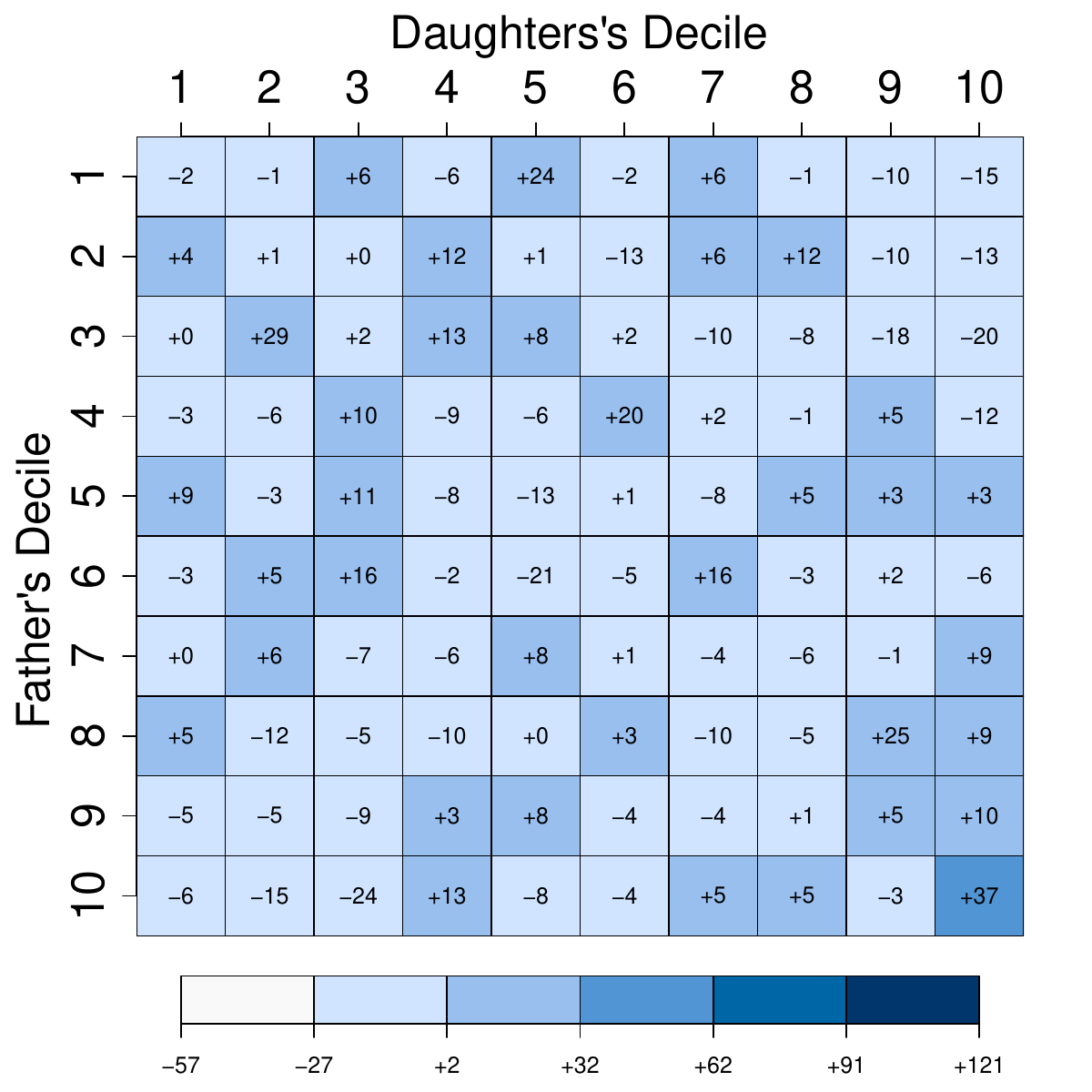}
		\caption{Conditional ranks}
	\end{subfigure} 
	\caption*{\footnotesize \textit{Notes:  Entries are in percent deviations from $0.01$. Sample size is $10,363$ for father-son and $9,581$ for father-daughter.} }
\end{center}
\end{scriptsize}
\end{figure}

Figures \ref{fig:tm-fs} and \ref{fig:tm-fd} show heatmaps of transition matrices for father-son and father-daughter, respectively. These matrices are a parsimonious representation of the joint distribution of income for father and child discretized in cells defined by deciles. They are commonly used in intergenerational mobility studies to provide a more granular measure of persistence than the rank-rank regressions. We report all the entries in percent deviations from $0.01$ because all the entries should be equal to $0.01$ under perfect mobility, that is when the income of the child is independent of the income of the father. Panels (A) report transition matrices based on marginal ranks, similar to previous studies. Panels (B) report conditional transition matrices based on conditional ranks, which are new to this paper and capture within-group dependence. For father-son, we find that the highest values in panel (A) are concentrated on the diagonal, which is consistent with the positive RRR estimate in Table \ref{table:crrr}. The results in panel (B) show a less clear pattern once we control for covariates, consistent with the lower CRRR estimates in Table \ref{table:crrr}.  The results for father-daughter show similar but weaker patterns as we expect from the smaller correlation estimates in Table \ref{table:crrr}. Interestingly, for both sons and daughters the highest probability occurs at the bottom right corner of the very top deciles conditionally and unconditionally.

% \begin{figure}[h]
% \caption{Conditional Density} \label{f:cpdf}
% \vspace{.2cm}
% \begin{scriptsize}
% \begin{center}
% 	\begin{subfigure}{0.45\textwidth}
% 		\centering
% 		\includegraphics[scale=0.33,keepaspectratio]{Figures/cpdf_fs.pdf}
% 		\caption{Sons}
% 	\end{subfigure}%
% 	\begin{subfigure}{0.45\textwidth}
% 		\centering
% 		\includegraphics[scale=0.33,keepaspectratio]{Figures/cpdf_fd.pdf}
% 		\caption{Daughters}
% 	\end{subfigure} 
% 	\caption*{\footnotesize \textit{Notes:} }
% \end{center}
% \end{scriptsize}
% \end{figure}

% \begin{figure}[h]
% \begin{scriptsize}
% \begin{center}
% 	\begin{subfigure}{0.5\textwidth}
% 		\centering
% 		\includegraphics[scale=0.6,keepaspectratio]{Figures/bpdf_fs.pdf}
% 		\caption{Sons}
% 	\end{subfigure}%
% 	\begin{subfigure}{0.5\textwidth}
% 		\centering
% 		\includegraphics[scale=0.6,keepaspectratio]{Figures/bpdf_fd.pdf}
% 		\caption{Daughters}
% 	\end{subfigure} 
% 	\vspace{-8.5cm}
% 	\caption{Bivariate PDF (kernel)} \label{f:bpdf}
% 	\vspace{7.5cm}
% 	\caption*{\footnotesize \textit{Notes:} }
% \end{center}
% \end{scriptsize}
% \end{figure}

\subsection{Rank-Rank Regressions Excluding Child's Covariates} One concern about the CRRR results in Table \ref{table:crrr} is that the child's covariates might be picking up indirect sources of intergenerational persistence of income. For example, fathers might invest in the child's education to increase the child's income prospects. To deal with this concern, Table \ref{table:nochild} reports CRRR results where the child's covariates, other than year and canton fixed effects, are excluded from the covariate set $X$. These results are obtained using the correlation-based estimator with a logistic link function with the same parameter choices as in Table \ref{table:crrr}.

As expected, not accounting for the child's covariates increases the importance of within-group persistence with the estimates increasing to about 80\% of overall persistence for father-son and 69\% for father-daughter. In both cases the increase is about 17-18 percentage points. The other conclusions remain unchanged. In particular, we still find a significant gender gap in intergenerational transmission of income, and relatively less persistence for sons in large families and more for daughters of high educated fathers. 

\begin{table}[h!]
\begin{center}
\begin{threeparttable}[b]
\setlength{\tabcolsep}{0pt}
\caption{Estimates Excluding Child's Covariates} \label{table:nochild}
\begin{tabular*}{15cm}{ @{\extracolsep{\fill}} lcccccccc} %
\toprule\toprule
&  \multicolumn{4}{c}{Father-Son}  & \multicolumn{4}{c}{Father-Daughter}  \\
\cmidrule{2-5}  \cmidrule{6-9} 
 & Coef. & SE & \multicolumn{2}{c}{95\% CI} & Coef. & SE & \multicolumn{2}{c}{95\% CI}  \\
\midrule
% & \multicolumn{8}{c}{Probit} \\
% \primitiveinput{"Tables/crrr_p_noson1"}
% \multicolumn{9}{l}{CRRR, by Father:}\\ 
% \primitiveinput{"Tables/crrr_p_noson_2"}
%  \\

% & \multicolumn{8}{c}{Logit} \\
\input{"Tables/crrr_l_cor_200_nochild1"}
\multicolumn{9}{l}{CRRR, by Father:}\\ 
\input{"Tables/crrr_l_cor_200_nochild2"}
\bottomrule\bottomrule

\end{tabular*}
\begin{tablenotes}[flushleft]
\tiny 
\item \textit{Notes: Correlation-based estimator with logistic link function  and a mesh of 200 points.  SE and 95\% CI obtained by empirical bootstrap with 500 repetitions. Covariates include father's months of experience, higher education, Swiss citizenship, single and number of own children; father's age at birth; and child's year and canton fixed effects. Sample size is $10,363$ for father-son  and $9,581$ for  father-daughter data.}
\end{tablenotes}
\end{threeparttable}
\end{center}
\end{table}

\subsection{Robustness to Link Function}\label{subsec:probit} Table \ref{table:crrr-probit} reports the results of CRRR using the correlation-based estimator with a Gaussian or probit link function. The estimates, standard errors and confidence intervals are almost identical to Table \ref{table:crrr} showing the robustness of the results to the use of the logistic versus Gaussian link functions. 

\begin{table}[h!]
\begin{center}
\begin{threeparttable}[b]
\setlength{\tabcolsep}{0pt}
\caption{Robustness to Link Function: Probit Estimates}\label{table:crrr-probit} 
\begin{tabular*}{15cm}{ @{\extracolsep{\fill}} lcccccccc} %
\toprule\toprule
&  \multicolumn{4}{c}{Father-Son}  & \multicolumn{4}{c}{Father-Daughter}  \\
\cmidrule{2-5}  \cmidrule{6-9} 
 & Coef. & SE & \multicolumn{2}{c}{95\% CI} & Coef. & SE & \multicolumn{2}{c}{95\% CI}  \\
\midrule
%& \multicolumn{8}{c}{Probit} \\
\input{"Tables/crrr_p_cor_2001"}
\multicolumn{9}{l}{CRRR, by Father:}\\ 
\input{"Tables/crrr_p_cor_2002"}
 % \\

% & \multicolumn{8}{c}{Logit} \\
% \primitiveinput{"Tables/crrr_l_base1"}
% \multicolumn{9}{l}{CRRR, by Father:}\\ 
% \primitiveinput{"Tables/crrr_l_base2"}
\bottomrule\bottomrule

\end{tabular*}
\begin{tablenotes}[flushleft]
\tiny 
\item \textit{Notes: Correlation-based estimator with Gaussian link function  and a mesh of 200 points.  SE and 95\% CI obtained by empirical bootstrap with 500 repetitions. Covariates include father's and child's months of experience, higher education, Swiss citizenship, single and number of own children; father's age at birth; and child's year and canton fixed effects. Sample size is $10,363$ for father-son  and $9,581$ for  father-daughter data.}
\end{tablenotes}
\end{threeparttable}
\end{center}
\end{table}

\section{Asymptotic Theory}\label{sec:theory}

In this section we provide asymptotic theory for the estimators of the CRRR slope $\rho_C$. We focus on the  correlation-based and fully-restricted estimators of Algorithm \ref{alg:crrr-reg}. We derive their asymptotic distributions by the delta method. For example, we take the following steps for the correlation-based estimator:
\begin{enumerate}
    \item Express the parameter $\rho_C$ as a correlation-based functional of the function-valued  inputs $F_{Y \mid X}$, $F_{W \mid X}$ and $F_Z$, where $Z = (Y,W,X')'$. That is: 
    \begin{equation}\label{eq:crrr-functional}
      \rho_C = \phi(F_{Y\mid X}, F_{W\mid X}, F_Z) := \frac{\int [F_{Y \mid X}(y\mid x) - .5] [F_{W \mid X}(w\mid x) - .5]  \dd F_{Z}(z)}{\sqrt{\int [F_{Y \mid X}(y\mid x) - .5]^2  \dd F_{Z}(z) \int [F_{W \mid X}(w\mid x) - .5]^2  \dd F_{Z}(z) }}.  
    \end{equation}
    \item Show that the plug-in estimator of $\rho_C$ using $\phi$, $\tilde \rho_C$, is a restricted correlation-based estimator:
    $$
    \tilde \rho_C = \phi(\hat F_{Y\mid X}, \hat F_{W\mid X}, \hat F_Z) =  \frac{\sum_{i=1}^n (\hat U_i - .5)(\hat V_i - .5) }{\sqrt{\sum_{i=1}^n (\hat V_i - .5)^2 \sum_{i=1}^n (\hat U_i - .5)^2}},
    $$
    where $\hat F_{Y \mid X}$ and $\hat F_{W \mid X}$  are the DR estimators of $F_{Y \mid X}$, $F_{W \mid X}$  and $\hat F_Z$ is the empirical distribution function of $Z$.
    \item Establish that the map $\phi$ is Hadamard differentiable in the relevant functional spaces at $(F_{Y\mid X}, F_{W\mid X}, F_Z)$ with the affine and continuous derivative operator:
    $$
    (z_Y,z_W,g_Z) \mapsto\phi'_{F_{Y\mid X}, F_{W\mid X}, F_Z}(z_Y,z_W,g_Z),
    $$
    where $z_Y$, $z_W$ and $g_Z$ are the limits of converging deviations from  $F_{Y\mid X}$, $F_{W\mid X}$ and $F_Z$. 
    
    \item Apply the functional delta method to obtain the limit of $\tilde \rho_C$ from the limit of the  deviations of the estimators of the inputs $\sqrt{n}(\hat F_{Y\mid X}-F_{Y\mid X})$, $\sqrt{n}(\hat F_{W\mid X}-F_{W\mid X})$, and $\sqrt{n}(\hat F_Z-F_Z)$,
    $$
    \sqrt{n}(\tilde \rho_C - \rho_C) \rightsquigarrow \phi'_{F_{Y \mid X},F_{W \mid X},F_{Z}}(Z_Y,Z_W,G_{Z}),
    $$
    where $\rightsquigarrow$ and $(Z_Y,Z_W,G_{Z})$ are defined below. 
    \item Show that $\hat \rho_C$ has the same asymptotic distribution as $\tilde \rho_C$,
    $$
    \sqrt{n}(\hat \rho_C - \tilde \rho_C) \to_P 0.
    $$
\end{enumerate}
The distribution of the fully-restricted estimator is derived following steps (1)-(4), and replacing the correlation-based functional in step (1) by the fully-restricted functional:
\begin{equation}\label{eq:crrr-functional-fr}
      \rho_C = 12 \ \phi_1(F_{Y\mid X}, F_{W\mid X}, F_Z) := 12 \int [F_{Y \mid X}(y\mid x) - .5] [F_{W \mid X}(w\mid x) - .5]  \dd F_{Z}(z).
    \end{equation}

\begin{remark}[Regression-Based Estimators] The limit distribution of the regression-based estimators is derived following analogous steps to the correlation-based estimator, and replacing the correlation-based representation of the functional in step (1) by the regression-based representation:
    \begin{equation}\label{eq:crrr-functional2}
      \rho_C = \varphi(F_{Y\mid X}, F_{W\mid X}, F_Z) := \frac{\int [F_{Y \mid X}(y\mid x) - .5] [F_{W \mid X}(w\mid x) - .5]  \dd F_{Z}(z)}{\int [F_{W \mid X}(w\mid x) - .5]^2  \dd F_{Z}(z) }.  
    \end{equation}
We provide the corresponding results in Appendix \ref{app:reg-estimators}. 
% The limit distribution of the reverse regression-based estimator can be trivially obtained from the regression-based case by relabeling the variables $Y$ and $W$. We omit the corresponding results for the sake of brevity.
\end{remark}

Before stating formally the main results, we review the existing theory for the estimator of the RRR slope. The purpose of this review is to explain why the existing results do not cover the estimators of the CRRR slope. \cite{hoeffding1948class} first derived the asymptotic distribution of the RRR slope estimator using the theory of U-statistics. We cannot follow the same approach because none of our estimators has a U-statistic representation. \cite{ren1995hadamard} alternatively derived the asymptotic distribution of the RRR slope estimator using the delta method. \cite{ren1995hadamard} used analogous steps to our procedure described above. The following remarks explain each step of our procedure and point out the challenges and differences with respect to \cite{ren1995hadamard}.

\begin{remark}[Functional Representation of $\rho_C$] The functional and the inputs of the functional representation of the RRR slope, $\rho$, are different from $\rho_C$. In particular, \cite{ren1995hadamard} showed that:
$$
\rho = \tilde \phi(F_{Y,W}) = 12 \int [F_{Y,W}(y,+\infty) - .5] [F_{Y,W}(+\infty,w) - .5]  \dd F_{Y,W}(y,w). 
$$
The functional  $\tilde \phi$ is a special case of the fully-restricted functional $12 \phi_1$ in \eqref{eq:crrr-functional-fr} where there are no covariates $X$. In the case of RRR, despite being a regression-based estimator, the denominator simplifies because the sample variances of the estimated marginal ranks are deterministic when $Y$ and $W$ are continuous.\footnote{Indeed, these sample variances are equal to  $(n^2-1)/(12n^2)$, see \cite{ren1995hadamard}; and the regression-based and correlation-based versions of the RRR estimator are numerically identical if there are no ties in the observations of $W$ and $Y$.} This simplification is not available for the correlation-based and regression-based estimators of  $\rho_C$ because the sample variances of the estimated conditional ranks are random. 
\end{remark}

\begin{remark}[Plug-in Estimator] The plug-in estimator of $\rho_C$ using $\phi$ is:
$$
\phi(\hat F_{Y\mid X}, \hat F_{W\mid X}, \hat F_Z) = \frac{\int [\Lambda(x'\hat \beta_Y(y)) - .5] [\Lambda(x'\hat \beta_W(w)) - .5]  \dd \hat F_{Z}(z)}{\sqrt{\int [\Lambda(x'\hat \beta_Y(y)) - .5]^2  \dd \hat F_{Z}(z) \int [\Lambda(x'\hat \beta_W(w)) - .5]^2  \dd \hat F_{Z}(z)}} = \tilde \rho_C,
$$
where the second equality follows from the properties of the empirical distribution function $\hat F_Z$.  This proves Step (2) above.   
\end{remark}

\begin{remark}[Hadamard Differentiability of $\phi$] The argument to establish differentiability of the RRR functional $\tilde \phi$ does not apply to the CRRR functional $\phi$ for several reasons. First, the expression of $\phi$ is different from $\tilde \phi$. Second, the inputs and their estimators are also different. Moreover, the estimators of the inputs of the CRRR functional live in more complicated spaces than the estimators of the inputs of the RRR functional. Thus, while the estimator of $F_Z$ lives in the space of c\`adl\`ag functions, the estimators of $F_{Y \mid X}$ and $F_{W \mid X}$ live in the space of bounded functions, but have limits in the space of continuous functions, once properly recentered and rescaled. Because of these differences, we need to establish Hadamard differentiability in the space of bounded functions, tangentially to the space of continuous functions. 
\end{remark}

\begin{remark}[Limit Process of Input Estimators] To apply the delta method, we need to characterize the limit process for the estimator of the inputs. This characterization is much more challenging for CRRR than RRR. Thus, for example, \cite{ren1995hadamard} can rely on existing functional central limit theorems for the empirical distribution to establish the limit process over the entire support of $Y$ and $W$. Unfortunately, the existing functional central limit theorems for  DR estimators of conditional distributions have only been established on a compact strict subset of the support of $Y$ and $W$; see, for example, \cite{chernozhukov+13inference}. We deal with this challenge by imposing restrictions on the DR model of the conditional distributions at the tails. These restrictions allow us to extend the central limit theorems to the entire support of $Y$ and $W$. In numerical simulations, however, we find that estimators with and without imposing the tail restrictions perform similarly in terms of bias, standard deviation and root mean squared error. 
\end{remark}

We formally state now the main results from the steps (4) and (5). The result from step (3) is relegated to Appendix \ref{app:hd} because it is of more technical nature. We state all the results for the logistic link function because it produces analytically simpler expressions, but it can be readily extended to the Gaussian link at the cost of more cumbersome notation. 

We start by imposing some conditions on the DR model. 

\begin{assumption}[DR Model]\label{ass:dr}  For $R \in \{Y,W\}$: (a) The conditional distribution function
takes the form $F_{R \mid X}(r \mid x)=\Lambda(x^{\prime}\beta_{R}(r))$ for all $r \in \mathcal{R}$ and $x \in \mathcal{X}$, where $\Lambda(u) = (1+\exp(-u))^{-1}$, the  standard logistic distribution. (b) The support $\mathcal{R}$
is an open interval in $\mathbb{R}$ and the conditional density function $
f_{R \mid X}(r \mid x)$ exists and  is positive  in $(r,x)$ on  $(\RR,\mX)$; it is  uniformly bounded and uniformly continuous
in $(r,x)$ on  $(\RR,\mX)$. (c) $E\|X \|^{2} < \infty$ and
the minimum eigenvalue of: 
\begin{equation*}
J_R(r) := \Ep \left[ \lambda(X^{\prime}\beta_R(r))
X X^{\prime}\right] ,
\end{equation*}
is bounded away from zero uniformly over $r \in \bar{\mathcal{R}}$, where $\lambda = \Lambda(1-\Lambda)$ is the derivative of $\Lambda$ and $\bar{\mathcal{R}}$ is a closed subinterval of the interior of $\RR$. (d) Let $\beta_R(r)$ be partitioned as $(\beta_{R,1}(r),\beta_{R,-1}(r)')'$ where $\beta_{R,1}(r)$ is the intercept and $\beta_{R,-1}(r)$ includes the rest of the components. Then, for $r \in \mathcal{R}\setminus\bar{\mathcal{R}}$,  
$\beta_{R,1}(r) = \beta_{R,1}(\bar r) + (r - \bar r)\alpha_{R}(\bar r)$ for $\bar r := \arg \min_{r' \in \bar{\mathcal{R}}} |r-r'|$ and some $\alpha_R(\bar r) > 0$,  and $\beta_{R,-1}(r) = \beta_{R,-1}(\bar r)$.
\end{assumption}

\begin{remark}[DR Model] The conditions in Assumption \ref{ass:dr}(a)-(c) are the same as in \cite{chernozhukov+13inference}. They are used to obtain a functional central limit for the DR estimator of the conditional distribution $F_{R \mid X}(r \mid x)$ on $\bar{\mathcal{R}}$. Assumption \ref{ass:dr}(d) imposes restrictions on the tails that allow us to extend the functional central limit theorem to $\RR$. 
\end{remark}

In order to state the result about the limit process for the inputs, we define, for $R \in \{Y,W\}$, 
\begin{multline*}
\ell_{r,x}(R,X) = 1\{r \in  \bar{\mathcal{R}}\}   \lambda(x^{\prime}\beta_R(r))x^{\prime}J_{R}^{-1}(r) \left[ 1 \{R \leq r\} - \Lambda(X^{\prime}%
\beta_R(r)) \right]X \\ 
+ 1\{r \in  \mathcal{R}\setminus\bar{\mathcal{R}}\}  \lambda(x^{\prime}\beta_R(r)) \left\{ \frac{r - \bar r}{r_0 - \bar r}   \frac{ 1 \{R \leq r_0\} - \Lambda(X^{\prime}
\beta_R(r_0))  }{\Ep[\lambda(X^{\prime} 
\beta_R(r_0))]} \right.
\\ 
+   \left. \left[x - \frac{r - \bar r}{r_0 - \bar r}  \frac{\Ep[\lambda(X^{\prime}
\beta_R(r_0))X]}{\Ep[\lambda(X^{\prime}
\beta_R(r_0))]}\right]^{\prime} 
 J_{R}^{-1}(\bar r) \left[1 \{R \leq \bar r\} - \Lambda(X^{\prime}%
\beta_R(\bar r))  \right] X \right\}.
\end{multline*}
% \begin{eqnarray*}
% \ell_{r,x}(R,X) = \lambda(x^{\prime}\beta_R(r))x^{\prime}\psi
% _{r}(R,X) , \quad \psi_{r}(R,X):=  J_R^{-1}(r) [1 \{R \leq r\} - \Lambda(X^{\prime}\beta_R(r)) ] X.
% \end{eqnarray*}
Consider the empirical processes
$
(r,x) \mapsto \hat Z_R(r,x) := \sqrt{n}\left(\hat F_{R \mid X}(r \mid x) - F_{R \mid X}(r \mid x) \right)$, $R \in \{Y,W\}$, and $f \mapsto \hat G_Z(f) := \sqrt{n} \int f \mathrm{d} (\hat F_Z - F_Z),$ where $\hat F_{R \mid X}(r \mid x) := \Lambda(x'\hat \beta_R(r))$, $\hat F_Z$ is the empirical distribution function of $Z = (Y,W,X)$, and $\mathcal{F}$ is a class of measurable functions that (i) consists of the multiplication of monomials of total degree at most two in $(F_{Y \mid X}-.5)$ and $(F_{W \mid X}-.5)$ with indicators of all the rectangles in $\bar{\mathbb{R}}^{d_x+2}$, where $\overline{\R} := \R \cup \{-\infty, \infty\}$  is the extended real line, and (ii) is totally bounded under the metric: 
$$\lambda(f,\tilde f) = \left[\int (f-\tilde f)^2 \mathrm{d} F_Z \right]^{1/2}, \quad f,\tilde f \in \mathcal{F}.$$ 
Let $Z_n\rightsquigarrow Z $ in $\mathbb{E}$ denote weak convergence of a
stochastic process $Z_n$ to a random element $Z$ in a normed space $\mathbb{E
}$, as defined in \cite{van1996weak}.

\begin{lemma}[Limit Processes for Inputs]\label{lemma:crrr} Assume that Assumption \ref{ass:dr} holds,  the support $\mathcal{X}$ is a compact subset of $\mathbb{R}^{d_x}$, and $\{Z_i=(Y_i,W_i,X_i)\}_{i=1}^n$ is a random sample of $Z=(Y,W,X)$. Then, in the metric space $\ell^{\infty}(\mathcal{Y}\mathcal{W}\mathcal{X}\mathcal{F})$,
$$
(\hat Z_Y(y,x), \hat Z_W(w,x), \hat G_Z(f)) \rightsquigarrow (Z_Y(y,x), Z_W(w,x),  G_Z(f))
$$    
as stochastic processes indexed by $(y,w,x,f)$. The limit process is a zero-mean tight Gaussian process such that:
\begin{equation*}
Z_{R}(r,x) =  \mathbb{G}( \ell_{r,x}), \ \ R \in \{Y,W\}, \text{ and } \ G_Z(f) = {\mathbb{G}}(f),
\end{equation*}
where ${\mathbb{G}}$ is a $P$-Brownian bridge. 
%(ii) Exchangeable bootstrap consistently estimates the limit law of these processes under Assumption \ref{ass:eb}
\end{lemma}

The next result states a central limit theorem for $\tilde \rho_C$ and $\breve \rho_C$, and the asymptotic equivalence between $\tilde \rho_C$ and $\hat \rho_C$.

\begin{theorem}[Limit Distribution of $\hat \rho_C$, $\tilde \rho_C$ and $\breve \rho_C$] 
\label{theorem:main} Under the conditions of Lemma \ref{lemma:crrr}: (1)  in $\mathbb{R}$,
\begin{equation*}  
\sqrt{n} \left( \tilde \rho_{C} - \rho_{C} \right) \rightsquigarrow Z_{\rho} := 12 \ [Z_{1,\rho} - \rho_{C} (Z_{2,\rho} + Z_{3,\rho})/2] \ \ \text{ and } \ \ \sqrt{n} \left( \breve \rho_{C} - \rho_{C} \right) \rightsquigarrow  12 Z_{1,\rho},
\end{equation*}
where $Z_{1,\rho}$, $Z_{2,\rho}$ and $Z_{3,\rho}$ are zero-mean Gaussian random variables defined by 
\begin{multline*}
Z_{1,\rho} :=  \int  \left\{ Z_{Y}(y,x)[F_{W\mid X}(w\mid x) - .5] + Z_{W}(w,x)[F_{Y\mid X}(y\mid x) - .5] \right\} \mathrm{d} F_{Z}(y,w,z) \\ + G_Z\left([F_{Y\mid X} - .5][F_{W\mid X} - .5]\right), 
\end{multline*} 
\begin{equation*}
Z_{2,\rho} :=  2  \int Z_{W}(w,x)[F_{W\mid X}(w \mid x) - .5] \mathrm{d} F_{Z}(y,w,z) + G_Z\left([F_{W\mid X} - .5]^2\right), 
\end{equation*}  
and
\begin{equation*}
Z_{3,\rho} :=  2  \int Z_{Y}(y,x)[F_{Y\mid X}(y \mid x) - .5] \mathrm{d} F_{Z}(y,w,z) + G_Z\left([F_{Y\mid X} - .5]^2\right). 
\end{equation*}  
(2)  $\hat \rho_{C}$ has the same limit distribution as $\tilde \rho_{C}$ because 
$$
\sqrt{n} \left( \hat \rho_{C} - \tilde \rho_{C} \right) \to_{\Pr} 0.
$$
% (2)  In $\mathbb{R}$,
% $$\sqrt{n} \left( \hat \rho^{*}_{Y,W \mid X} - \hat \rho_{Y,W \mid X} \right) \rightsquigarrow
% _{\Pr } 12 \ Z_{Y,W \mid X},$$ 
% that is, exchangeable bootstrap consistently estimates
% the law of the limit process $12\ Z_{Y,W \mid X}$.
\end{theorem}

The variances of the limit processes $Z_{1,\rho}$ and $Z_{\rho}$ have complicated expressions that might be difficult to estimate analytically. To avoid this difficulty, we propose the use of bootstrap to make inference. We show that the exchangeable bootstrap draws of Algorithm \ref{alg:eb} have the same asymptotic distribution as the CRRR estimators under the following assumption on the weights:

\begin{assumption}[Exchangeable Bootstrap]\label{ass:eb} For each $n$, $(\omega_{n1}, ..., \omega_{nn})$ is an exchangeable,\footnote{A sequence of random variables $X_1, X_2, ..., X_n$ is exchangeable if for any finite permutation $\sigma$ of the indices $1,2, ..., n$ the joint
distribution of the permuted sequence $X_{\sigma(1)}, X_{\sigma(2)},
...,X_{\sigma(n)} $ is the same as the joint distribution of the original
sequence.} nonnegative random vector, which is independent of the data, such that for some $\epsilon> 0:$ 
\begin{equation}  \label{eq: assumptions weighted bootstrap}
\begin{split}
\sup_{n} \Ep[\omega_{n1}^{2+\epsilon}] < \infty, \ \ n^{-1}\sum
_{i=1}^{n} \left( \omega_{ni} - \bar{\omega}_n \right)^{2} \to_{\Pr} 1, \ \
\bar \omega_n \to_{\Pr} 1,
\end{split}
\end{equation}
where $\bar \omega_n = n^{-1} \sum_{i=1}^{n} \omega_{ni} $. 
\end{assumption}

% \begin{remark}[Common bootstrap schemes]
% As pointed out in \cite{van1996weak}, by appropriately
% selecting the distribution of the weights, exchangeable bootstrap covers the
% most common bootstrap schemes as special cases. The empirical bootstrap
% corresponds to the case where $(w_{n1},...,w_{nn})$ is a multinomial
% vector with parameter $n$ and probabilities $(1/n,...,1/n)$. The
% weighted bootstrap corresponds to the case where $w_{n1},...,w_{nn}$ are
% i.i.d. nonnegative random variables with $E[w_{n1}]=Var[w_{n1}]=1$, e.g.
% standard exponential. 
% %The Bayesian or weighted bootstrap corresponds to the case where $%
% %U_{k1}, ..., U_{kn_{k}}$ are i.i.d. nonnegative random variables, e.g. unit
% %exponential, with $E| U_{k1}^{2+\epsilon}] < \infty$ for some $%
% %\epsilon> 0$, and $w_{ki} = U_{ki} / \bar U_{k},$ where $\bar U_{k} = {%
% %n_{k}}^{-1} \sum_{i=1}^{n_{k}} U_{ki}$.
% The wild bootstrap corresponds to the case where $w_{n1}, ...,
% w_{nn}$ are i.i.d. vectors with $E[w_{n1}^{2+\epsilon}] < \infty$
% for some $\epsilon > 0$, and $Var[w_{n1}]=1$.
% The $m$ out of $n$ bootstrap corresponds to letting $(w_{n1},...,w_{nn})$
% be equal to $\sqrt{n_/m}$ times multinomial vectors with parameter $%
% m$ and probabilities $(1/n,...,1/n)$. The subsampling bootstrap
% corresponds to letting $(w_{n1},...,w_{nn})$ be a row in which the
% number $n(n-m)^{-1/2}m^{-1/2}$ appears $m$ times and 0
% appears $n-m$ times ordered at random, independent of the data. 
% \end{remark}

To state the results about bootstrap validity formally, we follow the notation and
definitions in \cite{van1996weak}. Let $D_{n}$ denote the data
vector and $M_{n}$ be the vector of random variables used to generate
bootstrap draws given $D_{n}$. Consider the random element $%
\mathbb{Z}^{*}_{n} = \mathbb{Z}_{n}(D_{n}, M_{n})$ in a normed space $%
\mathbb{E}$. We say that the bootstrap law of $\mathbb{Z}^{*}_{n}$
consistently estimates the law of some tight random element $\mathbb{Z}$ and
write $\mathbb{Z}^{*}_{n} \rightsquigarrow_{\Pr} \mathbb{Z} $ in $\mathbb{E}$
if: 
\begin{equation}  \label{boot1}
\begin{array}{r}
\sup_{h \in\text{BL}_{1}(\mathbb{E})} \left| \Ep_{M_{n}} h \left( \mathbb{Z}%
^{*}_{n}\right) - \Ep h(\mathbb{Z})\right| \rightarrow_{\Pr} 0,%
\end{array}%
\end{equation}
where $\text{BL}_{1}(\mathbb{E})$ denotes the space of functions with
Lipschitz norm at most 1 and $\Ep_{M_{n}}$ denotes the conditional expectation
with respect to $M_{n}$ given the data $D_{n}$; and $ \rightarrow_{\Pr}$ denotes
convergence in (outer) probability.

We now provide a bootstrap central limit theorem for the estimators of the CRRR slope. This result follows from a functional central limit theorem for the input processes, which we establish in Lemma \ref{lemma:crrr-boot} in Appendix \ref{app:theory}, and  the functional delta method for the bootstrap.   
\begin{theorem}[Exchangeable Bootstrap Consistency]\label{thm:main-boot} 
Under the conditions of Lemma \ref{lemma:crrr} and Assumption \ref{ass:eb}: in $\mathbb{R}$,
$$\sqrt{n} \left( \tilde \rho^{*}_{C} - \tilde \rho_{C} \right) \rightsquigarrow
_{\Pr }  Z_{\rho}, \ \ \ \ \sqrt{n} \left( \hat \rho^{*}_{C} - \hat \rho_{C} \right) \rightsquigarrow
_{\Pr }  Z_{\rho}, \ \ \text{ and } \ \ \sqrt{n} \left( \breve \rho^{*}_{C} - \breve \rho_{C} \right) \rightsquigarrow
_{\Pr }  12 Z_{1,\rho}$$ 
that is, exchangeable bootstrap consistently estimates
the law of the limit processes $Z_{\rho}$ and $12Z_{1,\rho}$. In particular, if $B=B_n \to \infty$ and $\sigma_{\rho} >0$,
$$
\hat \sigma_{\rho} \to_{\Pr} \sigma_{\rho} \ \ \text{ and } \ \ \Pr\left\{ \rho_C \in \textrm{ACI}_{1-\alpha}(\rho_{C}) \right\} \to 1-\alpha \ \ \text{ as } \ \  n \to \infty,
$$
where $\sigma_{\rho}$ is the standard deviation of the limit process $Z_{\rho}$, and $\hat \sigma_{\rho}$ and $\textrm{ACI}_{1-\alpha}(\rho_{C})$ are defined in Algorithm \ref{alg:infer}.
\end{theorem}

\section{Conclusion}\label{sec:conclusion}

This paper introduces the conditional rank-rank regression (CRRR) as an alternative to traditional rank-rank regressions with covariates (RRRX) for measuring within-group mobility and persistence. The CRRR uses conditional ranks of the variables of interest given covariates, in contrast to RRRX which uses marginal ranks net of covariate effects. We show that the CRRR slope preserves an intuitive interpretation as the average conditional rank correlation between the variables, similar to RRR without covariates. In contrast, the slope of RRRX loses the rank correlation interpretation and can take on 
%extreme 
values outside the interval $[-1, 1].$ The CRRR is also suitable for subgroup analysis, where the CRRR slopes maintain a rank correlation interpretation conditional on the groups.

We propose a distribution regression estimator for CRRR where the conditional distributions are modeled flexibly using parametric link functions. The estimator is easy to implement and computationally tractable. We derive asymptotic theory for the estimator based on the functional delta method. The analytic asymptotic variance is cumbersome, so we propose an exchangeable bootstrap procedure for inference. The bootstrap procedure is also used to construct confidence intervals. We illustrate the usefulness of CRRR in an empirical application to intergenerational income mobility in Switzerland. The application reveals stronger intergenerational persistence between fathers and sons than fathers and daughters, where the within-group persistence represents between $52\%$ and $80\%$ of the overall persistence. We also find some evidence of heterogeneity across groups defined by father's education and family size. The results are robust to the exclusion of child's covariates and the use of logistic or Gaussian link functions. 

In summary, CRRR provides a well-grounded measure of within-group mobility and persistence. It can also be readily adapted to predict marginal ranks while ensuring that the predicted values lie in the interval $[0,1]$, unlike RRRX.
%It also allows us to decompose the overall persistence captured by RRR into within-group persistence captured by CRRR plus a remainder term interpretable as between-group persistence. 
The distribution regression estimator, coupled with exchangeable bootstrap inference, provides a practical and flexible way to implement CRRR in empirical applications. We expect CRRR will be a useful addition to the toolkit of methods for studying mobility and persistence. A natural next step is to consider high dimensional settings where there might be many covariates. An extension where we derive orthogonal moment conditions for CRRR to apply double/debiased machine learning (DML) is underway.

\bibliographystyle{ecta}
\bibliography{bibliography}

\appendix 

\section{Regression-Based Estimators}\label{app:reg-estimators}

\begin{algorithm}[Regression-based Estimators]\label{alg:crrr-cor} Steps (1)--(3) are the same as in Algorithm \ref{alg:crrr-reg}. In step (4) estimate $
\rho_{C} 
$ as either (a) the slope of the linear regression of $\hat U_i$ on $\hat V_i$, that is
$$
\hat \varrho_{C}  = \frac{\sum_{i=1}^n \hat U_i (\hat V_i - \overline{\hat V}) }{\sum_{i=1}^n (\hat V_i - \overline{\hat V})^2}, \quad \overline{\hat V} = \frac{1}{n} \sum_{i=1}^n \hat V_i;
$$
or (b) the slope of the restricted linear regression of $\hat U_i$ on $\hat V_i$, that is
$$
\tilde \varrho_{C}  = \frac{\sum_{i=1}^n (\hat U_i - .5)(\hat V_i - .5) }{\sum_{i=1}^n (\hat V_i - .5)^2}.
$$
\end{algorithm}

% \begin{algorithm}[Correlation-based Restricted Estimator]\label{alg:crrr4} Steps (1)--(3) are the same as in Algorithm \ref{alg:crrr}. In step (4) estimate $\rho_C$ as the restricted sample correlation between $\widehat U_i$ and $\widehat V_i$, that is
% $$
% \tilde \varrho_{C}  = \frac{\sum_{i=1}^n (\hat U_i - .5)(\hat V_i - .5) }{\sqrt{\sum_{i=1}^n (\hat V_i - .5)^2 \sum_{i=1}^n (\hat U_i - .5)^2}}.
% $$
% \end{algorithm}

\begin{theorem}[Limit Distribution of $\tilde \varrho_C$ and $\hat \varrho_C$] 
\label{theorem:main-corr} Under the conditions of Lemma \ref{lemma:crrr}: (1)  in $\mathbb{R}$,
\begin{equation*}  
\sqrt{n} \left( \tilde \varrho_{C} - \rho_{C} \right) \rightsquigarrow 12 \ [Z_{1,\rho} - \rho_{C} Z_{2,\rho}],
\end{equation*}
where $Z_{1,\rho}$ and $Z_{2,\rho}$  are zero-mean Gaussian random variables defined in Theorem \ref{theorem:main}. 
% Theorem by 
% \begin{multline*}
% Z_{1,\rho} :=  \int  \left\{ Z_{Y}(y,x)[F_{W\mid X}(w\mid x) - .5] + Z_{W}(w,x)[F_{Y\mid X}(y\mid x) - .5] \right\} \mathrm{d} F_{Z}(y,w,z) \\ + G_Z\left([F_{Y\mid X} - .5][F_{W\mid X} - .5]\right), 
% \end{multline*} 
% \begin{equation*}
% Z_{2,\rho} :=  2  \int Z_{W}(w,x)[F_{W\mid X}(w \mid x) - .5] \mathrm{d} F_{Z}(y,w,z) + G_Z\left([F_{W\mid X} - .5]^2\right),
% \end{equation*}  
% and
% \begin{equation*}
% Z_{3,\rho} :=  2  \int Z_{Y}(y,x)[F_{W \mid X}(y \mid x) - .5] \mathrm{d} F_{Z}(y,w,z) + G_Z\left([F_{Y\mid X} - .5]^2\right). 
% \end{equation*}  
(2)  $\hat \varrho_{C}$ has the same limit distribution as $\tilde \varrho_{C}$ because 
$$
\sqrt{n} \left( \hat \varrho_{C} - \tilde \varrho_{C} \right) \to_{\Pr} 0.
$$
% (2)  In $\mathbb{R}$,
% $$\sqrt{n} \left( \hat \rho^{*}_{Y,W \mid X} - \hat \rho_{Y,W \mid X} \right) \rightsquigarrow
% _{\Pr } 12 \ Z_{Y,W \mid X},$$ 
% that is, exchangeable bootstrap consistently estimates
% the law of the limit process $12\ Z_{Y,W \mid X}$.
\end{theorem}

\begin{remark}[Reverse Regression-Based Estimators]
    The limit distribution of the reverse regression-based estimator can be trivially obtained from the regression-based case by relabeling the variables $Y$ and $W$. Thus, let $\tilde r_C$ denote the reverse regression-based restricted estimator, that is
$$
\tilde r_C = \frac{\sum_{i=1}^n (\hat U_i - .5)(\hat V_i - .5) }{\sum_{i=1}^n (\hat U_i - .5)^2}.
$$
By Theorem \ref{theorem:main-corr}, switching the roles of $Y$ and $W$,
$$
\sqrt{n}(\tilde r_C - \rho_C) \rightsquigarrow 12 \ [Z_{1,\rho} - \rho_{C} Z_{3,\rho}] \text{ in } \mathbb{R},
$$
where $Z_{3,\rho}$  is a zero-mean Gaussian random variable defined in Theorem \ref{theorem:main}. 
\end{remark}

\begin{remark}[Correlation-based vs. Regression-based Estimators] The correlation-based estimators have the same first-order limiting distribution as the average of the regression-based and reversed regression-based restricted estimators. To see this equivalence, we combine  $\sqrt{n}(\tilde \varrho_C - \rho_C) \rightsquigarrow 12 \ [Z_{1,\rho} - \rho_{C} Z_{2,\rho}] $ with $\sqrt{n}(\tilde r_C - \rho_C) \rightsquigarrow 12 \ [Z_{1,\rho} - \rho_{C} Z_{3,\rho}]$ to get
$$
\sqrt{n}\left((\tilde \varrho_C + \tilde r_C)/2 - \rho_C \right) \rightsquigarrow 12 \ [Z_{1,\rho} - \rho_{C} (Z_{2,\rho}+Z_{3,\rho})/2] \text{ in } \mathbb{R}.
$$
The same result applies for the average of the regression-based and reversed regression-based unrestricted estimators.
\end{remark}

\begin{remark}[Relative Efficiency] The relative asymptotic efficiency of the different estimators depends on the variances of the components of the limit processes and the correlations between them. For example, the fully-restricted estimator $\breve \rho_C$ is relatively more efficient than the regression-based restricted estimator $\tilde \varrho_C$ if
$$
\Corr(Z_{1,\rho},\rho_C Z_{2,\rho}) \leq \frac{1}{2} \sqrt{\frac{\Var(\rho_C Z_{2,\rho})}{\Var(Z_{1,\rho})}},
$$    
and relative to the correlation-based estimator $\tilde\rho_C$ if
$$
\Corr(Z_{1,\rho},\rho_C (Z_{2,\rho}+Z_{3,\rho})/2) \leq \frac{1}{2} \sqrt{\frac{\Var(\rho_C (Z_{2,\rho}+Z_{3,\rho})/2)}{\Var(Z_{1,\rho})}}.
$$    
The correlation-based estimator  is relatively no less efficient than the regression-based restricted estimator  if $\Var(Z_{2,\rho}) = \Var(Z_{3,\rho})$ and $\Cov(Z_{1,\rho},Z_{2,\rho}) = \Cov(Z_{1,\rho},Z_{3,\rho})$.
\end{remark}

\section{Proofs of Section \ref{sec:theory}}\label{app:theory}

\subsection{Hadamard Differentiability of CRRR Functionals}\label{app:hd}
We start by establishing the Hadamard differentiability of the fully-restricted functional $\phi_1$ defined in \eqref{eq:crrr-functional-fr} and characterizing the expression of the corresponding derivative.  Next, we establish the Hadamard differentiability of the correlation-based and regression-based functionals $\phi$ and $\varphi$ defined in \eqref{eq:crrr-functional} and \eqref{eq:crrr-functional2}, respectively,  and characterize the corresponding derivatives. We provide a brief proof for the results for $\phi$ and $\varphi$  because they  follow by similar arguments as the proof for $\phi_1$. 

% Let $\mathcal{Y}$, $\mathcal{W}$ and $\mathcal{X}$ denote the supports of $Y$, $W$ and $X$, respectively, and $\mathcal{F}$  a $F_Z$-Donsker class, where $Z := (Y,W,X)$.
% In order to state the next result, we define the pseudometric
% $\rho_{L^2(P)}$ on $\mathcal{Y}\times\mathcal{R},$ and on
% $\mathcal{F}$ by
%  $$
%  \rho_{L^2(P)}  ((r,x),(\tilde{r},\tilde{x})) =  \left[\Ep\Big \{ Z(r,x) - Z(\tilde{r}, \tilde{x})\Big
%  \}^2\right]^{1/2}, 
% $$
%   $$
%   \rho_{L^2(P)}  (f,\tilde{f}) =  \left[\Ep \left\{ G_{Z}(f)  - G_{Z}(\tilde{f})\right\}^2
%   \right]^{1/2},
%  $$
% where $\mathcal{R} = \{\mathcal{Y}, \mathcal{W}\}$. It follows from Lemma 18.15 in van der Vaart (1998) that
% $\mathcal{R}\times\mathcal{X}$ is totally bounded under
% $\rho_{L^2(P)}$  and $Z$ has continuous paths with
% respect to $\rho_{L^2(P)}$. Moreover, the completion of
% $\mathcal{R}\times\mathcal{X}$, denoted
% $\overline{\mathcal{R}\times\mathcal{X}}$,  with respect to either
% of the pseudometrics is compact. Likewise, $\mathcal{F}$ is totally
% bounded under $\rho_{L^2(P)}$.

We need some setup and preliminary observations. For $\RR \in \{\mathcal{Y},\mW\}$, let $\ell^{\infty}_m(\RR\mathcal{X})$ denote the set of all bounded and measurable
mappings $\mathcal{R}\mathcal{X} \mapsto \mathbb{R}$. Let $\overline{\R} := \R \cup \{-\infty, \infty\}$  be the extended real line. We consider $\mathcal{R}\mathcal{X}$ as a subset of $%
\overline{\mathbb{R}}^{d_x+1}$, with relative topology. Let $\rho$ denote a
standard metric on $\overline{\mathbb{R}}^{d_x+1}$. The closure of $\mathcal{%
R}\mathcal{X} $ under $\rho$, denoted $\overline{\mathcal{R}\mathcal{X}}$,
is compact in $\overline{\mathbb{R}}^{d_x+1}$. Let $UC(\mathcal{R}\mathcal{X}, \rho)$ be the set of functions
mapping $\mathcal{R}\mathcal{X}$ to the real line that are uniformly
continuous with respect to the metric $\rho$ , and can be continuously extended
to $\overline{\mathcal{R}\mathcal{X}}$, so that $UC(\mathcal{R}\mathcal{X},
\rho) \subset\ell^{\infty}_m(\mathcal{R}\mathcal{X})$. For a class of functions $\mathcal{F}$, let $UC(\mathcal{F}, \lambda)$ be the set of functionals mapping $\mathcal{F}$ to the real line that are uniformly continuous with respect to the (semi) metric  $\lambda(f,\tilde{f}) = [\Pr (f -
\tilde f)^2]^{1/2}$.

\begin{lemma}[Hadamard differentiability of $\phi_1$]
\label{lemma: Hadamard dif of CRRR FR} Let $\mathcal{RX} \subseteq 
\mathbb{R}^{d_x+1}$, $\RR \in \{\mathcal{Y},\mW\}$, and $\mathcal{F}$ be the class of bounded functions,
mapping $\overline{\mathbb{R}}^{d_x+2}$ to $\mathbb{R}$, that contains the multiplication of monomials of total degree at most two in $(F_{Y \mid X}-.5)$ and $(F_{W \mid X}-.5)$ with indicators of all the rectangles in $\bar{\mathbb{R}}^{d_x+2}$, such that $\mathcal{F}$ is
totally bounded under $\lambda$. Let $\mathbb{D}_{\phi}$ be the product of
the spaces of measurable functions $\Gamma_Y: \mathcal{Y}\mathcal{X} \mapsto[-.5,.5
]$ defined by $(y,x) \mapsto \Gamma_Y(y,x)$ and $\Gamma_W: \mathcal{W}\mathcal{X} \mapsto[-.5,.5%
]$ defined by $(w,x) \mapsto \Gamma_W(w,x)$, and the bounded maps $\Pi: 
\mathcal{F} \mapsto\mathbb{R}$ defined by $f \mapsto\int f \dd\Pi$, where $\Pi$
is restricted to be a probability measure on $\mathcal{Z} := \mathcal{YWX}$. Consider the map 
$\phi_1: \mathbb{D}_{\phi} \subset \mathbb{D} = \ell^{\infty}_m(\mathcal{YX}) \times \ell^{\infty}_m(\mathcal{W}
\mathcal{X}) \times\ell^{\infty }( \mathcal{F}) \to\mathbb{E} \subset
\R$, defined by 
\begin{equation*}
(\Gamma_Y, \Gamma_W, \Pi) \mapsto \phi_1(\Gamma_Y, \Gamma_W, \Pi) : =  \int \Gamma_Y(y, x) \Gamma_W(w, x) \dd \Pi (z).
\end{equation*}
Then the map $\phi_1$ is well defined. Moreover, the map $\phi_1$ is
Hadamard-differentiable at $(\Gamma_Y,\Gamma_W,\Pi) = (F_{Y\mid X}-.5,F_{W\mid X}-.5, F_{Z})$, tangentially
to the subset $\mathbb{D}_{0} = UC(\mathcal{YX}, \rho) \times UC(\mathcal{WX}, \rho) \times UC(\mathcal{F}%
, \lambda) $, with the derivative map $(\gamma_Y, \gamma_W, \pi)
\mapsto\phi^{\prime}_{1,F_{Y\mid X},F_{W\mid X}, F_{Z}}(\gamma_Y,\gamma_W, \pi)$ mapping $\mathbb{D} $ to $%
\mathbb{E}$ defined by 
\begin{multline*}\label{eq:dphi-num}
\phi'_{1, F_{Y\mid X},F_{W\mid X},F_{Z}}(\gamma_Y, \gamma_W, \pi)  :=   \int
\gamma_Y (y, x) [F_{W \mid X}(w \mid x) - 1/2] \mathrm{d}F_{Z}(y,w, x) \\ +
  \int
\gamma_W(w,  x) [F_{Y \mid X}(y \mid x) - 1/2] \mathrm{d}F_{Z}(y,w, x) \\ +
 \int
[F_{Y \mid X}(y \mid x) - 1/2][F_{W \mid X}(w \mid x) - 1/2] \mathrm{d}\pi(y,w,x),
\end{multline*}
and  the derivative is defined and is continuous on $\mathbb{D}$.
\end{lemma}

\noindent \textbf{Proof of Lemma \ref{lemma: Hadamard dif of CRRR FR}.} 
First we show that the map $\phi_1$ is well defined. Any probability measure $\Pi$
on $\mathcal{Z}$ is determined by the values $\int f \dd \Pi$ for $f \in 
\mathcal{F}$, since $\mathcal{F}$ contains all the indicators of the
rectangles in $\overline{\mathbb{R}}^{d_x+2}$. By Caratheodory's extension theorem $%
\Pi(A) = \Pi 1_A$ is well defined on all Borel subsets $A$ of $\mathbb{R}%
^{d_x+2}$. Since $z \mapsto \Gamma_Y(y,x)\Gamma_W(w,x)$ is Borel measurable and takes values
in $[-.5^2,.5^2]$, it follows that $\int \Gamma_Y(y, x)\Gamma_W(w, x) \dd \Pi(z) $ is well defined as a
Lebesgue integral, and $\int \Gamma_Y(y, x) \Gamma_W(w, x) \dd \Pi (z) \in \R$. 

Next we show the main claim. We establish the Hadamard differentiability of $\phi_1$.
 Consider any sequence $(\Gamma_Y^{t},\Gamma_W^{t}, \Pi^{t}) \in%
\mathbb{D}_{\phi}$ such that for $\gamma_Y^{t} := (\Gamma_Y^{t} - F_{Y \mid X} + .5)/t,$ $\gamma_W^{t} := (\Gamma_W^{t} - F_{W \mid X} + .5)/t,$
and $\pi^{t} (f) := \int f \dd(\Pi^{t}- F_{Z})/t,$ 
\begin{equation*}
\begin{array}{lll}
(\gamma_Y^{t},\gamma_W^{t}, \pi^{t}) \to(\gamma_Y,\gamma_W, \pi), \ \ \text{in } \ \ \ell_m^{\infty }(%
\mathcal{Y}\mathcal{X}) \times \ell_m^{\infty }(%
\mathcal{W}\mathcal{X}) \times\ell^{\infty}(\mathcal{F}), \text{ where }
(\gamma_Y,\gamma_W, \pi) \in\mathbb{D}_{0}. &  & 
\end{array}%
\end{equation*}
We want to show that as $t \searrow0$ 
\begin{equation*}
\frac{\phi_1(\Gamma_Y^{t},\Gamma_W^{t}, \Pi^{t})- \phi_1(F_{Y\mid X}-.5,F_{W\mid X}-.5,F_{Z})}{t} - \phi^{\prime
}_{1,F_{Y\mid X},F_{W\mid X}, F_{Z}}(\gamma_Y, \gamma_W, \pi) \to 0 \text{ in } \R.
\end{equation*}
Write the difference above as: 
\begin{multline}
 \int (\gamma_Y^t - \gamma_Y)[F_{W\mid X} - .5] \mathrm{d}F_{Z} + \int (\gamma_W^t - \gamma_W)[F_{Y\mid X} - .5] \mathrm{d}F_{Z} +
 \int [F_{Y\mid X} - .5][F_{W \mid X} - .5](\mathrm{d}\pi^t - \mathrm{d} \pi) \\ 
 + \int \gamma_Y [F_{W\mid X} - .5] t \mathrm{d}\pi^t + \int \gamma_W [F_{Y\mid X} - .5] t \mathrm{d}\pi^t + \int (\gamma^t_Y - \gamma_Y) [F_{W\mid X} - .5] t \mathrm{d}\pi^t \\ + \int (\gamma^t_W - \gamma_W) [F_{Y\mid X} - .5] t \mathrm{d}\pi^t  
 + \int \gamma_Y^t \gamma_W^t t \mathrm{d}F_{Z} + \int \gamma_Y^t \gamma_W^t t^2 \mathrm{d}\pi^t 
%\sqrt{\lambda_j \lambda_k} \int (\gamma_j^t - \gamma_j) t d\pi_k^t.
\label{four terms}    
\end{multline}

The first two terms of (\ref{four terms}) are bounded by $ \|\gamma_R^t -
\gamma_R \|_{\mathcal{RX}} \int \mathrm{d}F_{Z} \to 0$, $R \in \{Y,W\}$. The third term
vanishes, since for any  $f \in \mathcal{F}$, $ \int
f \mathrm{d}\pi^t \to \int f \mathrm{d}\pi \text{ in }
\ell^{\infty}(\mathcal{F}), $ and $[F_{Y \mid X}-.5][F_{W \mid X}-.5] \in \mathcal{F}$  by assumption. The fourth and fifth terms
vanish by the argument provided below. The sixth and seventh term vanish,
since $ | \int(\gamma_R^t -\gamma_R) t \mathrm{d}\pi^t | \leq \|
\gamma_R^t - \gamma_R \|_{\mathcal{RX}} \int |t \mathrm{d} \pi^t| \leq 2
\|\gamma_R^t - \gamma_R\|_{\mathcal{RX}} \to 0$ for $R \in \{Y,W\}$, where $\int |\dd \mu|$ is the total variation of the signed measure $\mu$. The eighth term is bounded by $C t \int \mathrm{d}F_{Z} \to 0$, for some $C > 0$.  The last term can be bounded as:
$$
2 t \| \gamma^t_Y \|_{\mathcal{YX}} \|
\gamma^t_W \|_{\mathcal{WX}} 
= 2 t \{\| \gamma_Y \|_{\mathcal{YX}} +o(1)\}
\{\|\gamma_W \|_{\mathcal{WX}} +o(1)\} \to 0.
$$

Here we consider the fourth term $\int \gamma_Y [F_{W\mid X} - .5] t \mathrm{d}\pi^t$ and show that it vanishes. The argument for the fifth term is analogous. Since $\gamma_Y$ is continuous on the compact semi-metric space
$(\overline{\mathcal{YX}}, \rho)$, there exists a finite partition of $\overline{
\R}^{d_x+1}$  into non-overlapping rectangular regions $(R_{im}: 1
\leq i \leq m) $ (rectangles are allowed not to include their sides to make
them non-overlapping) such that $\gamma_Y$ varies at most $\epsilon$ on $%
\mathcal{Y}\mathcal{X} \cap R_{im}$. Let $p_{m}(y,x) := (y_{im},x_{im})$ if $%
(y,x) \in \mathcal{Y}\mathcal{X} \cap R_{im}$, where $(y_{im},x_{im})$ is an
arbitrarily chosen point within $\mathcal{YX} \cap R_{im}$ for each $i$;
also let $\chi_{im}(z) := 1\{ (y,x) \in R_{im}\}$. Then, as $t \to 0$, 
\begin{multline*}
\left| \int\gamma_Y [F_{W\mid X} - .5]  t d \pi
^{t}\right| \leq\left| \int(\gamma_Y- \gamma_Y\circ p_{m} )
[F_{W\mid X} - .5] t d \pi^{t} \right| + \left| \int(\gamma_Y\circ{%
p_{m}}) [F_{W\mid X} - .5] t d \pi^{t} \right| \\
 \leq\| \gamma_Y- \gamma_Y\circ{p_{m}}\|_{\mathcal{YX}} \int| t d \pi^{t} | +
\sum_{i=1}^{m} |\gamma_Y(y_{im},x_{im})| t \left |\pi^{t}
([F_{W\mid X} - .5] \chi_{im})\right|\\
 \leq 2 \| \gamma_Y- \gamma_Y\circ{p_{m}}\|_{\mathcal{YX}}  + t m \|\gamma_Y\|_{%
\mathcal{YX}} \max_{1 \leq i \leq m} \left|\pi^{t} ([F_{W\mid X} - .5]\chi_{im})\right
| \\ \leq 2\epsilon+t m \|\gamma_Y\|_{\mathcal{YX}}
\left\|\pi^t\right \|_{\mathcal{F}} 
 \leq2\epsilon+ t m \left[ \|\gamma_Y\|_{\mathcal{YX}} \left \| \pi \right
\|_{\mathcal{F}} + o(1) \right] \leq2\epsilon+ O(t) \to2 \epsilon,
\end{multline*}
since $\|F_{W\mid X} - .5\|_{\mathcal{WX}} < 1$ and $\{[F_{W\mid X} - .5]\chi_{im}: 1\leq i \leq m\} \subset\mathcal{F}$%
, so that $\max_i | \pi^{t} ([F_{W\mid X} - .5]\chi_{im}) | \leq \|
\pi^t\|_{\mathcal{F}} \to \|\pi\|_{\mathcal{F}} < \infty$.\footnote{%
The set $\mathcal{F}$ is allowed to include zero, the indicator  of an empty
rectangle.} The constant $\epsilon$ is arbitrary, so that the right hand side vanishes as $t \to 0$.

The derivative is well-defined over the entire $\mathbb{D}$ and is
in fact continuous with respect to the norm on $\mathbb{D}$ given by $%
\|\cdot\|_{\mathcal{YX}} \vee \|\cdot\|_{\mathcal{WX}} \vee\| \cdot\|_{\mathcal{F}}$. The third component
of the derivative map is trivially continuous with respect to $\| \cdot\|_{%
\mathcal{F}}$. The first component is continuous with respect to $\|\cdot\|_{%
\mathcal{YX}}$ since: 
\begin{equation*}
\left| {\int(\gamma_Y - \tilde\gamma_Y) [F_{W\mid X} - .5]  \dd F_{Z}(z)} \right| \leq\| \gamma_Y- \tilde\gamma_Y\|_{\mathcal{Y}\mathcal{X}} \int \dd
F_{Z}(z) .
\end{equation*}
The second component is continuous with respect to $\|\cdot\|_{%
\mathcal{WX}}$ by an analogous argument.  Hence the derivative map is continuous. \qed

\begin{lemma}[Hadamard differentiability of $\phi$]
\label{lemma: Hadamard dif of CRRR COR} Let $\mathcal{RX} \subseteq 
\mathbb{R}^{d_x+1}$, $\RR \in \{\mathcal{Y},\mW\}$, and $\mathcal{F}$ be the class of bounded functions,
mapping $\overline{\mathbb{R}}^{d_x+2}$ to $\mathbb{R}$, that contains the multiplication of monomials of total degree at most two in $(F_{Y \mid X}-.5)$ and $(F_{W \mid X}-.5)$ with indicators of all the rectangles in $\bar{\mathbb{R}}^{d_x+2}$, such that $\mathcal{F}$ is
totally bounded under $\lambda$. Let $\mathbb{D}_{\phi}$ be the product of
the spaces of measurable functions $\Gamma_Y: \mathcal{Y}\mathcal{X} \mapsto[-.5,.5
]$ defined by $(y,x) \mapsto \Gamma_Y(y,x)$ and $\Gamma_W: \mathcal{W}\mathcal{X} \mapsto[-.5,.5%
]$ defined by $(w,x) \mapsto \Gamma_W(w,x)$, and the bounded maps $\Pi: 
\mathcal{F} \mapsto\mathbb{R}$ defined by $f \mapsto\int f \dd\Pi$, where $\Pi$
is restricted to be a probability measure on $\mathcal{YWX}$, $\int \Gamma_Y(y,x)^2 \dd\Pi > 0$ and  $\int \Gamma_W(w,x)^2 \dd\Pi > 0$. Consider the map 
$\phi: \mathbb{D}_{\phi} \subset \mathbb{D} = \ell^{\infty}_m(\mathcal{YX}) \times \ell^{\infty}_m(\mathcal{W}
\mathcal{X}) \times\ell^{\infty }( \mathcal{F}) \to\mathbb{E} \subset
\R$, defined by 
\begin{equation*}
(\Gamma_Y, \Gamma_W, \Pi) \mapsto\phi(\Gamma_Y, \Gamma_W, \Pi) : = \frac{\int \Gamma_Y(y, x) \Gamma_W(w, x) \dd \Pi (z)}{\sqrt{\int \Gamma_W(w, x)^2 \dd \Pi (z)\int \Gamma_Y(y, x)^2 \dd \Pi (z)}}.
\end{equation*}
Then the map $\phi$ is well defined. Moreover, the map $\phi$ is
Hadamard-differentiable at $(\Gamma_Y,\Gamma_W,\Pi) = (F_{Y\mid X}-.5,F_{W\mid X}-.5, F_{Z})$, tangentially
to the subset $\mathbb{D}_{0} = UC(\mathcal{YX}, \rho) \times UC(\mathcal{WX}, \rho) \times UC(\mathcal{F}%
, \lambda) $, with the derivative map $(\gamma_Y, \gamma_W, \pi)
\mapsto\phi^{\prime}_{F_{Y\mid X},F_{W\mid X}, F_{Z}}(\gamma_Y,\gamma_W, \pi)$ mapping $\mathbb{D} $ to $%
\mathbb{E}$ defined by 
\begin{multline*}
    \phi'_{F_{Y\mid X},F_{W\mid X},F_{Z}}(\gamma_Y, \gamma_W, \pi)  \\ :=  12[\phi'_{1,F_{Y\mid X},F_{W\mid X},F_{Z}}(\gamma_Y, \gamma_W, \pi) - \rho_C (\phi'_{2,F_{W\mid X},F_{Z}}(\gamma_W, \pi)+\phi'_{3,F_{Y\mid X},F_{Z}}(\gamma_Y, \pi))/2],
\end{multline*}
with $\phi'_{1, F_{Y\mid X},F_{W\mid X},F_{Z}}(\gamma_Y, \gamma_W, \pi)$ defined as in Lemma \ref{lemma: Hadamard dif of CRRR FR},
% \begin{multline*}\label{eq:dphi-num}
% \phi'_{1, F_{Y\mid X},F_{W\mid X},F_{Z}}(\gamma_Y, \gamma_W, \pi)  :=  \int
% \gamma_Y (y, x) [F_{W \mid X}(w \mid x) - .5] \mathrm{d}F_{Z}(y,w, x) \\ +
%  \int
% \gamma_W(w,  x) [F_{Y \mid X}(y \mid x) - .5] \mathrm{d}F_{Z}(y,w, x) \\ +
% \int
% [F_{Y \mid X}(y \mid x) - .5][F_{W \mid X}(w \mid x) - .5] \mathrm{d}\pi(y,w,x),
% \end{multline*}
\begin{multline*}\label{eq:dphi-den1}
\phi'_{2,F_{W\mid X},F_{Z}}(\gamma_W, \pi)  :=  
 2 \int
\gamma_W(w, x) [F_{W \mid X}(w \mid x) - .5] \mathrm{d}F_{Z}(y,w, x) \\ +
\int
[F_{W \mid X}(w \mid x) - .5]^2 \mathrm{d}\pi(y,w,x),
\end{multline*}
and
\begin{multline*}
\phi'_{3,F_{Y\mid X},F_{Z}}(\gamma_Y, \pi)  :=  
 2 \int
\gamma_Y(y, x) [F_{Y \mid X}(y \mid x) - .5] \mathrm{d}F_{Z}(y,w, x) \\ +
\int
[F_{Y \mid X}(y \mid x) - .5]^2 \mathrm{d}\pi(y,w,x);
\end{multline*}
where the derivative is defined and is continuous on $\mathbb{D}$.
\end{lemma}

\noindent \textbf{Proof of Lemma \ref{lemma: Hadamard dif of CRRR COR}.} 
It is convenient to express 
$$
\phi(\Gamma_Y, \Gamma_W, \Pi)  =  \frac{\phi_1(\Gamma_Y, \Gamma_W, \Pi) }{\sqrt{\phi_2(\Gamma_W, \Pi)\phi_3(\Gamma_Y, \Pi)} },
$$
where $\phi_1$ defined as in Lemma \ref{lemma: Hadamard dif of CRRR FR},
$$
\phi_2(\Gamma_W, \Pi)  := \int \Gamma_W(w, x)^2 \dd \Pi (z) \ \ \text{ and } \ \ 
\phi_3(\Gamma_Y, \Pi)  := \int  \Gamma_Y(y, x)^2 \dd \Pi (z).
$$

First note that the maps $\phi_2$ and $\phi_3$ are well defined by a similar argument to the proof of Lemma \ref{lemma: Hadamard dif of CRRR FR} that shows that  $\phi_1$ is well-defined. The map $\phi$ is also well-defined because $\phi_2(\Gamma_W, \Pi)  > 0$ and $\phi_3(\Gamma_Y, \Pi)  > 0$ by assumption.

Next we show the main claim. The Hadamard differentiability  of $\phi_1$  is establish in Lemma \ref{lemma: Hadamard dif of CRRR FR}. The Hadamard differentiability of $\phi_2$ and $\phi_3$ can be established by analogous arguments. In particular, the maps $\phi_2$ and $\phi_3$  in the denominator are Hadamard differentiable at $(F_{W\mid X}-.5,F_{Z})$ and $(F_{Y\mid X}-.5,F_{Z})$, respectively, with derivatives $\phi'_{2,F_{W\mid X},F_{Z}}$ and $\phi'_{3,F_{Y\mid X},F_{Z}}$. Indeed, we can show that as $t \searrow0$ 
\begin{equation*}
\frac{\phi_2(\Gamma_W^{t}, \Pi^{t})- \phi_2(F_{W\mid X}-.5,F_{Z})}{t} - \phi^{\prime
}_{2,F_{W\mid X}, F_{Z}}(\gamma_W, \pi) \to 0 \text{ in } \R
\end{equation*}
and
\begin{equation*}
\frac{\phi_3(\Gamma_Y^{t}, \Pi^{t})- \phi_3(F_{Y\mid X}-.5,F_{Z})}{t} - \phi^{\prime
}_{3,F_{Y\mid X}, F_{Z}}(\gamma_Y, \pi) \to 0 \text{ in } \R
\end{equation*}
following an analogous argument as for $\phi_1$ in the proof of Lemma \ref{lemma: Hadamard dif of CRRR FR}. It can also be shown that $\phi^{\prime
}_{2,F_{W\mid X}, F_{Z}}$ and $\phi^{\prime
}_{3,F_{Y\mid X}, F_{Z}}$ are  well-defined over the entire $\mathbb{D}$ and are continuous. We omit the proof for the sake of brevity.

The final result then follows by the chain-rule for Hadamard differentiable maps using that
$$
\Var(F_{W \mid X}(W \mid X)) = \Var(F_{Y \mid X}(Y \mid X))  = 1/12.
$$
Continuity of the derivative with respect to the norm on $\mathbb{D}$ given by $%
\|\cdot\|_{\mathcal{YX}} \vee \|\cdot\|_{\mathcal{WX}} \vee\| \cdot\|_{\mathcal{F}}$ follows by continuity of $\phi^{\prime
}_{1,F_{Y\mid X},F_{W\mid X}, F_{Z}}$, $\phi^{\prime
}_{2,F_{W\mid X}, F_{Z}}$  and $\phi^{\prime
}_{3,F_{Y\mid X}, F_{Z}}$.
\qed

\begin{lemma}[Hadamard differentiability of $\varphi$]
\label{lemma: Hadamard dif of CRRR REG} Let $\mathcal{RX} \subseteq 
\mathbb{R}^{d_x+1}$, $\RR \in \{\mathcal{Y},\mW\}$, and $\mathcal{F}$ be the class of bounded functions,
mapping $\overline{\mathbb{R}}^{d_x+2}$ to $\mathbb{R}$, that contains the multiplication of monomials of total degree at most two in $(F_{Y \mid X}-.5)$ and $(F_{W \mid X}-.5)$ with indicators of all the rectangles in $\bar{\mathbb{R}}^{d_x+2}$, such that $\mathcal{F}$ is
totally bounded under $\lambda$. Let $\mathbb{D}_{\phi}$ be the product of
the spaces of measurable functions $\Gamma_Y: \mathcal{Y}\mathcal{X} \mapsto[-.5,.5
]$ defined by $(y,x) \mapsto \Gamma_Y(y,x)$ and $\Gamma_W: \mathcal{W}\mathcal{X} \mapsto[-.5,.5%
]$ defined by $(w,x) \mapsto \Gamma_W(w,x)$, and the bounded maps $\Pi: 
\mathcal{F} \mapsto\mathbb{R}$ defined by $f \mapsto\int f \dd\Pi$, where $\Pi$
is restricted to be a probability measure on $\mathcal{YWX}$ and  $\int \Gamma_W(w,x)^2 \dd\Pi > 0$. Consider the map 
$\varphi: \mathbb{D}_{\phi} \subset \mathbb{D} = \ell^{\infty}_m(\mathcal{YX}) \times \ell^{\infty}_m(\mathcal{W}
\mathcal{X}) \times\ell^{\infty }( \mathcal{F}) \to\mathbb{E} \subset
\R$, defined by 
\begin{equation*}
(\Gamma_Y, \Gamma_W, \Pi) \mapsto\varphi(\Gamma_Y, \Gamma_W, \Pi) : = \frac{\int \Gamma_Y(y, x) \Gamma_W(w, x) \dd \Pi (z)}{\int \Gamma_W(w, x)^2 \dd \Pi (z)}.
\end{equation*}
Then the map $\varphi$ is well defined. Moreover, the map $\varphi$ is
Hadamard-differentiable at $(\Gamma_Y,\Gamma_W,\Pi) = (F_{Y\mid X}-.5,F_{W\mid X}-.5, F_{Z})$, tangentially
to the subset $\mathbb{D}_{0} = UC(\mathcal{YX}, \rho) \times UC(\mathcal{WX}, \rho) \times UC(\mathcal{F}%
, \lambda) $, with the derivative map $(\gamma_Y, \gamma_W, \pi)
\mapsto\varphi^{\prime}_{F_{Y\mid X},F_{W\mid X}, F_{Z}}(\gamma_Y,\gamma_W, \pi)$ mapping $\mathbb{D} $ to $%
\mathbb{E}$ defined by 
\begin{equation*}
    \varphi'_{F_{Y\mid X},F_{W\mid X},F_{Z}}(\gamma_Y, \gamma_W, \pi)  :=  12[\phi'_{1,F_{Y\mid X},F_{W\mid X},F_{Z}}(\gamma_Y, \gamma_W, \pi) - \rho_C \phi'_{2,F_{W\mid X},F_{Z}}(\gamma_W, \pi)],
\end{equation*}
with $\phi'_{1, F_{Y\mid X},F_{W\mid X},F_{Z}}(\gamma_Y, \gamma_W, \pi)$ defined as in Lemma \ref{lemma: Hadamard dif of CRRR FR} and $\phi'_{2, F_{W\mid X},F_{Z}}(\gamma_W, \pi)$ defined as in Lemma \ref{lemma: Hadamard dif of CRRR COR}, where the derivative is defined and is continuous on $\mathbb{D}$.
\end{lemma}

\noindent \textbf{Proof of Lemma \ref{lemma: Hadamard dif of CRRR REG}.} The result follows by an analogous argument to the proof of Lemma \ref{lemma: Hadamard dif of CRRR COR}. We omit the proof for the sake of brevity. \qed

\subsection{Proof of Lemma \ref{lemma:crrr}} We start by stating a Lemma with a bootstrap functional central limit theorem for the bootstrap draws of the inputs needed to establish Theorem \ref{thm:main-boot}. We shall prove this lemma together with Lemma \ref{lemma:crrr}. 

For  $R \in \{Y,W\}$, let 
$
(r,x) \mapsto \hat Z^*_R(r,x) := \sqrt{n}\left(\hat F^*_{R \mid X}(r \mid x) - \hat F_{R \mid X}(r \mid x) \right)$ and $f \mapsto \hat G^*_Z(f) := \sqrt{n} \int f \mathrm{d} (\hat F^*_Z - \hat F_Z),$ where $\hat F^*_{R \mid X}(r \mid x) := \Lambda(x'\hat \beta^*_R(r))$, $\hat \beta^*_R(r)$ is the bootstrap draw of $\hat \beta_R(r)$ defined in Algorithm \ref{alg:eb} and $\hat F^*_Z$ is the bootstrap draw of the empirical distribution function of $Z$, be exchangeable bootstrap draws of the empirical processes $(r,x) \mapsto \hat Z_R(r,x)$ and $f \mapsto \hat G_Z(f)$.
\begin{lemma}[Bootstrap Limit Processes for Inputs]\label{lemma:crrr-boot} Under the conditions of Lemma \ref{lemma:crrr} and Assumption \ref{ass:eb},  in the metric space $\ell^{\infty}(\mathcal{Y}\mathcal{W}\mathcal{X}\mathcal{F})$,
$$
(\hat Z^*_Y(y,x), \hat Z^*_W(w,x), \hat G^*_Z(f)) \rightsquigarrow_{\Pr} (Z_Y(y,x), Z_W(w,x),  G_Z(f)),
$$    
as stochastic processes indexed by $(y,w,x,f)$, where $(Z_Y(y,x), Z_W(w,x),  G_Z(f))$ has the same distribution as the limit process in Lemma \ref{lemma:crrr}. 
\end{lemma}

The proofs of Lemmas \ref{lemma:crrr} and \ref{lemma:crrr-boot} follow similar steps to the proof of Theorem 5.2 in \cite{chernozhukov+13inference}, suitably modified to extend the process to the tails.
The main differences are highlighted in Steps 1, 2, and 3 below.

\textsc{Step 1.}(Results for coefficients and empirical measures). 
Application of the Hadamard differentiability results for Z-processes in \cite{chernozhukov+13inference} gives that, in $\ell^{\infty}({\bar{\mathcal{Y}}}%
)^{d_x}\times\ell^{\infty}({\bar{\mathcal{W}}}%
)^{d_x}\times\ell^{\infty}(\mathcal{F})$, 
\begin{equation}\label{eq:fclt_coeffs}
(\sqrt{n}(\hat{\beta}_{Y}(\cdot)-\beta_{Y}(\cdot)),\sqrt{n_{{}}}(\hat{\beta}_{W}(\cdot)-\beta_{W}(\cdot)),\hat G_{Z}) \rightsquigarrow(H_Y(\cdot),H_W(\cdot),G_Z), 
\end{equation}
where $\bar{\mathcal{Y}}$ and $\bar {\mathcal{W}}$ are any compact strict subsets of $\mathcal{Y}$ and $\mathcal{W}$, respectively, and $$r \mapsto H_R(r) :=  J_R(r)^{-1} \mathbb{G} (\varphi_{r,\beta}), \quad \varphi_{r,\beta}(R,X) := \left[1 \{R \leq r\} - \Lambda(X^{\prime}
\beta_R(r))  \right]X,$$ has continuous
paths a.s., for $R \in \{Y,W\}$.\footnote{\cite{chernozhukov+13inference} gives detailed arguments on how H-differentiability of Z-processes implies that $
(\sqrt{n}(\hat{\beta}_{Y}(y)-\beta_{Y}(y)),\hat G_{X}(f))\rightsquigarrow(H_Y(y),G_X(f)) 
$ in $\ell^{\infty}({\bar{\mathcal{Y}}}%
)^{d_x}\times\ell^{\infty}(\mathcal{F})$, where $\hat G_{X}(f)$ is the empirical process induced by the marginal distribution of $X$. The extension to stacking another Z-process is straightforward, implying the result \eqref{eq:fclt_coeffs}.
}

We extend the process $\hat \beta(r)$ to the tails as: $$\hat \beta_R(r) = \hat \beta_R(\bar r) + (r-\bar r)\hat \alpha_R(\bar r) e_1, \quad r \in \RR\setminus \bar{\RR},$$ where $e_1$ is a unitary $d_x$-vector with a one in the first component.  Likewise, the estimands are given by: $$\beta_R(r) = \beta_R(\bar r) + (r - \bar r)\alpha_{R}(\bar r) e_1 \quad r \in \RR\setminus \bar{\RR},$$ by assumption.  

In what follows it is convenient to analyze the estimator for the lower tail, the analysis for estimators for upper tails follows exactly the same steps, switching the signs on the dependent variables, $ R \in \{ Y, W, -Y, -W\}$. The estimators $(\hat \beta_R(\bar r),\hat \alpha_R(\bar r) )$  can be seen as Z-estimators with moment function:
$$
\bar \varphi_{\beta,\alpha}(R,X) = \left(
     \varphi_{\bar r,\beta}(R,X)',
     \varphi_{\alpha}(R,X)( r_0 - \bar r) \right)', \ \ \varphi_{\alpha}(R,X) := 1 \{R \leq  r_0\} -\Lambda(X^{\prime}
\beta_R(\bar r) + (r_0 - \bar r)\alpha_{R}(\bar r)).
$$
Invoking Z-process theory again but this time for the simple case of finite-dimensional space $\R^{d_x+1}$,
we have that jointly in $R \in \{Y, -Y, W, -W\}$,
\begin{multline}
 \sqrt{n} \left(\hat{\beta}_{R}(\bar r)'-\beta_{R}(\bar r)', \hat \alpha_R(\bar r) - \alpha_R(\bar r)\right)'  \\ 
 \rightsquigarrow  \left[\begin{array}{cc}
    J_R(\bar r) & 0 \\
    (r_0 - \bar r) \Ep[\lambda(X^{\prime}
\beta_R(r_0))X'] & (r_0 - \bar r)^2 \Ep[\lambda(X^{\prime}
\beta_R(r_0))]
\end{array}\right]^{-1} \mathbb{G}(\bar \varphi_{\beta,\alpha})  \\
= \left[\begin{array}{cc}
    J^{-1}_R(\bar r) \mathbb{G}(\varphi_{\bar r,\beta}) \\
    \frac{\mathbb{G}\left(\varphi_{\alpha}\right) - \Ep[\lambda(X^{\prime}
\beta_R(r_0))X]' J^{-1}_R(\bar r) \mathbb{G}(\varphi_{\bar r,\beta})}{(r_0 - \bar r) \Ep[\lambda(X^{\prime}
\beta_R(r_0))]} 
\end{array} \right].
\label{eq:tail_coeffs}\end{multline}
In fact using Hadamard differentiability results for Z-processes given in \cite{chernozhukov+13inference}, we conclude that convergence results (\ref{eq:fclt_coeffs}) and (\ref{eq:tail_coeffs}) for all $R \in \{Y, W, - Y, -W\}$ hold jointly.\footnote{Of course, it is cumbersome to put this joint convergence statement into one display, so we state this verbally.}

The Hadamard differentiability results for Z-processes also imply
that the bootstrap analogs of the results (\ref{eq:fclt_coeffs}) and (\ref{eq:tail_coeffs}) are also valid and hold jointly as well. We omit writing the formulas for these convergence results, since they are analogous to (\ref{eq:fclt_coeffs}) and (\ref{eq:tail_coeffs}).

%$$
%r \mapsto H_R(r) := H_R(\bar r) + \frac{r-\bar r}{r_0 - \bar r} %\frac{\mathbb{G}\left[\Lambda(X^{\prime}
%\beta_R(r_0)) - 1 \{R \leq  r_0\} \right] - \Ep[\lambda(X^{\prime}
%\beta_R(r_0))X]' H_R(\bar r)}{\Ep[\lambda(X^{\prime}
%\beta_R(r_0))]}e_1
%$$

\textsc{Step 2.}(Main: Results for conditional cdfs).
Here we show that, 
\begin{align*}
& (\hat Z_Y, \hat Z_W, \hat G_Z) \rightsquigarrow (Z_Y, Z_W,  G_Z) \text{ in } \ell^{\infty}({\mathcal{Y}}{\mathcal{W}}\mathcal{X}\mathcal{F}),\\
& (\hat Z^*_Y, \hat Z^*_W, \hat G^*_Z) \rightsquigarrow_{\Pr} (Z_Y, Z_W,  G_Z) \text{ in } \ell^{\infty}({\mathcal{Y}}{\mathcal{W}}\mathcal{X}\mathcal{F}).    
\end{align*}

%The uniform $\epsilon$-covering numbers for this class
%can be bounded independently of $F_{X_{}}$ by the previous argument and ..., and
%so the Pollard's entropy integral is finite. Hence we can construct a class
%of functions $\mathcal{F}$ containing the union of all the families %$%
%\mathcal{F}_Y$, $\mathcal{F}_W$ and the indicators of
%all rectangles in $\overline{\mathbb{R}}^{d_x+2}$. Note that these indicators
%form a VC class. The final set $\mathcal{F}$ therefore is a DKP class.[VICTOR: THIS STEP NEEDS TO BE UPDATED]

For the body part, $r \in \overline{\mathcal{R}}$,
consider the mapping $\nu: \mathbb{D}_\nu \subset \ell^{\infty}(\overline{\mathcal{R}}^{d_x}) \to \ell^{\infty}(\overline{\mathcal{R}} \mathcal{X})$, defined as: $$b \mapsto \nu(b), \quad \nu(b)(x, y):= \Lambda\left(x^{\prime} b(y)\right).$$ 
It is straightforward to deduce that this map is Hadamard differentiable at $b(\cdot)=$ $\beta_R(\cdot)$ tangentially to $UC(\mathcal{R},\rho)^{d_x}$ with the derivative map given by: $$v \mapsto \nu_{}^{\prime}(v), \quad \nu_{}^{\prime}(v)(r, x)= \lambda\left(x^{\prime} \beta_R(r)\right) x^{\prime} v(r).$$

For the tail part $ r \in \mathcal{R}\setminus \overline{\mathcal{R}}$, we consider the mapping
$\mu: \Bbb{R}^{d_x+1} \to \ell^\infty( (\mathcal{R} \setminus \overline{\mathcal{R}}) \times \mX)$ defined by:
$$
(d,a) \mapsto \mu(d,a), \quad  \mu(d,a)(r,x) := \Lambda(x'd + (r-\bar r) a)
$$
This is also Hadamard differentiable  at $(d,a) =
(\beta_R(\bar r), \alpha_R(\bar r))$ tangentially to the entire domain with the derivative:
$$
(h,u) \mapsto \mu'(h, u), \quad \mu'(h, u)(r,x) = \lambda(x'\beta_R(r)) (x'h + (r-\bar r) u).
$$
The derivative is a bounded (continuous) linear operator (note that as $r \to \pm \infty$, the derivative vanishes, with the linear growth factor $r- \bar r$ being dominated by the term $\lambda(x'\beta_R(r))$ with exponential tails).

We can now define the "extended map" that combines the body and tail pieces: 
$$
(b, d, a) \mapsto \bar \nu(b,d,a);  \quad \bar \nu(b,d,a)(r,x):= \nu(b)(r,x) 
1( r \in \overline{\mathcal{R}}) + \mu(d,a)(r,x) 1 ( r \in \RR \setminus \overline{\mathcal{R}}).
$$
By combining the two differentiability results above, we can deduce that this map is Hadamard differentiable with the derivative map:
$$
(v, h, u) \mapsto \bar \nu_{}^{\prime}(v,h,u), \quad 
\bar \nu_{}^{\prime}(v,h,u) (r,x) :=
\nu_{}^{\prime}(v)(r,x) 1( r \in \overline{\mathcal{R}}) + \mu'(h, u)(r,x) 1 ( r  \in \RR \setminus \overline{\mathcal{R}}).
$$

Then, the claim follows by the functional delta method, and we find that the limit process is given by:
$$
Z_R(r,x) = \lambda(x'\beta_R(r))x'H_R(r), \quad R \in \{Y,W\}.
$$
where we now define the extended version of $H_R$ as:
$
H_R(r) = \mathbb{G} (\psi_{R,r,\beta_R}) 
$ 
where:
\begin{multline*}
\psi_{R,r,\beta_R}(R,X) =  1\{r \in  \bar{\mathcal{R}}\} J_R(r)^{-1} \varphi_{r,\beta}(R,X)    
+ 1\{r \in  \RR\setminus \bar{\mathcal{R}}\} \Big[ J_R(\bar r)^{-1} \varphi_{\bar r,\beta}(R,X)  \\
+ \frac{r-\bar r}{r_0 - \bar r} \frac{\varphi_{\alpha}(R,X)  - \Ep[\lambda(X^{\prime}
\beta_R(r_0))X]' J_R(\bar r)^{-1} \varphi_{\bar r,\beta}(R,X)}{\Ep[\lambda(X^{\prime}
\beta_R(r_0))]}e_1 \Big],
\end{multline*}
and letting $\ell_{r,x}(R,X) := \lambda(x'\beta_R(r))x'\psi_{R,r,\beta_R}(R,X) $, after some algebra.

\textsc{Step 3.} (Auxiliary: Donskerness). One key ingredient for the result is to show that all the functions involved in the stochastic processes are in Dudley-Koltchinskii-Pollard (DKP) classes, namely they have bounded uniform covering entropy integrals and obey standard measurability conditions (Dudley's image-admissible Suslin condition). We omit any discussion of measurability in this paper, but we note that it trivially holds.  We split the proof in the functions involved in the conditional processes and the functions involved in the covariate process.

\paragraph{\textbf{Conditional processes $\hat Z_Y$ and $\hat Z_W$.} } Here we show that the class $\mathcal{F}_R = \{F_{R \mid X}(r \mid \cdot): r \in\mathcal{R}\}$, $R \in \{Y,W\}$, where  $F_{R \mid X}$ satisfies Assumption \ref{ass:dr}, is DKP. The proof in \cite{chernozhukov+13inference} relies on compactness of the set $\bar{\mathcal{R}}\mathcal{X}$ and does not apply immediately. We extend the result to $\mathcal{RX}$. For $R \in \{Y,W\}$, note that $\mathcal{F}_R$
is a uniformly bounded ``parametric" family indexed by $r \in\mathcal{R}_{}$
that obeys $|F_{R_{} \mid X_{}}(r \mid \cdot) - F_{R_{} \mid X_{}}(r^{\prime} \mid \cdot)| \leq
L| r - r^{\prime}|$, given the assumption that the density function $%
f_{R_{} \mid X_{}}$ is uniformly bounded by some constant $L$. 
This was enough to bound the covering numbers for the index set $\bar{\mathcal{R}}$, but is not enough
to bound the covering number over the unbounded set $\mathcal{R}$.

Under our modeling hypotheses, there exists a small enough constant $C>0$ such that:
$$
F_{R \mid X}(r\mid \cdot) \leq \exp(r C); \ \ r<0; \quad 1-F_{R \mid X}(r\mid \cdot) \leq \exp(-r C); \ \ r>0; 
$$
Let $R_{j} = -M(\eps) - \eps/(2L) + j(\eps/L)$, with $j = 0,..., J$, where $J= \lceil 2 M(\eps) L/\eps \rceil +1 $,  $M(\eps) = \log (1/\eps)/C$ and $\lceil \cdot \rceil$ is the ceiling function. Let $R_{-1} = -\infty$ and $R_{J+1}=+ \infty$. The sets $B_j = \{ F_{R \mid X} (r \mid X):  R_{j} \leq  r \leq R_{j+1} \}$
for $j \in \{-1,..., J\}$ have the $L^2$ diameter of at most $\eps$ independently of the distribution of $F_X$:

\begin{itemize}
\item Indeed by the previous paragraph, if $j \in \{0,..., J-1\}$, then the diameter of the set $B_j$ is at most $ L (\eps/L) = \eps$.
\item For $j = -1$ or $J$, then any pair of conditional cdfs in the same ball obey:
$$
|F_{R_{} \mid X_{}}(r \mid \cdot) - F_{R_{} \mid X_{}}(r^{\prime} \mid \cdot)| \leq \exp (- M(\eps)C) = \exp ( -[\log (1/\eps)/C] C ) \leq \epsilon.
$$
\end{itemize}

The number of sets is at most $2 \log (1/\eps)L/(C\eps) + 5$. It follows that the uniform covering number of the function set $\mathcal{F}_R = \{F_{R \mid X}(r\mid X): r \in \mathcal{R}\}$
is bounded by $(1+ (1/\eps)^{2})$ up to a constant that does not depend on the distribution of $X$. Hence $\mathcal{F}_R $ is DKP, $R \in \{Y,W\}$

% Further if we take $\mathcal{F}$ as generated by union of products of $\mathcal{F}_R$ over different labels $R$, and the union of rectangles, the resulting set is still a DKP class, by standard uniform covering entropy calculus. \qed

%\newpage
%\textsc{Step 3.} (Auxiliary: Donskerness). One key ingredient for the result is to show that $\mathcal{F}$ is a Dudley-Koltchinskii-Pollard (DKP) class, namely it has bounded uniform covering entropy integral and obeys standard measurability condition (Dudley's image-admissible Suslin condition). 
%We omit any discussion of measurability in this paper, but we note that it trivially holds. 

\paragraph{\textbf{Covariate process $\hat G_Z$}} Here we show that the class $\mathcal{F}$ defined before Lemma \ref{lemma:crrr} is DKP. Let
$$
a(z) := F_{Y\mid X}(y\mid x)-.5,
\qquad
b(z) := F_{W\mid X}(w\mid x)-.5,
\qquad z=(y,w,x),
$$
and let $\mathcal I$ denote the class of rectangle indicators on the
joint domain. By definition,
$$
\mathcal F
=
\bigcup_{g\in\mathcal G}g\mathcal I,
\qquad
\mathcal G:=\{1,a,b,a^2,ab,b^2\},
\qquad
g\mathcal I:=\{g1_A:1_A\in\mathcal I\}.
$$
Each multiplier $g\in\mathcal G$ is fixed, Borel measurable, and bounded
in absolute value by one. For every probability measure $Q$ and every
pair of rectangles $A,B$,
$$
\|g1_A-g1_B\|_{Q,2}
\leq
\|g\|_\infty\|1_A-1_B\|_{Q,2}
\leq
\|1_A-1_B\|_{Q,2}.
$$
Consequently,
$$
N(\epsilon,\mathcal F,L^2(Q))
\leq
\sum_{g\in\mathcal G}N(\epsilon,g\mathcal I,L^2(Q))
\leq
6N(\epsilon,\mathcal I,L^2(Q)).
$$
Since $\mathcal I$ is a VC class, there exist constants $C\geq1$ and
$v<\infty$, independent of $Q$, such that
$$
\sup_Q N(\epsilon,\mathcal F,L^2(Q))
\leq 6C\epsilon^{-v},
\qquad 0<\epsilon<1.
$$
It follows that
$$
\int_0^1
\sqrt{\log\sup_Q N(\epsilon,\mathcal F,L^2(Q))}
\,d\epsilon
\leq
\int_0^1
\sqrt{\log(6C)+v\log(1/\epsilon)}
\,d\epsilon
<\infty.
$$
%Moreover, $\mathcal F$ has envelope one and satisfies the
%image-admissible Suslin condition: rectangle indicators admit a jointly
%Borel measurable parametrization by their endpoints and
%endpoint-inclusion conventions, and this property is preserved by
%multiplication by the fixed Borel functions in $\mathcal G$ and by
%finite unions. 
Hence $\mathcal F$ is a 
DKP class.\qed

% \textsc{Step 4. }(Auxiliary: Verification of Conditions of Lemma E.1 in \cite{chernozhukov+13inference}). We
% verify conditions (a)-(c) of Lemma Lemma E.1 in \cite{chernozhukov+13inference}. Conditions (a) and
% (b) are immediate by the assumptions. To verify (c), a straightforward
% computation gives that for $(b, y)$ in the neighborhood of $(\beta(y), y)$, $%
% \frac{\partial}{\partial(b^{\prime}, y)} \Psi(b, y) = [J(b, y), R(b, y)], $
% where, for $H(z) = \lambda(z)/\{\Lambda(z)[1 - \Lambda(z)]\}$ and $h(z) =
% dH(z)/dz$, 
% \begin{equation*}
% J(b, y) := E \left[ \{h(X^{\prime}b) [\Lambda(X^{\prime}b) - 1 (Y \leq y) ]
% + H(X^{\prime}b) \lambda(X^{\prime}b) \} XX^{\prime}\right] ,
% \end{equation*}
% and $R(b, y) = - E \left[ H(X^{\prime}b) f_{Y|X}(y|X)X \right] .$ Both terms
% are continuous in $(b, y)$ at $(\beta(y), y)$ for each $y \in\mathcal{Y}$.
% The computation above as well as the verification of continuity follows from
% using the dominated convergence theorem, and the following ingredients: (1)
% a.s. continuity of the map $(b, y) \mapsto\frac{\partial}{\partial b^{\prime}%
% } \varphi_{b, y}(Y,X)$, (2) domination of $\| \frac{\partial}{\partial
% b^{\prime}} \varphi_{b,y}(X,Y)\|$ by a square-integrable function $\text{%
% const} \|X\|$, (3) a.s. continuity and uniform boundedness of the
% conditional density function $y \mapsto f_{Y|X}(y|X)$, and (4) $H(X^{\prime
% }b)$ being bounded uniformly on $b \in\mathbb{R}^{d_x}$, a.s. By assumption $%
% J(y) = J(\beta(y), y) $ is positive-definite uniformly in $y \in{\mathcal{Y}}
% $. \qed

\subsection{Proof of Theorem \ref{theorem:main}} Part (1) follows by the functional delta method (see, e.g., Lemma B.1 of \cite{chernozhukov+13inference}). Indeed, in the notation of Lemma \ref{lemma: Hadamard dif of CRRR COR}, 
\begin{multline*}
    \tilde \rho_{C} = \phi(\hat F_{Y \mid X}-.5, \hat F_{W \mid X}-.5, \hat F_{Z}) \\ = \frac{ \int [\hat F_{Y \mid X}(y \mid x) - .5] [\hat F_{W \mid X}(w \mid x) - .5] \dd
\hat F_{Z}(y,w,x)}{\sqrt{\int   [\hat F_{W \mid X}(w \mid x) - .5]^2 \dd
\hat F_{Z}(y,w,x) \int   [\hat F_{Y \mid X}(y \mid x) - .5]^2 \dd
\hat F_{Z}(y,w,x)}}, 
\end{multline*}
and $\rho_{C} = \phi(F_{Y \mid X}-.5, F_{W \mid X}-.5, F_{Z})$. By Lemmas \ref{lemma: Hadamard dif of CRRR COR} and \ref{lemma:crrr}, together with the functional delta method,
$$
\sqrt{n}(\tilde \rho_C - \rho_C) \rightsquigarrow \phi'_{F_{Y\mid X}, F_{W\mid X}, F_Z}(Z_Y,Z_W,G_Z) \text{ in } \R.
$$
Similarly, in the notation of Lemma \ref{lemma: Hadamard dif of CRRR FR}, 
$$
\breve \rho_{C} = 12 \phi_1(\hat F_{Y \mid X}-.5, \hat F_{W \mid X}-.5, \hat F_{Z}) =  12 \int [\hat F_{Y \mid X}(y \mid x) - .5] [\hat F_{W \mid X}(w \mid x) - .5] \dd
\hat F_{Z}(y,w,x), 
$$ and $\rho_{C} = 12 \phi_1(F_{Y \mid X}-.5, F_{W \mid X}-.5, F_{Z})$. By Lemmas \ref{lemma: Hadamard dif of CRRR FR} and \ref{lemma:crrr}, together with the functional delta method,
$$
\sqrt{n}(\breve \rho_C - \rho_C) \rightsquigarrow 12 \ \phi'_{1,F_{Y\mid X}, F_{W\mid X}, F_Z}(Z_Y,Z_W,G_Z) \text{ in } \R.
$$

To show part (2), write:
$$
\frac{1}{n} \sum_{i=1}^n (\hat V_i - \overline{\hat V})^2  = \frac{1}{n}  \sum_{i=1}^n (\hat V_i - .5)^2  - (\overline{\hat V} - .5)^2,  \quad \overline{\hat V} = \frac{1}{n} \sum_{i=1}^n \hat V_i, 
$$
$$
\frac{1}{n} \sum_{i=1}^n (\hat U_i - \overline{\hat U})^2  = \frac{1}{n}  \sum_{i=1}^n (\hat U_i - .5)^2  - (\overline{\hat U} - .5)^2, \quad \overline{\hat U} = \frac{1}{n} \sum_{i=1}^n \hat U_i,
$$
and
$$
\frac{1}{n} \sum_{i=1}^n \hat U_i (\hat V_i - \overline{\hat V})  = \frac{1}{n}  \sum_{i=1}^n (\hat U_i - .5) (\hat V_i - .5)  - (\overline{\hat U} - .5)(\overline{\hat V} - .5).
$$
% $$
% \sqrt{n}(\tilde \rho_C - \hat \rho_C) = \frac{\sqrt{n}(\overline{\hat U} - .5)(\overline{\hat V} - .5)}{\sum_{i=1}^n (\hat V_i - \overline{\hat V})^2/n } - \tilde \rho_C \frac{\sqrt{n}(\overline{\hat V} - .5)^2}{\sum_{i=1}^n (\hat V_i - \overline{\hat V})^2/n }, \quad \overline{\hat V} = \frac{1}{n} \sum_{i=1}^n \hat V_i, \quad \overline{\hat U} = \frac{1}{n} \sum_{i=1}^n \hat U_i.
% $$

To analyze the second components of the previous expressions, note that
$$
|\overline{\hat V} - .5| \leq \frac{1}{n} \sum_{i=1}^n |\hat V_i - V_i| + \left|\frac{1}{n} \sum_{i=1}^n (V_i - .5)  \right| = O_{\Pr}(n^{-1/2}), \quad V_i = F_{W \mid X}(W_i \mid X_i), 
$$
by the central limit theorem applied to the second term and
\begin{equation}\label{eq:uconsist-v}
\max_{i \in \{1,\ldots,n\}} |\hat V_i - V_i| = \max_{i \in \{1,\ldots,n\}} |\hat F_{W \mid X}(W_i \mid X_i) - F_{W \mid X}(W_i \mid X_i)| \leq \|\hat F_{W \mid X} - F_{W \mid X}\|_{\mathcal{WX}}  = O_{\Pr}(n^{-1/2}), 
\end{equation}
where the last equality follows by Lemma \ref{lemma:crrr}.  A similar argument gives 
$
|\overline{\hat U} - .5| = O_{\Pr}(n^{-1/2}),
$
and
\begin{equation}\label{eq:uconsist-u}
\max_{i \in \{1,\ldots,n\}} |\hat U_i - U_i| =  O_{\Pr}(n^{-1/2}),  \quad U_i = F_{Y \mid X}(Y_i \mid X_i).
\end{equation}

% Similarly, we can show
% $$
% \frac{1}{n} \sum_{i=1}^n (\hat U_i -.5)(\hat V_i  -.5) = \frac{1}{n} \sum_{i=1}^n (U_i -.5)(V_i  -.5)+ o_{\Pr}(1) \text{ and } \frac{1}{n} \sum_{i=1}^n (\hat V_i  -.5)^2 = \frac{1}{n} \sum_{i=1}^n (V_i  -.5)^2 + o_{\Pr}(1),
% $$
% using  \eqref{eq:uconsist-v} and  \eqref{eq:uconsist-u}. Then, 
% $$
% \tilde \rho_C = \frac{\sum_{i=1}^n (U_i -.5)(V_i  -.5)/n}{\sum_{i=1}^n (V_i  -.5)^2/n} + o_{\Pr}(1) = \rho_C + o_{\Pr}(1),
% $$
% by the law of large numbers and the continuous mapping theorem. 

Combining the previous results with the continuous mapping theorem, and using a mean value expansion, we conclude that:
$$
\hat \rho_C = \frac{\frac{1}{n}  \sum_{i=1}^n (\hat U_i - .5) (\hat V_i - .5)  + O_{\Pr}(n^{-1})}{\sqrt{\left[\frac{1}{n}  \sum_{i=1}^n (\hat V_i - .5)^2 + O_{\Pr}(n^{-1})\right]\left[\frac{1}{n}  \sum_{i=1}^n (\hat U_i - .5)^2 + O_{\Pr}(n^{-1})\right]}} = \tilde \rho_C + O_{\Pr}(n^{-1}).
$$ \qed

\subsection{Proof of Theorem \ref{thm:main-boot}} The results for $\tilde \rho_C^*$ and $\breve \rho_C^*$ follow by Lemmas \ref{lemma: Hadamard dif of CRRR COR}, \ref{lemma: Hadamard dif of CRRR FR} and \ref{lemma:crrr-boot}, together with the functional delta method for bootstrap (see, e.g., Lemma B.3 of \cite{chernozhukov+13inference}). 

The result for  $\hat \rho_C^*$ follows from:
$$
\sqrt{n} (\hat \rho_C^* - \hat \rho_C) = \sqrt{n} (\tilde \rho_C^* - \tilde \rho_C)   + \sqrt{n} (\tilde \rho_C - \hat \rho_C)+ \sqrt{n} (\hat \rho_C^* - \tilde \rho_C^*) = \sqrt{n} (\tilde \rho_C^* - \tilde \rho_C) + o_{\Pr}(1),
$$
because $\sqrt{n} (\tilde \rho_C - \hat \rho_C) = o_{\Pr}(1)$ by part (2) of Theorem \ref{theorem:main}, and $\sqrt{n} (\hat \rho_C^* - \tilde \rho_C^*) = o_{\Pr}(1)$ by the same argument as in the proof of part (2) of Theorem \ref{theorem:main} replacing Lemma \ref{lemma:crrr} by Lemma \ref{lemma:crrr-boot}.%[VICTOR: THIS ARGUMENT NEEDS TO BE CHECKED]

\section{Simulations}\label{sec:simul}
We show that the asymptotic theory provides a good approximation to the behavior of the CRRR estimator through a small sample Monte Carlo simulation. In particular, we document that the CRRR estimator converges at the expected rate and that the corresponding confidence interval has coverage close to its nominal level. We focus on the correlation-based estimator, but we find very similar performance for the regression-based and fully-restricted estimators in results not reported.\footnote{The results for the regression-based and fully-restricted estimators are available from the authors upon request.} 

% We do not impose any restriction at the tails because we obtain good results for the unrestricted estimator even for small sample sizes. In preliminary results not reported, we find similar results imposing the restrictions of the tail model.

\medskip

We consider a bivariate normal design for analytical convenience. In particular, we draw data from the process:\footnote{Note that the distribution of $X$ does not satisfy the compact support assumption of Lemma \ref{lemma:crrr}.}
\begin{equation}\label{eq:mc_design}
\left(\begin{array}{c}
     Y  \\
     W 
\end{array}\right) \mid X = x \sim N_2 \left( \left(\begin{array}{c}
     x \\
     x 
\end{array}\right),  \left(\begin{array}{cc}
     1 & c  \\
     c & 1 
\end{array}\right)\right), \quad X \sim N(0,1).
\end{equation}
We consider 3 different values for the correlation parameter $c \in \{0.25,0.50,0.75\}$ and 3 sample sizes $n \in \{625, 2500, 10000\}$. We choose these sample sizes because they correspond to $\sqrt{n} \in \{25,50,100\}$, where $\sqrt{n}$ is the theoretical rate of convergence of the CRRR estimator. The true value of the CRRR slope is obtained from $c$ using the expression of the rank correlation of the bivariate normal, $\rho_C = 6 \arcsin(c/2)/\pi$ \citep[e.g.,][]{cramer1999mathematical}.

Table \ref{table:simulation1} shows the result from $2,000$ simulations using the correlation-based estimator of Algorithm \ref{alg:crrr-reg} with Gaussian or probit link function, a mesh of 500 points located at sample quantiles in a sequence of orders from 0.005 to 0.995 with increments of 0.99/499, and $m=30$ to estimate the tail parameters. We report root mean squared error (RMSE), bias, standard deviation (SD) and coverage of 95\% confidence intervals (Cover.). The confidence intervals are obtained from Algorithm \ref{alg:infer} by empirical bootstrap with 200 repetitions. The results indicate that (i) the estimator converges at the expected rate of $\sqrt{n}$ and (ii) the coverage of the confidence intervals floats around the nominal level of $.95$.\footnote{Note that the simulation standard error for coverage is about $0.5\%$.} 

\begin{table}[ht]
\begin{center}
\begin{threeparttable}[b]
\setlength{\tabcolsep}{0pt}
\caption{Properties of CRRR Estimator ($m=30$)} \label{table:simulation1} 
\begin{tabular*}{9cm}{ @{\extracolsep{\fill}}crccccc} %
\toprule\toprule 
$c$ & \multicolumn{1}{c}{$n$} & RMSE & Bias & SD & Cover.   \\ 
\midrule 
0.25 & 625 & 0.038 & 0.002 & 0.038 & 0.94 \\ 
 & 2,500 & 0.020 & 0.001 & 0.020 & 0.94 \\ 
 & 10,000 & 0.009 & 0.001 & 0.009 & 0.94 \medskip \\ 
0.5 & 625 & 0.032 & 0.003 & 0.032 & 0.95 \\ 
 & 2,500 & 0.016 & 0.001 & 0.016 & 0.94 \\ 
 & 10,000 & 0.008 & 0.001 & 0.008 & 0.95 \medskip \\ 
0.75 & 625 & 0.021 & 0.005 & 0.020 & 0.96 \\ 
 & 2,500 & 0.010 & 0.001 & 0.010 & 0.95 \\ 
 & 10,000 & 0.005 & 0.001 & 0.005 & 0.95 \\ 
\bottomrule\bottomrule

\end{tabular*}
\begin{tablenotes}[flushleft]
\tiny 
\item {\textit{Notes: Results based on 2,000 simulations of the DGP in \eqref{eq:mc_design} for the correlation-based estimator with probit link function and a mesh of 500 points. The nominal level for coverage is $0.95$.} } 
\end{tablenotes}
\end{threeparttable}
\end{center}
\end{table}

Table \ref{table:simulation2} reports results using the same design as in table \ref{table:simulation1} but with $m=0$, that is, without imposing any restriction at the tails. We find that the results are not sensitive to the treatment of the tails, even for the smallest sample size.

\begin{table}[ht]
\begin{center}
\begin{threeparttable}[b]
\setlength{\tabcolsep}{0pt}
\caption{Properties of CRRR Estimator ($m=0$)} \label{table:simulation2} 
\begin{tabular*}{9cm}{ @{\extracolsep{\fill}}crccccc} %
\toprule\toprule 
$c$ & \multicolumn{1}{c}{$n$} & RMSE & Bias & SD & Cover.   \\ 
\midrule 
0.25 & 625 & 0.038 & 0.002 & 0.038 & 0.95 \\ 
 & 2,500 & 0.020 & 0.001 & 0.020 & 0.94 \\ 
 & 10,000 & 0.009 & 0.001 & 0.009 & 0.94 \medskip \\ 
0.5 & 625 & 0.032 & 0.004 & 0.032 & 0.95 \\ 
 & 2,500 & 0.016 & 0.001 & 0.016 & 0.94 \\ 
 & 10,000 & 0.008 & 0.001 & 0.008 & 0.95 \medskip \\ 
0.75 & 625 & 0.021 & 0.005 & 0.020 & 0.96 \\ 
 & 2,500 & 0.010 & 0.001 & 0.010 & 0.95 \\ 
 & 10,000 & 0.005 & 0.001 & 0.005 & 0.95 \\ 
\bottomrule\bottomrule

\end{tabular*}
\begin{tablenotes}[flushleft]
\tiny 
\item {\textit{Notes: Results based on 2,000 simulations of the DGP in \eqref{eq:mc_design} for the correlation-based estimator with probit link function and a mesh of 500 points. The nominal level for coverage is $0.95$.} } 
\end{tablenotes}
\end{threeparttable}
\end{center}
\end{table}

\section{Additional Empirical Results}\label{app:empirics}
We repeat the analysis using marginal child's ranks (CRRR') instead of conditional child's ranks (CRRR), and interpret the results accordingly. 

\subsection{Rank-Rank Regressions} 
Table \ref{table:crrr'} reports the results of RRR and CRRR'. The CRRR' results are obtained using the regression-based estimator with a logistic link function and a mesh of 200 points located at sample quantiles in a sequence of orders from $0.01$ to $0.99$ with increments of $0.98/199$. We use linear interpolation to obtain estimates of the conditional father's ranks corresponding to intermediate points in the mesh. The standard errors (SE) and 95\% confidence intervals (95\% CI) are computed by empirical bootstrap with 500 repetitions. We show the robustness of the results to the choice of  link function in Section \ref{subsec:probit}.

We find significant positive income persistence in both father-son and father-daughter relationships, with and without covariates. However, the persistence is much stronger for sons than for daughters suggesting the presence of a gender gap in intergenerational transmission of income  even after controlling for the father's and child's characteristics. 
% To compare RRR and CRRR, it is useful to decompose the overall income persistence ($\rho$) as follows:
% $$
% \rho = \rho_C + (\rho - \rho_C),
% $$
% where $\rho_C$ represents the average \textit{within-group} or unexplained income persistence, and the remainder term $(\rho - \rho_C)$ can be interpreted as \textit{between-group} income persistence or income persistence explained by covariates. Thus, the overall income persistence is equal to the sum of the average within-group income persistence and the between-group income persistence. 
Comparing RRR and CRRR',  we find that covariates account for 44\% of the overall income persistence for sons and about 57\% for daughters. These results highlight the substantial role of both father's income and covariate differences in explaining unconditional intergenerational mobility. The results of the subgroup analysis are similar to the conditional case.

\begin{table}[h!]
\begin{center}
\begin{threeparttable}[b]
\setlength{\tabcolsep}{0pt}
\caption{Intergenerational Income Mobility in Switzerland}\label{table:crrr'} 
\begin{tabular*}{15cm}{ @{\extracolsep{\fill}} lcccccccc} %
\toprule\toprule
&  \multicolumn{4}{c}{Father-Son}  & \multicolumn{4}{c}{Father-Daughter}  \\
\cmidrule{2-5}  \cmidrule{6-9} 
 & Coef. & SE & \multicolumn{2}{c}{95\% CI} & Coef. & SE & \multicolumn{2}{c}{95\% CI}  \\
\midrule
%& \multicolumn{8}{c}{Probit} \\
\input{"Tables/marg_rank_table_l1"}
\multicolumn{9}{l}{CRRR', by Father:}\\ 
\input{"Tables/marg_rank_table_l2"}
 % \\

\bottomrule\bottomrule

\end{tabular*}
\begin{tablenotes}[flushleft]
\tiny 
\item \textit{Notes: Regression-based estimator with logistic link function  and a mesh of 200 points.  SE and 95\% CI obtained by empirical bootstrap with 500 repetitions. Covariates include father's and child's months of experience, higher education, Swiss citizenship, single and number of own children; father's age at birth; and child's year and canton fixed effects. Sample size is $10,363$ for father-son  and $9,581$ for  father-daughter data.}
\end{tablenotes}
\end{threeparttable}
\end{center}
\end{table}

\subsection{Transition Matrices} 
\begin{figure}[h]
\caption{Transition matrices: Father-Son} \label{fig:tm-fs'}
\vspace{.2cm}
\begin{scriptsize}
\begin{center}
	\begin{subfigure}{0.5\textwidth}
		\centering
		\includegraphics[scale=0.43,keepaspectratio]{Figures/pmf_percent_fs.pdf}
		\caption{Marginal father's ranks}
	\end{subfigure}%
	\begin{subfigure}{0.5\textwidth}
		\centering
		\includegraphics[scale=0.43,keepaspectratio]{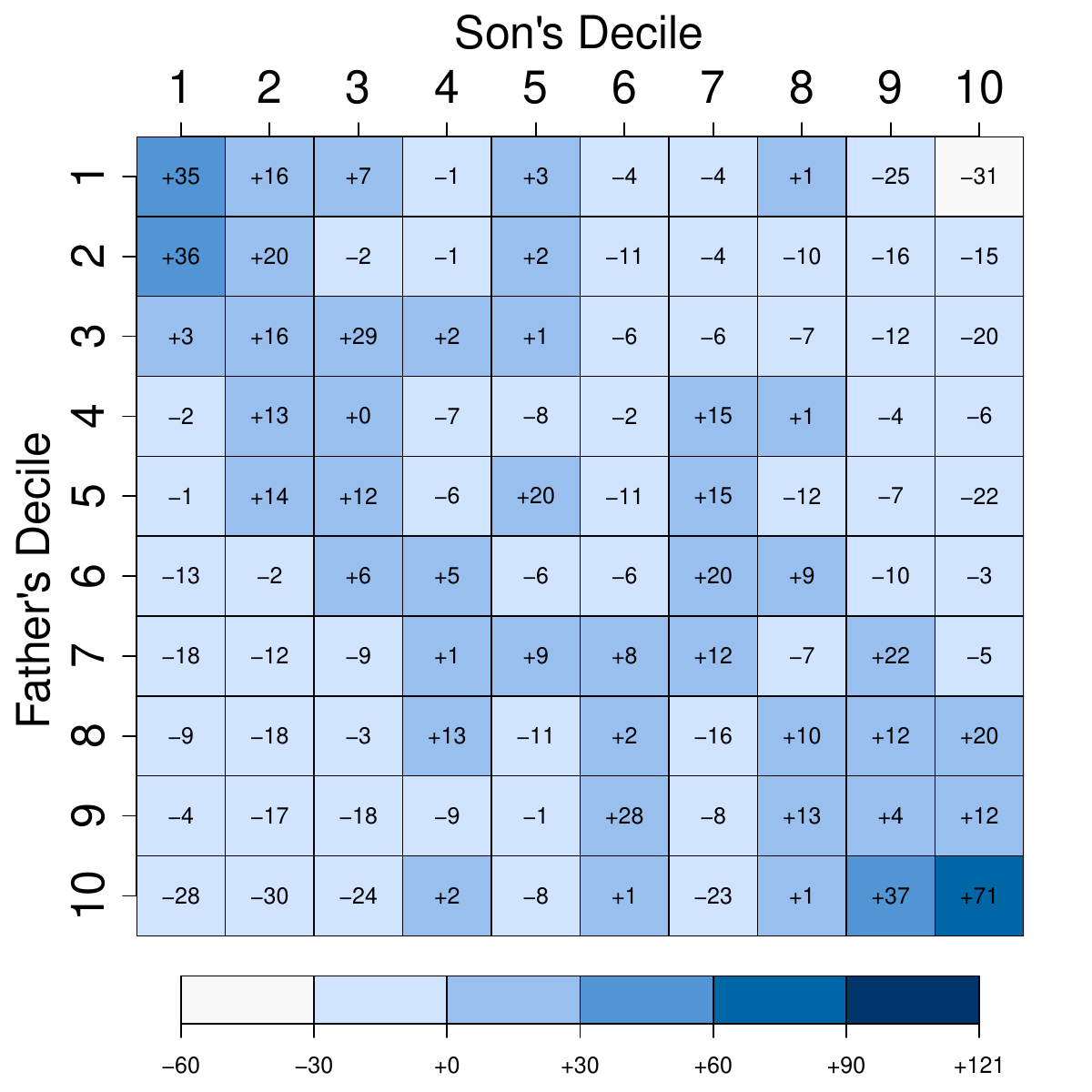}
		\caption{Conditional father's ranks}
	\end{subfigure} 
	\caption*{\footnotesize \textit{Notes:  Entries are in percent deviations from $0.01$. Sample size is $10,363$ for father-son and $9,581$ for father-daughter.} }
\end{center}
\end{scriptsize}
\end{figure}

\begin{figure}[h]
\caption{Transition matrices: Father-Daughter} \label{fig:tm-fd'}
\vspace{.2cm}
\begin{scriptsize}
\begin{center}
	\begin{subfigure}{0.5\textwidth}
		\centering
		\includegraphics[scale=0.43,keepaspectratio]{Figures/pmf_percent_fd.pdf}
		\caption{Marginal father's ranks}
	\end{subfigure}%
	\begin{subfigure}{0.5\textwidth}
		\centering
		\includegraphics[scale=0.43,keepaspectratio]{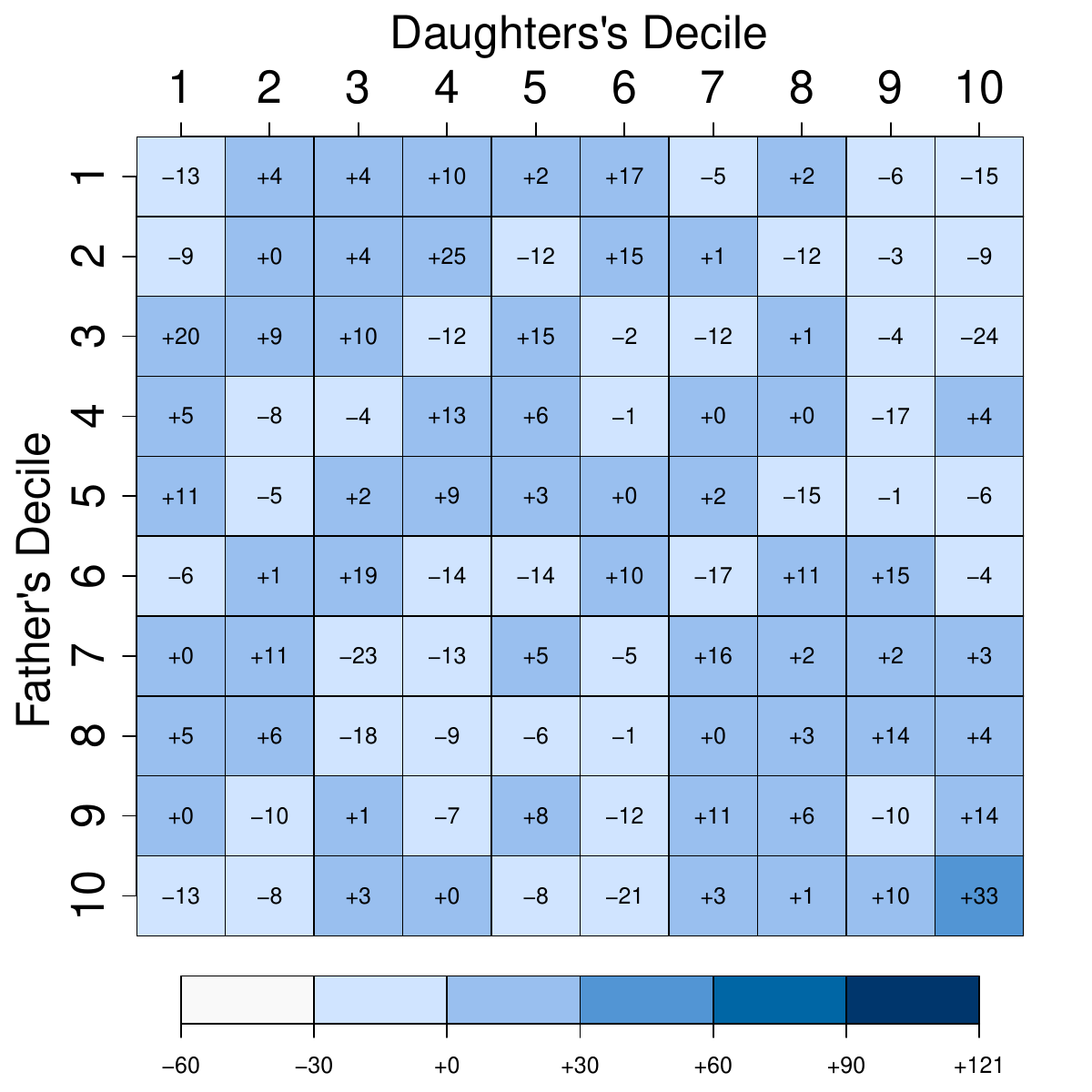}
		\caption{Conditional father's ranks}
	\end{subfigure} 
	\caption*{\footnotesize \textit{Notes:  Entries are in percent deviations from $0.01$. Sample size is $10,363$ for father-son and $9,581$ for father-daughter.} }
\end{center}
\end{scriptsize}
\end{figure}

Figures \ref{fig:tm-fs'} and \ref{fig:tm-fd'} show heatmaps of transition matrices for father-son and father-daughter, respectively.  We report all the entries in percent deviations from $0.01$ because all the entries should be equal to $0.01$ under perfect mobility, that is when the income of the child is independent of the income of the father. Panels (A) report transition matrices based on marginal father's ranks, similar to previous studies. Panels (B) report conditional transition matrices based on conditional father's ranks, which are new to this paper. For father-son, we find that the highest values in panel (A) are concentrated on the diagonal, which is consistent with the positive RRR estimate in Table \ref{table:crrr'}. The results in panel (B) show a less clear pattern once we control for covariates, consistent with the lower CRRR' estimates in Table \ref{table:crrr'}.  The results for father-daughter show similar but weaker patterns as we expect from the smaller correlation estimates in Table \ref{table:crrr'}. Interestingly, for both sons and daughters the highest probability occurs at the bottom right corner of the very top deciles with and without accounting for covariates.

% \begin{figure}[h]\caption{Transition matrices: Father-Son} \label{fig:tm-fs}
% \vspace{.2cm}
% \begin{scriptsize}
% \begin{center}
% 	\begin{subfigure}{0.5\textwidth}
% 		\centering
% 		\includegraphics[scale=0.43,keepaspectratio]{Figures/pmf_percent_fs.pdf}
% 		\caption{Marginal ranks}
% 	\end{subfigure}%
% 	\begin{subfigure}{0.5\textwidth}
% 		\centering
% 		\includegraphics[scale=0.43,keepaspectratio]{Figures/pmf_percent_cond_fs.pdf}
% 		\caption{Conditional ranks}
% 	\end{subfigure} 
% \caption*{\footnotesize \textit{Notes: entries are in percent deviations from $0.01$. Sample size is $10,363$ for father-son and $9,581$ for father-daughter. } }
% \end{center}
% \end{scriptsize}
% \end{figure}

% \begin{figure}[h]
% \caption{Transition matrices: Father-Daughter} \label{fig:tm-fd}
% \vspace{.2cm}
% \begin{scriptsize}
% \begin{center}
% 	\begin{subfigure}{0.5\textwidth}
% 		\centering
% 		\includegraphics[scale=0.43,keepaspectratio]{Figures/pmf_percent_fd.pdf}
% 		\caption{Marginal ranks}
% 	\end{subfigure}%
% 	\begin{subfigure}{0.5\textwidth}
% 		\centering
% 		\includegraphics[scale=0.43,keepaspectratio]{Figures/pmf_percent_cond_fd.pdf}
% 		\caption{Conditional ranks}
% 	\end{subfigure} 
% 	\caption*{\footnotesize \textit{Notes:  entries are in percent deviations from $0.01$. Sample size is $10,363$ for father-son and $9,581$ for father-daughter.} }
% \end{center}
% \end{scriptsize}
% \end{figure}

% \begin{figure}[h]
% \caption{Conditional Density} \label{f:cpdf}
% \vspace{.2cm}
% \begin{scriptsize}
% \begin{center}
% 	\begin{subfigure}{0.45\textwidth}
% 		\centering
% 		\includegraphics[scale=0.33,keepaspectratio]{Figures/cpdf_fs.pdf}
% 		\caption{Sons}
% 	\end{subfigure}%
% 	\begin{subfigure}{0.45\textwidth}
% 		\centering
% 		\includegraphics[scale=0.33,keepaspectratio]{Figures/cpdf_fd.pdf}
% 		\caption{Daughters}
% 	\end{subfigure} 
% 	\caption*{\footnotesize \textit{Notes:} }
% \end{center}
% \end{scriptsize}
% \end{figure}

% \begin{figure}[h]
% \begin{scriptsize}
% \begin{center}
% 	\begin{subfigure}{0.5\textwidth}
% 		\centering
% 		\includegraphics[scale=0.6,keepaspectratio]{Figures/bpdf_fs.pdf}
% 		\caption{Sons}
% 	\end{subfigure}%
% 	\begin{subfigure}{0.5\textwidth}
% 		\centering
% 		\includegraphics[scale=0.6,keepaspectratio]{Figures/bpdf_fd.pdf}
% 		\caption{Daughters}
% 	\end{subfigure} 
% 	\vspace{-8.5cm}
% 	\caption{Bivariate PDF (kernel)} \label{f:bpdf}
% 	\vspace{7.5cm}
% 	\caption*{\footnotesize \textit{Notes:} }
% \end{center}
% \end{scriptsize}
% \end{figure}

\subsection{Rank-Rank Regressions Excluding Child's Covariates} One concern about the CRRR' results in Table \ref{table:crrr'} is that the child's covariates might be picking up indirect sources of intergenerational mobility of income. For example, fathers might invest in child's education to increase the child's income prospects. To deal with this concern, Table \ref{table:nochild'} reports CRRR' results where the child's covariates, other than year and canton fixed effects, are excluded from the covariate set $X$. These results are obtained using the regression-based estimator with a logistic link function with the same parameter choices as in Table \ref{table:crrr'}.

As expected, not accounting for the child's covariates reduces the importance of the covariates with their effect on persistence estimates decreasing to about 21\% for father-son and 35\% for father-daughter. In both cases the decrease is about 22-23 percentage points. The other conclusions remain unchanged. In particular, we still find a significant gender gap in intergenerational transmission of income, and relatively less persistence for sons in large families and for daughters of high educated fathers. 

% \begin{table}[h!]
% \begin{center}
% \begin{threeparttable}[b]
% \setlength{\tabcolsep}{0pt}
% \caption{Estimates Excluding Child's Covariates} \label{table:nochild}
% \begin{tabular*}{15cm}{ @{\extracolsep{\fill}} lcccccccc} %
% \toprule\toprule
% &  \multicolumn{4}{c}{Father-Son}  & \multicolumn{4}{c}{Father-Daughter}  \\
% \cmidrule{2-5}  \cmidrule{6-9} 
%  & Coef. & SE & \multicolumn{2}{c}{95\% CI} & Coef. & SE & \multicolumn{2}{c}{95\% CI}  \\
% \midrule
% % & \multicolumn{8}{c}{Probit} \\
% % \primitiveinput{"Tables/crrr_p_noson1"}
% % \multicolumn{9}{l}{CRRR, by Father:}\\ 
% % \primitiveinput{"Tables/crrr_p_noson_2"}
% %  \\

% % & \multicolumn{8}{c}{Logit} \\
% \primitiveinput{"Tables/crrr_l_cor_200_nochild1"}
% \multicolumn{9}{l}{CRRR, by Father:}\\ 
% \primitiveinput{"Tables/crrr_l_cor_200_nochild2"}
% \bottomrule\bottomrule

% \end{tabular*}
% \begin{tablenotes}[flushleft]
% \tiny 
% \item \textit{Notes: Correlation-based estimator with logistic link function  and a mesh of 200 points.  SE and 95\% CI obtained by empirical bootstrap with 500 repetitions. Covariates include father's months of experience, higher education, Swiss citizenship, single and number of own children; father's age  birth; and child's year and canton fixed effects. Sample size is $10,363$ for father-son  and $9,581$ for  father-daughter data.}
% \end{tablenotes}
% \end{threeparttable}
% \end{center}
% \end{table}

\begin{table}[h!]
\begin{center}
\begin{threeparttable}[b]
\setlength{\tabcolsep}{0pt}
\caption{Estimates Excluding Child's Covariates} \label{table:nochild'}
\begin{tabular*}{15cm}{ @{\extracolsep{\fill}} lcccccccc} %
\toprule\toprule
&  \multicolumn{4}{c}{Father-Son}  & \multicolumn{4}{c}{Father-Daughter}  \\
\cmidrule{2-5}  \cmidrule{6-9} 
 & Coef. & SE & \multicolumn{2}{c}{95\% CI} & Coef. & SE & \multicolumn{2}{c}{95\% CI}  \\
\midrule

% & \multicolumn{8}{c}{Logit} \\
\input{"Tables/marg_rank_table_nochild_l1"}
\multicolumn{9}{l}{CRRR', by Father:}\\ 
\input{"Tables/marg_rank_table_nochild_l2"}
 % \\
\bottomrule\bottomrule

\end{tabular*}
\begin{tablenotes}[flushleft]
\tiny 
\item \textit{Notes: Regression-based estimator with logistic link function  and a mesh of 200 points.  SE and 95\% CI obtained by empirical bootstrap with 500 repetitions. Covariates include father's months of experience, higher education, Swiss citizenship, single and number of own children; father's age at birth; and child's year and canton fixed effects. Sample size is $10,363$ for father-son  and $9,581$ for  father-daughter data.}
\end{tablenotes}
\end{threeparttable}
\end{center}
\end{table}

\subsection{Robustness to Link Function}\label{subsec:probit} Table \ref{table:crrr'-probit} reports the results of CRRR using the regression-based estimator with a Gaussian or probit link function. The estimates, standard errors and confidence intervals are almost identical to Table \ref{table:crrr} showing the robustness of the results to the use of the logistic versus Gaussian link functions.

% \begin{table}[h!]
% \begin{center}
% \begin{threeparttable}[b]
% \setlength{\tabcolsep}{0pt}
% \caption{Robustness to Link Function: Probit Estimates}\label{table:crrr-probit} 
% \begin{tabular*}{15cm}{ @{\extracolsep{\fill}} lcccccccc} %
% \toprule\toprule
% &  \multicolumn{4}{c}{Father-Son}  & \multicolumn{4}{c}{Father-Daughter}  \\
% \cmidrule{2-5}  \cmidrule{6-9} 
%  & Coef. & SE & \multicolumn{2}{c}{95\% CI} & Coef. & SE & \multicolumn{2}{c}{95\% CI}  \\
% \midrule
% %& \multicolumn{8}{c}{Probit} \\
% \primitiveinput{"Tables/crrr_p_cor_2001"}
% \multicolumn{9}{l}{CRRR, by Father:}\\ 
% \primitiveinput{"Tables/crrr_p_cor_2002"}
%  % \\

% % & \multicolumn{8}{c}{Logit} \\
% % \primitiveinput{"Tables/crrr_l_base1"}
% % \multicolumn{9}{l}{CRRR, by Father:}\\ 
% % \primitiveinput{"Tables/crrr_l_base2"}
% \bottomrule\bottomrule

% \end{tabular*}
% \begin{tablenotes}[flushleft]
% \tiny 
% \item \textit{Notes: Correlation-based estimator with Gaussian link function  and a mesh of 200 points.  SE and 95\% CI obtained by empirical bootstrap with 500 repetitions. Covariates include father's and child's months of experience, higher education, Swiss citizenship, single and number of own children; father's age  birth; and child's year and canton fixed effects. Sample size is $10,363$ for father-son  and $9,581$ for  father-daughter data.}
% \end{tablenotes}
% \end{threeparttable}
% \end{center}
% \end{table}

\begin{table}[h!]
\begin{center}
\begin{threeparttable}[b]
\setlength{\tabcolsep}{0pt}
\caption{Robustness to Link Function: Probit Estimates}\label{table:crrr'-probit} 
\begin{tabular*}{15cm}{ @{\extracolsep{\fill}} lcccccccc} %
\toprule\toprule
&  \multicolumn{4}{c}{Father-Son}  & \multicolumn{4}{c}{Father-Daughter}  \\
\cmidrule{2-5}  \cmidrule{6-9} 
 & Coef. & SE & \multicolumn{2}{c}{95\% CI} & Coef. & SE & \multicolumn{2}{c}{95\% CI}  \\
\midrule
%& \multicolumn{8}{c}{Probit} \\
\input{"Tables/marg_rank_table_p1"}
\multicolumn{9}{l}{CRRR', by Father:}\\ 
\input{"Tables/marg_rank_table_p2"}
 % \\

\bottomrule\bottomrule

\end{tabular*}
\begin{tablenotes}[flushleft]
\tiny 
\item \textit{Notes: Regression-based estimator with Gaussian link function  and a mesh of 200 points.  SE and 95\% CI obtained by empirical bootstrap with 500 repetitions. Covariates include father's and child's months of experience, higher education, Swiss citizenship, single and number of own children; father's age at birth; and child's year and canton fixed effects. Sample size is $10,363$ for father-son  and $9,581$ for  father-daughter data.}
\end{tablenotes}
\end{threeparttable}
\end{center}
\end{table}

\end{document}

%% file: Tables/crrr_l_cor_2001.tex
RRR & 0.202 & 0.010 & 0.182 & 0.222 & 0.088 & 0.010 & 0.069 & 0.107 \\ 
CRRR & 0.126 & 0.011 & 0.106 & 0.147 & 0.046 & 0.010 & 0.027 & 0.066 \\ 

%% file: Tables/crrr_l_cor_2002.tex
High education & 0.131 & 0.017 & 0.095 & 0.167 & 0.085 & 0.016 & 0.014 & 0.156 \\ 
Age \textgreater  26 at birth & 0.130 & 0.015 & 0.100 & 0.160 & 0.054 & 0.014 & 0.021 & 0.087 \\ 
Swiss citizen & 0.128 & 0.012 & 0.105 & 0.152 & 0.045 & 0.011 & 0.024 & 0.065 \\ 
More \textgreater  2 children & 0.109 & 0.016 & 0.070 & 0.149 & 0.044 & 0.016 & 0.012 & 0.077 \\ 

%% file: Tables/crrr_l_cor_200_nochild1.tex
RRR & 0.202 & 0.010 & 0.182 & 0.222 & 0.088 & 0.010 & 0.069 & 0.107 \\ 
CRRR & 0.161 & 0.010 & 0.140 & 0.182 & 0.061 & 0.010 & 0.041 & 0.081 \\ 

%% file: Tables/crrr_l_cor_200_nochild2.tex
High education & 0.166 & 0.016 & 0.132 & 0.200 & 0.089 & 0.015 & 0.031 & 0.147 \\ 
Age \textgreater  26 at birth & 0.164 & 0.014 & 0.135 & 0.194 & 0.077 & 0.013 & 0.038 & 0.116 \\ 
Swiss citizen & 0.164 & 0.012 & 0.140 & 0.187 & 0.060 & 0.010 & 0.039 & 0.081 \\ 
More \textgreater  2 children & 0.142 & 0.017 & 0.100 & 0.183 & 0.064 & 0.016 & 0.031 & 0.098 \\ 

%% file: Tables/crrr_p_cor_2001.tex
RRR & 0.202 & 0.010 & 0.182 & 0.222 & 0.088 & 0.010 & 0.069 & 0.107 \\ 
CRRR & 0.127 & 0.011 & 0.107 & 0.148 & 0.047 & 0.010 & 0.028 & 0.067 \\ 

%% file: Tables/crrr_p_cor_2002.tex
High education & 0.132 & 0.017 & 0.096 & 0.168 & 0.086 & 0.016 & 0.015 & 0.157 \\ 
Age \textgreater  26 at birth & 0.131 & 0.014 & 0.102 & 0.161 & 0.055 & 0.015 & 0.023 & 0.087 \\ 
Swiss citizen & 0.129 & 0.012 & 0.106 & 0.153 & 0.045 & 0.011 & 0.025 & 0.066 \\ 
More \textgreater  2 children & 0.110 & 0.016 & 0.071 & 0.150 & 0.046 & 0.016 & 0.014 & 0.079 \\ 

%% file: Tables/marg_rank_table_l1.tex
RRR & 0.202 & 0.010 & 0.182 & 0.222 & 0.088 & 0.010 & 0.069 & 0.107 \\ 
CRRR' &  0.114 &  0.009 &  0.095 &  0.132 &  0.038 &  0.008 &  0.023 &  0.054 \\ 

%% file: Tables/marg_rank_table_l2.tex
High education &  0.123 &  0.015 &  0.087 &  0.159 &  0.022 &  0.013 & -0.016 &  0.059 \\ 
Age \textgreater  26 at birth &  0.113 &  0.012 &  0.089 &  0.136 &  0.024 &  0.011 & -0.006 &  0.054 \\ 
Swiss citizen &  0.113 &  0.011 &  0.094 &  0.132 &  0.035 &  0.008 &  0.019 &  0.051 \\ 
More \textgreater  2 children &  0.099 &  0.014 &  0.067 &  0.132 &  0.040 &  0.013 &  0.015 &  0.066 \\ 

%% file: Tables/marg_rank_table_nochild_l1.tex
RRR & 0.202 & 0.010 & 0.182 & 0.222 & 0.088 & 0.010 & 0.069 & 0.107 \\ 
CRRR' & 0.160 & 0.010 & 0.139 & 0.181 & 0.057 & 0.010 & 0.037 & 0.077 \\ 

%% file: Tables/marg_rank_table_nochild_l2.tex
High education & 0.173 & 0.017 & 0.128 & 0.219 & 0.047 & 0.017 & 0.010 & 0.083 \\ 
Age \textgreater  26 at birth & 0.162 & 0.013 & 0.134 & 0.190 & 0.046 & 0.012 & 0.016 & 0.075 \\ 
Swiss citizen & 0.161 & 0.010 & 0.137 & 0.184 & 0.056 & 0.010 & 0.036 & 0.077 \\ 
More \textgreater  2 children & 0.141 & 0.016 & 0.103 & 0.179 & 0.064 & 0.016 & 0.031 & 0.097 \\ 

%% file: Tables/marg_rank_table_p1.tex
RRR & 0.202 & 0.010 & 0.182 & 0.222 & 0.088 & 0.010 & 0.069 & 0.107 \\ 
CRRR' &  0.114 &  0.009 &  0.096 &  0.131 &  0.038 &  0.008 &  0.023 &  0.053 \\ 

%% file: Tables/marg_rank_table_p2.tex
High education &  0.123 &  0.015 &  0.087 &  0.158 &  0.022 &  0.013 & -0.016 &  0.060 \\ 
Age \textgreater  26 at birth &  0.113 &  0.012 &  0.089 &  0.136 &  0.024 &  0.011 & -0.006 &  0.054 \\ 
Swiss citizen &  0.113 &  0.011 &  0.094 &  0.132 &  0.035 &  0.008 &  0.018 &  0.052 \\ 
More \textgreater  2 children &  0.099 &  0.014 &  0.066 &  0.133 &  0.040 &  0.013 &  0.015 &  0.066 \\ 